\documentclass[twocolumn,amsmath,aps,prd,preprintnumbers,amssymb,nofootinbib,superscriptaddress,showpacs,floatfix]{revtex4-1}
\usepackage{amssymb}
\usepackage{amsmath}
\usepackage{epsfig}
\usepackage{subfigure}
\usepackage{mathrsfs}
\usepackage{hyperref}
\usepackage{longtable}
\usepackage[usenames,dvipsnames]{xcolor}
\usepackage{mathtools}
\usepackage{relsize}

\begin{document}
 \renewcommand{\thefigure}{\arabic{figure}}
\newcommand{\noj}{}

\newcommand{\apjl}{Astrophys. J. Lett.}
\newcommand{\aap}{Astron. Astrophys.}
\newcommand{\apjs}{Astrophys. J. Suppl. Ser.}
\newcommand{\sa}{Sov. Astron. Lett.}.
\newcommand{\jpb}{J. Phys. B.}
\newcommand{\natu}{Nature (London)}
\newcommand{\aaps}{Astron. Astrophys. Supp. Ser.}
\newcommand{\aj}{Astron. J.}
\newcommand{\aas}{Bull. Am. Astron. Soc.}
\newcommand{\mnras}{Mon. Not. R. Astron. Soc.}
\newcommand{\pasp}{Publ. Astron. Soc. Pac.}
\newcommand{\jcap}{JCAP.}
\newcommand{\jmat}{J. Math. Phys.}
\newcommand{\prep}{Phys. Rep.}
\newcommand{\jtep}{Sov. Phys. JETP.}
\newcommand{\plb}{Phys. Lett. B.}
\newcommand{\pla}{Phys. Lett. A.}
\newcommand{\jhep}{Journal of High Energy Physics}

\newcommand{\be}{\begin{equation}}
\newcommand{\ee}{\end{equation}}
\newcommand{\bea}{\begin{align}}
\newcommand{\eea}{\end{align}}
\newcommand{\lsim}{\mathrel{\hbox{\rlap{\lower.55ex\hbox{$\sim$}} \kern-.3em \raise.4ex \hbox{$<$}}}}
\newcommand{\gsim}{\mathrel{\hbox{\rlap{\lower.55ex\hbox{$\sim$}} \kern-.3em \raise.4ex \hbox{$>$}}}}
\newcommand{\grad}{\ensuremath{\vec{\nabla}}}
\newcommand{\adotoa}{\ensuremath{{\cal H}}} 
\newcommand{\Uc}{\ensuremath{{\cal U}}}
\newcommand{\Vc}{\ensuremath{{\cal V}}}
\newcommand{\Jc}{\ensuremath{{\cal J}}}
\newcommand{\Mc}{\ensuremath{{\cal M}}}

\newcommand{\unit}[1]{\ensuremath{\, \mathrm{#1}}}

\newcommand{\gb}{\gamma_{\rm b}}
\newcommand{\dx}{\delta x}
\newcommand{\dy}{\delta y}
\newcommand{\dz}{\delta z}
\newcommand{\dr}{\delta r}
\newcommand{\ds}{\delta s}
\newcommand{\dt}{\delta t}
\newcommand{\noi}{\noindent}
\newcommand{\uns}{\rmunderscore}
\newcommand{\chimin}{\langle \chi \rangle}

\title{A search for ultralight axions using precision cosmological data}

\author{Ren\'{e}e Hlozek}
\affiliation{Department of Astronomy, Princeton University, Princeton, NJ 08544, USA}
\author{Daniel Grin}
\affiliation{Kavli Institute for Cosmological Physics, Department of Astronomy and Astrophysics, University of Chicago, Chicago, Illinois, 60637, U.S.A.}
\author{David J.~E.~Marsh}\email{dmarsh@perimeterinstitute.ca}
\affiliation{Perimeter Institute, 31 Caroline Street N,  Waterloo, ON, N2L 6B9, Canada}
\author{Pedro G.~Ferreira}
 \affiliation{Astrophysics, University of Oxford, DWB, Keble Road, Oxford, OX1 3RH, UK}
\date{\today}

\begin{abstract}
Ultralight axions (ULAs) with masses in the range $10^{-33}~{\rm eV}\leq m_{a}\leq 10^{-20}~{\rm eV}$ are motivated by string theory and might contribute to either the dark-matter or dark-energy densities of the Universe. ULAs could suppress the growth of structure on small scales, lead to an altered integrated Sachs-Wolfe effect on cosmic microwave-background (CMB) anisotropies, and change the angular scale of the CMB acoustic peaks. In this work, cosmological observables over the full ULA mass range are computed and then used to search for evidence of ULAs using CMB data from the Wilkinson Microwave Anisotropy Probe (WMAP), \textit{Planck} satellite, Atacama Cosmology Telescope, and South Pole Telescope, as well as galaxy clustering data from the WiggleZ galaxy-redshift survey. In the mass range $10^{-32}~{\rm eV}\leq m_{a}\leq 10^{-25.5}~{\rm eV}$, the axion relic-density $\Omega_{a}$ (relative to the total dark-matter relic density $\Omega_{d}$) must obey the constraints $\Omega_{a}/\Omega_{d} \leq 0.05$ and $\Omega_{a}h^{2}\leq 0.006$ at $95\%$-confidence. For $m_{a}\gsim 10^{-24}~{\rm eV}$, ULAs are indistinguishable from standard cold dark matter on the length scales probed, and are thus allowed by these data. For $m_{a}\lsim 10^{-32}~{\rm eV}$, ULAs are allowed to compose a significant fraction of the dark energy.
\end{abstract}
\maketitle

\section{Introduction}
\label{intro}
A multitude of data supports the existence of dark matter (DM) \cite{bennett/etal:2012,hinshaw/etal:2012,hou/etal:2013, sievers/etal:2013, blake2011, anderson/etal:2012,  busca/etal:2012, reid/etal:2012, ho/etal:2012,hicken/etal:2009,kessler/etal:2009, conley/etal:2011}. The identity of the DM, however, remains elusive. Axions \cite{pecceiquinn1977,weinberg1978,wilczek1978} are a leading candidate for this DM component of the Universe \cite{wise1981,dine1981,Dine:1982ah,abbott1983,preskill1983,steinhardt1983,Kim:1986ax,berezhiani1992}. Originally proposed to solve the strong $CP$ problem \cite{pecceiquinn1977}, they are also generic in string theory \cite{witten1984,witten2006}, leading to the idea of an \emph{axiverse} \cite{axiverse2009}. In the axiverse there are multiple axions with masses spanning many orders of magnitude and composing \emph{distinct} DM components. 
\begin{figure}[ht]
\includegraphics[width=0.44\textwidth]{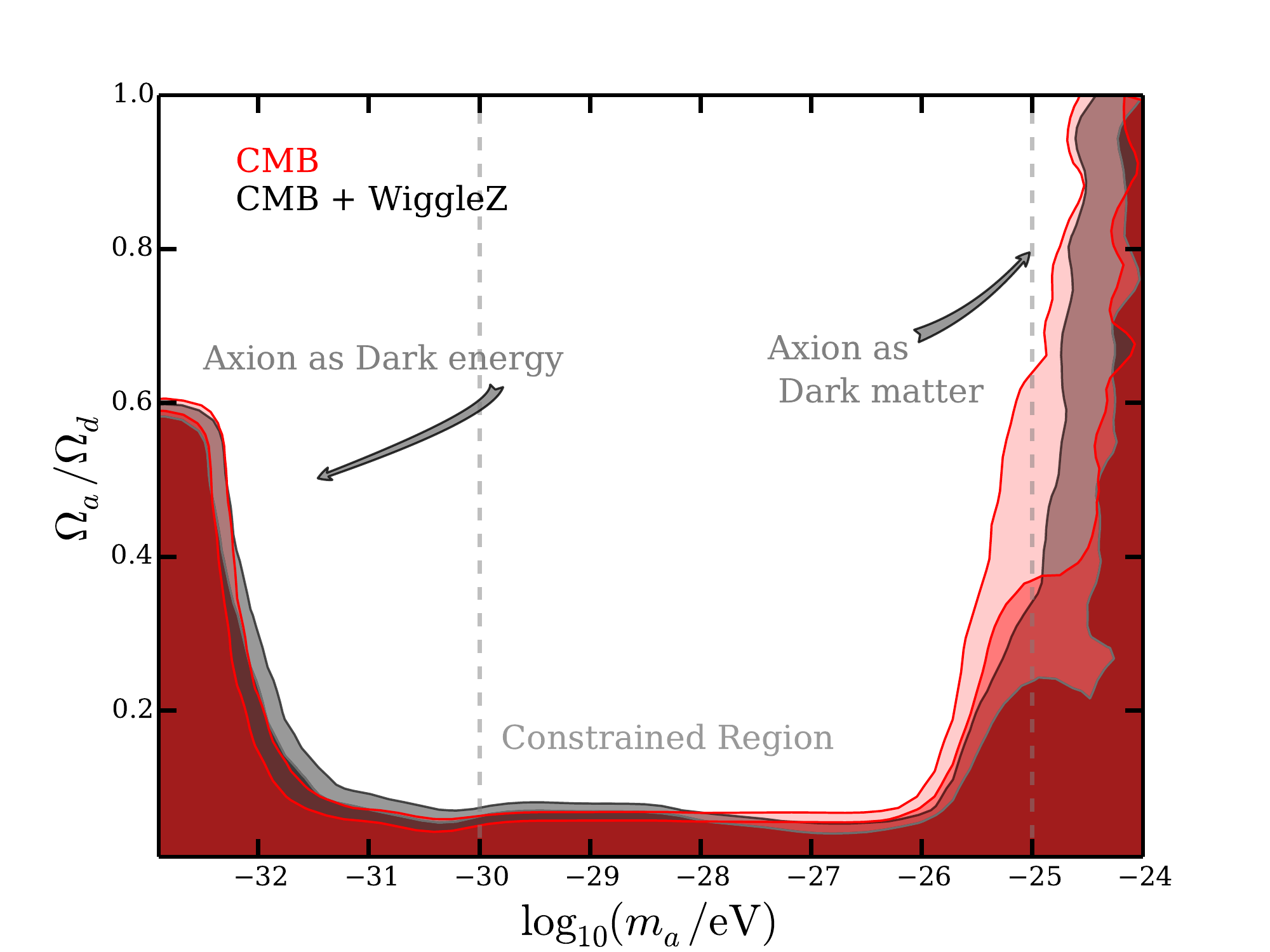}
\caption{\label{fig:money_plot}Marginalized $2$ and $3\sigma$ contours show limits to the ultralight axion (ULA) mass fraction $\Omega_{a}/\Omega_{d}$ as a function of ULA mass $m_{a}$, where $\Omega_{a}$ is the axion relic-density parameter today and $\Omega_{d}$ is the total dark-matter energy density parameter. The vertical lines denote our three sampling regions, discussed below. The mass fraction in the middle region is constrained to be $\Omega_{a}/\Omega_{d}\lsim 0.05$ at $95\%$ confidence. Red regions show CMB-only constraints, while grey regions include large-scale structure data.}
\end{figure}
For all axion masses $m_{a}\gsim 3H_{0} \sim10^{-33}{\rm eV}$, the condition $m_{a}>3H$ is first satisfied prior to the present day. When this happens, the axion begins to coherently oscillate with an amplitude set by its initial misalignment, leading to axion homogeneous energy densities that redshift as $a^{-3}$ (where $a$ is the cosmic scale factor). If $m_{a}\gsim 10^{-27}~{\rm eV}$, the axion energy-density dilutes just as nonrelativistic particles do after matter-radiation equality, making the axion a plausible DM-candidate.

The fact that axions can be so light places them, like neutrinos, in a unique and powerful position in cosmology. For as we shall show, unlike all other candidates for DM, axions lead to observational effects that are {\it directly} tied to their fundamental properties, namely the mass and field displacement. Signatures in the cosmic microwave background (CMB) and large-scale structure (LSS) can be used to pin down axion abundances to high precision as a function of the mass; these constraints can be used to place stringent limits on the mass of the axion as a candidate for DM. Furthermore, the nature of inhomogeneities in the axion distribution yield, as with primordial gravitational waves, a direct window on the very early Universe and, in particular, the energy scale of inflation. This state of affairs echoes the remarkable recent developments in constraining neutrino masses with weak lensing of the CMB \cite{plancklensing,dasetal:2013,hou/etal:2013} and places cosmological constraints on axions on par with current and future particle physics constraints.

For ultralight axions (ULAs) with masses $m_a\lesssim 10^{-20}\unit{eV}$, small-scale structure formation is suppressed \cite{Frieman:1995pm,Coble:1996te,hu2000,amendola2005,marsh2010,park2012} on astronomically observable length scales. This allows ULA DM to be distinguished from CDM using large-scale structure (LSS) data. Hot (H)DM (e.g. $\sim {\rm eV}$ or lighter neutrinos) and warm (W)DM (eg. $\sim~{\rm keV}$ sterile neutrinos) \cite{viel2013} exhibit a qualitatively similar effect. The physical origin of power suppression for ULAs, however, is distinct (see Ref. \cite{marsh2011b} and references therein), resulting from the macroscopic de Broglie wavelength of ULAs as opposed to thermal free-streaming. The detailed shape of the power spectrum on small scales thus distinguishes ULA DM from CDM, WDM, and HDM.

Additionally, in this window, ULAs change the matter content during the radiation era [behaving as dark energy (DE) before beginning to oscillate] \cite{Frieman:1995pm,2009ARNPS..59..397C}, thus changing the heights of the CMB acoustic peaks. Interestingly, because ULAs change the amplitude of the late-time Integrated Sachs-Wolfe (ISW) effect \cite{Coble:1996te} and alter the expansion history during radiation domination, the CMB is comparably sensitive to LSS measurements over the bulk of the mass range explored; this augments the tests of ULAs enumerated in Ref. \cite{axiverse2009}

For lower masses still ($m_{a}\lsim~10^{-27}~{\rm eV}$), axions would roll slowly and contribute to DE (as opposed to DM)\cite{Frieman:1995pm,2009ARNPS..59..397C,kim2009,kaloper2009,panda2010,gupta2011,amin2011,marsh2011,marsh2012,kim2013,kim2013b} for some period of time after matter-radiation equality, perhaps even explaining the current era of accelerated expansion. In this case, ULAs change the amplitude of the large-angle ISW plateau in the CMB.

In this work, we search for ULAs in the mass range $10^{-33}\unit{eV}\leq m_a\leq 10^{-22}\unit{eV}$ by comparing precision CMB and galaxy-clustering data to theoretical predictions from a self-consistent Boltzmann code (an appropriately modified version of \textsc{camb}). This code follows the evolution of ULA, standard fluid, and potential perturbations, including the effect of ULAs on the Hubble expansion-rate and recombination. This builds upon past work, in which the effect of ULAs was treated semi-analytically \cite{amendola2005}.

When $m_{a}\gg 3H$, the rapid oscillation of the ULA field requires a small timestep ($\sim m_{a}^{-1}$), making an exhaustive search of ULA parameter space computationally prohibitive. This bottleneck is addressed using an effective-fluid formalism, averaging over the fast oscillation time scale and following the evolution of the system containing standard cosmological fluids and ULAs coupled only through gravity. The CMB data used are the temperature anisotropy (TT) power spectrum measured by the \textit{Planck} \cite{planckfull,plancklikelihood} satellite, E-mode polarization data from the WMAP 9-year data release \cite{bennett/etal:2012}, as well as small-scale CMB data from the South Pole Telescope (SPT) \cite{spt} and Atacama Cosmology Telescope (ACT) \cite{dasetal:2013}. Finally, we use the galaxy power-spectrum measured in the WiggleZ survey \cite{blake2011,blakeetal:2011,Parkinson:2012vd}.

We explore both the low-mass ($m_{a}\leq 10^{-27}~{\rm eV}$) region of ULA parameter space, in which they are DE-like, and the higher-mass ($m_{a}\geq10^{-27}~{\rm eV}$) region of parameter space, in which they are DM-like. The parameter space is multimodal, requiring us to adapt the usual \textsc{CosmoMC} code \cite{cosmomc} using nested sampling, as implemented in the \textsc{MultiNest} code \cite{multinest}. We obtain marginalized constraints varying all the primary cosmological parameters, namely the baryon and CDM density parameters $\Omega_{b}h^{2}$ and $\Omega_{c}h^{2}$, the amplitude $\Delta_{\mathcal{R}}^{2}$ and logarithmic slope $n_{s}$ of the primordial power spectrum, the optical depth $\tau_{\rm re}$ to reionization, and the angular sound horizon $\theta_{\rm A}$ at baryon-photon decoupling, in addition to the ULA mass $m_{a}$ and $\Omega_{a}h^{2}$ (where $h$ is the dimensionless Hubble parameter today). We check that degeneracies with foreground parameters may be neglected.

These techniques allow a search for ULAs to be conducted with precision cosmological data, applying the structure-suppressing imprint of ULAs. As this effect is gravitational in origin, it is independent of model-dependent ULA couplings. Therefore our constraints are applicable to any coherently oscillating particle in this mass range, irrespective of its couplings. We find that in the mass range $10^{-32}~{\rm eV}\leq m_{a}\leq 10^{-25.5}~{\rm eV}$, the LSS and CMB data imply that the ULA relic-density must obey the constraint $ \Omega_{a}/\Omega_{d}\leq 0.05$, and that $\Omega_{a}h^{2}\leq 0.006$. Our key result is shown in Fig. \ref{fig:money_plot}, where upper limits to the axion mass fraction $\Omega_{a}/\Omega_{d}$ in the ``Dark-matter like", ``Dark-energy like", and highly constrained mass-regimes are shown.

This paper is organized as follows. We begin in Sec.~\ref{model} by introducing the ultralight axion scenario and its cosmology. We then present the effective fluid formalism for ULA perturbations in Sec.~\ref{axion_fluid_pert}, including discussion of initial conditions and implementation in the Boltzmann code \textsc{camb} \cite{camb}.  In Sec.~\ref{sec:observables} we discuss the effect of ULAs on LSS and CMB observables. In Sec.~\ref{results} we present our methodology and key results, which are constraints to the ULA  parameter-space. We interpret the constraints and conclude in Sec.~\ref{conclusions}. In Appendix \ref{hubble_debroglie}, we give a simple argument for the suppression of structure on small scales in the ULA dark-matter scenario. In Appendix \ref{icinit}, we derive the early-time power-series initial condition for the ULA+fluid system in the adiabatic mode, which is used to set initial conditions in \textsc{camb}. 
\section{Ultralight Axions}
\label{model}
\subsection{Axions in String Theory}
Axions are described by two energy scales: the Peccei-Quinn (PQ) symmetry-breaking scale, $f_a$, and the energy scale of nonperturbative physics, $\Lambda_a$, which gives rise to the axion mass $m_{a}$. In QCD, $\Lambda_a$ is fixed by the requirement that the axion solve the strong $CP$ problem, and so the axion mass is controlled by $f_a$ and QCD physics, in the form of the pion mass and decay constant, and the quark masses \cite{weinberg1978}. In the absence of fine-tuning and to avoid an axion relic-density so high that the Universe is overclosed, QCD axions must obey the constraint $f_a\lsim 10^{12}\unit{GeV}$ \cite{preskill1983,steinhardt1983} or $m_a \gsim 10^{-6}~{\rm eV}$; this is the classic QCD CDM window. When fine tuning of the initial misalignment is allowed, there is no upper bound on $f_a$ from relic density constraints, and this defines the anthropic axion window (e.g. Refs.~\cite{hertzberg2008,Wantz:2009it}).

It is possible for string theory to furnish us with the QCD axion and its solution to the strong $CP$ problem. Indeed axions will always arise in string-theory compactifications \cite{witten1984,witten2006} as Kaluza-Klein zero modes of antisymmetric tensor (form) fields analogous to the Maxwell tensor, $F_{\mu\nu}$. These terms appear when the form fields are compactified on closed cycles in the compact space. For example the heterotic string theories contain the so-called `model independent' axion arising from compactification of the antisymmetric partner of the metric, $B_{\mu\nu}$, on closed 2-cycles. The number of axions is fixed by the topology of the compactification. String theory compactifications on Calabi-Yau manifolds  \cite{candelas1985} capable of realizing realistic models of high energy physics can be highly complicated topologies, and the number of axions is given by the Hodge numbers of the Calabi-Yau manifold, which can be large (see e.g. \cite{he2013} and references therein). Such compactifications therefore give rise to \emph{many axions} \cite{candelas1997,dimopoulos2008}.

The relevant scales, $f_a$ and $\Lambda_a$, in string theory are both determined separately for each axion, and depend on the action, $S$, due to nonperturbative physics on the corresponding cycle:
\begin{align}
f_a &\sim \frac{M_{pl}}{S}\, , \label{faeq}\\
\Lambda_a^4 &= \mu^4 e^{-S} \, , \label{lameq}
\end{align}
where $M_{pl}$ is the reduced Planck mass: $M_{pl}^2=1/8\pi G$. The hard nonperturbative scale is $\mu$, which may be due to, for example, gauge-theory instantons (as is the case for QCD), world-sheet instantons, or Euclidean D-branes, and its value should be roughly given by the geometric mean of the Planck scale and the SUSY scale \cite{witten2006,axiverse2009}. Solving the strong $CP$ problem with one of the string axions requires $S\gtrsim 200$ \cite{witten2006,axiverse2009}, giving rise to stringy values of $f_a \approx 10^{16}\unit{GeV}$, near the GUT scale. The exact value of $S$, however, scales with the volume of the corresponding cycle (a dynamically distributed quantity in the landscape), so that small variations in the area lead to exponential variations in the scale of the potential, and thus the axion mass.

The scale of the decay constant is unknown. For the QCD axion one requires $10^9$ GeV$\lesssim f_a\lesssim 10^{17}$ GeV, where the lower bound comes from stellar cooling \cite{Grifols:1996id,Friedland:2012hj} and the upper bound comes from constraints from the spins of stellar mass black holes \cite{arvanitaki2010}. Neither bound applies to a general axion-like particle (ALP), since the coupling to the standard model is model-dependent, and the mass is not fixed by $f_a$. There is, however, a strong theoretical upper bound of $f_a<M_{pl}$, realized in string models \cite{banks2003}, which follows from the `weak gravity conjecture' (WGC) \cite{arkani-hamed2007,cheung2014} and bounds the instanton action $S\lesssim M_{pl}/f_a$. The periodicity of the axion field implies that $f_a$ bounds the maximum and natural field excursion, with implications for the DM abundance that we discuss further in subsequent sections. 

Our final constraints to $\phi_i$ (the initial, and therefore maximum necessary, axion field displacement), discussed in Sec.~\ref{results}, are unsurprisingly consistent with WGC. 
The value of $f_a$ can be further constrained if the energy scale of inflation is large, generating primordial CMB B-mode polarization of observable amplitude \cite{Ade:2014xna,mortonson2014,flauger2014,Adam:2014bub}. In this case, large isocurvature perturbations would result, violating \textit{Planck} limits and severely constraining ULA DM \cite{Marsh:2014qoa,Visinelli:2014twa}. A full analysis of ULA isocurvature constraints is in progress. In this work we fix the tensor and isocurvature perturbations to be zero, consistent with a low inflationary energy scale.

To date there are two explicit realizations of the axiverse idea within string/M-theory: the Type IIB Axiverse \cite{cicoli2012c} and the M-theory Axiverse \cite{acharya2010a}.\footnote{An accessible review of the Type IIB models, giving more details than we give here, is Ref.~\cite{ringwald2012b}. See also Ref.~\cite{cicoli2013b}.} The distribution of $f_a$ (across different axions) is different in each of these models. A discussion of the expected distribution for $f_a$ in the landscape is given in Ref.~\cite{long2014b}.

The Type IIB axiverse is constructed in the LARGE volume scenario (LVS) for moduli stabilization \cite{balasubramanian2005,Conlon:2005ki}, where axions can emerge from compactifying the $C_4$ 4-form of IIB supergravity. Within the LVS one requires the number of axions $n_{ax}\geq 2$ in order to maintain the natural value of the superpotential, $W_0\sim \mathcal{O}(1)$ while at the same time reproducing the visible sector GUT coupling, $\alpha_{GUT}$. The axions in the LVS that remain light are associated to moduli which are fixed perturbatively. The perturbative shift symmetry of axions protects them from acquiring mass via this mechanism, so that the masses come from higher order nonperturbative effects and are naturally small. The Type IIB axiverse has been constructed explicitly with a decay constant $f_a\approx 10^{10}$ GeV and axion masses ranging from an essentially massless axion (associated with the volume modulus) up to and beyond the QCD axion. The small values of the decay constant arise from the large volume.

The M-theory axiverse has $W_0\ll 1$ and this fixes just one axion with nonperturbative physics giving a high mass, corresponding to the small compactification volume on the $G_2$ manifold. All other axions are again fixed by higher-order effects giving small masses. Axions in these theories are compactified on closed 3-cycles. Again, achieving the correct value of $\alpha_{GUT}$ requires introducing a second axion, whose mass is fixed by $\alpha_{GUT}$ to be $m_{a,GUT}\approx 10^{-15}$ eV. The small compactification volume leads to GUT scale decay constants, $f_a\approx 10^{16}$ GeV, and also implies a \emph{maximum axion mass}, $m_{a,\rm max}=\mathcal{O}(1) (10^{-8}\rightarrow 1)$ eV, in order to maintain control over the framework.

Finally, it is worth mentioning the recent explicit construction of N-flation \cite{Liddle:1998jc} within Type IIB theory \cite{cicoli2014}. This construction not only allows for N-flation in the standard way \cite{Liddle:1998jc} (with $m_a\sim H_I$, where $H_I$ is the Hubble scale during inflation), but with only a small change in the volume of the compact space from $\mathcal{V}=\mathcal{O}(10^2)$ to $\mathcal{O}(10^3)$ (in string units) one can also realize N-quintessence (with $m_a\sim H_0\ll H_I$). It is therefore completely plausible within this model that one can realize all axion masses in between, in particular those we constrain, giving N-ULA models for DM with potentially large effective decay constants from alignment (e.g. Refs.~\cite{Kim:2004rp,long2014}).
 
 \subsection{Ultralight Axion Cosmology}
 \label{sec:cosmo_pheno}

The low-energy four-dimensional Lagrangian for a single axion field $\theta$ is [with metric signature $(-,+,+,+)$]:
\begin{equation}
\mathcal{L}=-\frac{1}{2}f_a^2(\partial \theta)^2-\Lambda_a^4 U(\theta) ,
\end{equation}
where $U(\theta)$ is any periodic potential, with $\theta$ chosen such that it is minimized at $\theta=0$. Canonically normalizing, we use the field $\phi=f_a\theta$. When the potential is expanded to leading order in $1/f_a$, only the mass term appears, with
\begin{equation}
m_{a}^{2}=\frac{\Lambda_{a}^{4}}{f_{a}^{2}}.\label{msp}
\end{equation}
The value of the mass depends exponentially on the nonperturbative action $S$, which we expect to be uniformly distributed, and so the axion mass spectrum can be taken as a uniform distribution on a logarithmic scale \cite{axiverse2009} (although see Ref. \cite{easther2006}), as we can see from Eqs.~(\ref{lameq}) and (\ref{msp}). In a Bayesian context, this Jeffreys prior is uninformative and thus natural.

We will work only with the mass term in the potential, since the form of the potential away from the minimum (the axion self interactions) is unknown without an explicit model for the nonperturbative physics. The Lagrangian we use is
\begin{equation}
\mathcal{L}=-\frac{1}{2}(\partial \phi)^2-\frac{1}{2}m_a^2 \phi^2,
\end{equation}later shown to be a valid approximation over the vast majority of observationally allowed parameter space if 
 $f_a<M_{pl}$.

In this work, our focus is on the effect of a single ULA, whose homogeneous energy-density and pressure are given (in a Friedmann-Robertson-Walker spacetime) by
\begin{eqnarray}
\rho_a =&~\frac{a^{-2}}{2}\dot{\phi}_0^2 + \frac{m_a^2}{2}\phi_0^2 \label{eqn:rhoa} \, , \\
P_a =&~\frac{a^{-2}}{2}\dot{\phi}_0^2 - \frac{m_a^2}{2}\phi_0^2 \label{eqn:pa},\end{eqnarray}
where $\phi_{0}(\tau)$ is the homogeneous value of the scalar field as a function of the conformal time $\tau$, $a$ is the cosmological scale factor, and dots denote derivatives with respect to conformal time. We restrict ourselves to a single ULA, as the effective fluid formalism described in Sec. \ref{axion_fluid_pert} has only been developed for this case. The mass independence of constraints in certain windows may mitigate this limitation. 

The equation of motion for the axion field is
\begin{align}
\ddot{\phi}_0+2\mathcal{H}\dot{\phi}_0+m_a^2 a^2 \phi_0&=0,\label{homo_eom}
\end{align}
where the conformal Hubble parameter is $\mathcal{H}=\dot{a}/a=aH$. 

At early times when $m_a\ll H$, the axion rolls slowly, and if its initial field-velocity $\dot{\phi}_{i,0}=0$, it has equation of state $w_{a}\equiv P_{a}/\rho_{a}\simeq -1$. The axion thus behaves as a DE component, with roughly constant energy density in time. As the Universe cools and $H$ falls, eventually the axion field begins to coherently oscillate about the potential minimum. This occurs when 
\begin{equation}
m_a\approx 3H(a_{\rm osc})\, ,
\end{equation} 
where this equation defines the scale factor $a_{\rm osc}$. The oscillation is on time scales $\delta t\sim m_a^{-1}$, with $\phi\propto a^{-3/2}$ on longer time scales. Thereafter, the number of axions is roughly conserved, yielding an axion energy-density that redshifts as matter, with $\rho_{\rm a}\propto a^{-3}$ \cite{turner1986}. The relic-density parameter $\Omega_{\rm a}$ is given by
\begin{equation}
\Omega_{\rm a}=\left[\frac{a^{-2}}{2}\dot{\phi}_0^2 + \frac{m_a^2}{2}\phi_0^2 \right]_{m_a=3H}a_{\rm osc}^{3}/\rho_{\rm crit},\label{homorelic}
\end{equation}
where $\rho_{\rm crit}$ is the cosmological critical density today. This production mode is known as the misalignment mechanism. 
When the ULA behaves as DE, it rolls slowly and sources the ISW effect (due to the decay of gravitational potentials wells) \cite{Coble:1996te}.

We can use Eq.~(\ref{homorelic}) to obtain a crude estimate for the relic density in axions. Assuming that $a^{-2}\dot{\phi}_0^2(a_{\rm osc})/2\ll (m_a^2/2)\phi_0^2(a_{\rm osc}) \approx m^2\phi_{0,i}^2/2$ (where $\phi_{0,i}$ is the initial homogeneous field displacement), and taking the background evolution to be described by either pure radiation or pure matter domination at $a=a_{\rm osc}$, one obtains \cite{marsh2010}:
\begin{widetext}
\begin{eqnarray}
\Omega_a = \left\{ 
\begin{array}{ll} 
\frac{1}{6}(9 \Omega_r)^{3/4} \left( \frac{m_a}{H_0} \right)^{1/2} \left( \frac{\phi_{0,i}}{M_{pl}} \right)^2\mbox{if $a_{\rm osc}< a_{\rm eq}$}\, ,\\
\frac{9}{6}\Omega_m \left( \frac{\phi_{0,i}}{M_{pl}} \right)^2\mbox{if $a_{\rm eq}<a_{\rm osc}\lesssim 1$} \, , \\
\frac{1}{6}\left( \frac{m_a}{H_0} \right)^2\left( \frac{\phi_{0,i}}{M_{pl}} \right)^2\mbox{if $a_{\rm osc}\gtrsim 1$} \, ,\label{simpledens}
\end{array}
\right. ,
\end{eqnarray}
\end{widetext}
where the final line accounts for axions that never oscillate.\footnote{Our mass prior terminates below $m_a=10^{-33}$ eV$\sim H_0$. For significantly lighter ULA masses, the early time ULA dark-energy behavior is trivial, and the final line of Eq.~(\ref{simpledens}) is exact, while for masses on the border of quintessence with $a_{\rm osc}\sim a_0$ the guess in the second line of Eq.~(\ref{simpledens}) is still very good through most of parameter space (since $\Omega_a\sim \Omega_m\sim \mathcal{O}(1)$ for quintessence).}

The expressions in Eq.~(\ref{simpledens}) are useful for estimates, but in our analysis we always compute the relic density numerically by solving the Klein-Gordon equation with an initial value $\phi_i$. We iterate this value to get the desired $\Omega_a$: Eq.~(\ref{simpledens}) is used as the first guess in this iteration. We find that independent of $\Omega_a$ our procedure returns $\Omega_a/\Omega_d$ to a relative precision of better than $10^{-4}$, within the limits set by the approximation to treat $w_a=0$ for $H<m_a/3$. The relic density can also receive other nonthermal and thermal contributions, but since the theoretical uncertainty and model dependence in such contributions is large, we take the vacuum-realignment production alone as the most conservative estimate \cite{hertzberg2008}. 

At fixed $\phi_i<f_a\ll M_{pl}$ Eq.~(\ref{simpledens}) restricts $\Omega_a<\Omega_d$ in certain parts of parameter space \cite{Frieman:1991qv}. For an axion respecting the residual shift symmetry $\theta\rightarrow\theta+2\pi$ there is a maximum value $\phi_{0,i}\sim \pi f_a$. This yields an `anthropic boundary': for axions beginning oscillation in the radiation dominated era, with $f_a\sim 0.01 M_{pl}$, it is impossible to have $\Omega_a>1$ for $m_a \lsim 10^{-19}\unit{eV}$ \cite{axiverse2009}. This is an anthropic boundary since axions above this mass must be fine tuned anthropically to satisfy DM-density (or closure) bounds \cite{Frieman:1991qv}. However, when $\Omega_a$ is observationally restricted lighter axions may start to be fine tuned in a nonanthropic way. On the other hand, for $a_{\rm osc}>a_{\rm eq}$ and $f_a\sim 0.01 M_{pl}$, when the shift symmetry is respected there is a maximum axion density of $\Omega_a\sim 10^{-5}$.
 
There are ways to obtain large $\Omega_a$ for low axion masses from the misalignment mechanism. The most obvious is to allow larger symmetry-breaking scale $f_a\sim M_{pl}$, which still respects the WGC. That this can give $\Omega_a\sim \mathcal{O}(1)$ even for the lightest axion we consider, with $m_a=10^{-33}\text{ eV}\sim H_0$ is obvious from Eq.~(\ref{simpledens}). For low individual $f_a$, as already mentioned, alignment of many axions can give an effective $f_a$ which is large. For a single axion with low $f_a$, anharmonic effects at $\phi_i\sim f_a$ flatten the potential and delay oscillations \cite{turner1986}, while broken shift-symmetry can allow $\phi_i>f_a$ \cite{silverstein2008,panda2010}. In light of these issues, we treat the axion abundance as a free model-parameter. 

The expressions in Eqs. (\ref{simpledens}) differ from classic QCD expressions (e.g. Ref.~\cite{Wantz:2009it}). The QCD axion has temperature-dependent corrections to its mass which are still relevant when it begins to coherently oscillate. For ULAs, however, the temperature dependence of the axion mass is negligible by the time the misalignment mechanism begins if the scale of nonperturbative physics is above the QCD scale, as in string theory \cite{acharya2010a}. Therefore one can use the constant, zero-temperature mass in all calculations, which simplifies the approximate expressions for the relic density, a simplification also present for large $f_a$ QCD axions \cite{Wantz:2009it}. The temperature-dependence of the axion mass depends on its couplings to standard-model particles, which in turn offer noncosmological tests of the axion hypothesis.

\subsection{Direct/indirect detection of axions, and Astrophysical Probes}
\label{sec:direct_searches}

Axions can only have perturbative couplings that respect the shift symmetry, $\theta\rightarrow \theta+{\rm const}$ (e.g., derivative couplings). Therefore ULAs are not subject to the same fifth-force constraints as other light bosons and do not require a screening mechanism. Axions can, however, have model-dependent couplings to topological gauge-theory interactions of the form $g_iF_i\tilde{F}_i$, where $F_i$ is the field-strength tensor, which for coupling to the standard model could be electromagnetism or QCD, $\tilde{F}_i$ is its dual, and $g_i$ is a model dependent coupling constant. The QCD axion has couplings of this form to both electromagnetism, via pions, and to QCD by virtue of it solving the strong $CP$ problem, and the value of $g_i$ is determined by $f_a$.

There are many experimental constraints to axions that couple to electromagnetism \cite{Frieman:1991qv}. There are three classic methods to constrain axions through such a coupling: RF-cavity searches (haloscopes), solar axion conversion to x-ray photons (helioscopes), and ``light shining through a wall" (LSW) experiments \cite{Sikivie:1983ip}. The QCD axion has only one free parameter, $f_a$, in such constraints and occupies a line in the mass-coupling plane, but constraints to general axion-like particles apply to regions of this parameter space. Current experiments include ADMX \cite{Asztalos:2009yp} (haloscope), CAST \cite{Arik:2011rx} (helioscope), and ALPS-I \cite{Ehret:2010mh} (LSW). 

Astrophysical constraints to axions largely follow from their electromagnetic coupling. If coupled to photons, axions would hasten the cooling of stars. For the QCD axion, this gives the lower limit to $f_a\gtrsim 10^9 \text{ GeV}$ \cite{Grifols:1996id,Brockway:1996yr}. The neutrino burst from Supernova 1987a would also have been shortened, yielding a similar constraint \cite{raffelt2001}. Constraints can also be derived from the dimming of supernovae and quasars \cite{Mortsell:2002dd,mortsell2003}, CMB spectral distortions \cite{mirizzi2009,tashiro2013} and various other astrophysical and cosmological processes \cite{dansearch,wd_a,wd_b}. It has also been proposed that a coupling of ULAs to electromagnetism might explain some features related to the CMB cold spot \cite{csaki2014}, and can act as a source of $B$-mode polarization via cosmological birefringence \cite{pospelov2008}. Reviews of axion searches can be found in Refs.~\cite{raffelt2001,ringwald2012,dias2014}.

Recently, a number of new experimental techniques to search for axions have been proposed. These include nuclear spin precession \cite{Budker:2013hfa}, using a LC circuit as a RF cavity \cite{Sikivie:2013laa}, and searching for axion-mediated forces \cite{arvanitaki2014,Stadnik:2014ala}. 

All the searches we have so far described constrain the axion coupling, $g_i$, to some standard-model field. Few existing experiments yet reach the sensitivity to detect the QCD axion, and it might be expected that a general axion couples more weakly, at least to nucleons, than the QCD axion \cite{blum2014}. Axion DM searches depend on all the axion parameters, $\{g_i, m_a,\Omega_a\}$, and constraints vanish if either of $g_i,\Omega_a$ go to zero. Constraints relating to axion production, such as LSW, do not depend on $\Omega_a$, but vanish if $g_i$ goes to zero. Many of the constraints we have mentioned apply to ULAs in the mass range we consider, but only if some $g_i$ are nonzero.

Only ULA constraints that depend on gravitational interactions alone apply when all $g_i$ go to zero. ULA masses, and indeed the mass of any light boson, can be constrained by the effect of the Penrose process leading to a super-radiant instability of rotating black holes \cite{axiverse2009,arvanitaki2010,pani2012}. The observation of spinning stellar-mass black holes constrains the QCD axion to have $f_a\lesssim 10^{17}\text{ GeV}$, excluding $m_a\sim 10^{-11}\text{ eV}$ for ULAs/ALPs. The observation of spinning supermassive black holes excludes ULAs with masses $10^{-18}\text{ eV}\lesssim m_a\lesssim 10^{-19}\text{ eV}$ \cite{pani_private}. 

These are the only constraints to axions that are independent of both $\Omega_a$ and $g_i$: since super-radiance is essentially a gravitational production of axions it applies even when $\Omega_a$ and all $g_i$ go to zero. It is therefore the only constraint that applies in a completely model-independent way to our search. Black hole super-radiance constraints provide an upper bound to the cosmological axion mass range, but do not extend to the lower masses probed in this work. As we now discuss, ULA DM or DE would change the growth of cosmological structure, providing an additional (and gauge-coupling independent) test of the ULA hypothesis.

\section{ULA perturbations}
\label{axion_fluid_pert}
So far, we have discussed the homogeneous cosmology of ULAs. We now discuss how the perturbed inhomogeneous Universe can be used to probe ULA DM and DE, beginning with a qualitative discussion here and moving on later in this section to formal developments and computational techniques. 

It is well known that a coherently oscillating gas of light axions (nearly all of which are in the ground state) manifests a new scale, the axion ``Jeans" scale, $k_{\rm J}\sim \sqrt{m_{a}H}$, below which axions cannot cluster \cite{khlopov_scalar,branden,peebles1988,nambu1990,ratra1991a,ratra1991b,Frieman:1995pm,Coble:1996te,hu2000,hwang_noh_a,axiverse2009,hwang_noh_b,marsh2010,sim_bec,Noh:2013coa}. This is the de Broglie wavelength of axions moving with the Hubble flow, as discussed in Appendix \ref{hubble_debroglie}, and manifests itself as a downward step in clustering power at small scales in the matter power-spectrum \cite{sin1994}.\footnote{This also applies to axions moving at the virial velocity inside halos, and implies the formation of density cores in axion halos \cite{hu2000}.} 

Depending on the axion mass, this scale could be macroscopic, and thus affect the CMB anisotropy and observed galaxy clustering power spectra.\footnote{Self-interactions of the field, however, can be important at low mass and affect the resulting Jeans scale \cite{Boyle:2001du}. The (model-dependent) form of these interaction terms can change the evolution of the DM-density and determine whether or not the DM ends up in a Bose-Einstein condensate \cite{ferreira1997,ferreira1998,matos2000,arbey2001,arbey2003,bernal2006,lee2009,matos2009,li2013,2012MNRAS.422..135R,suarez2013}.} For the classic QCD axion ($m_{a}\gsim 10^{-6}~{\rm eV}$), this scale is not cosmologically relevant, but for ULAs, this scale could be observationally relevant. 

In the effective fluid formalism developed here, $k_{\rm J}$ arises dynamically in the axion fluid, which has effective sound speed
\begin{eqnarray}
c_a^{2}=\left\{\begin{array}{ll}
\frac{k^2}{4m_a^2a^2}&\mbox{if $k\ll k_{m}\equiv 2m_{a}a$},\\
1&\mbox{if $k\gg k_{m}$}.\end{array}\right.\label{heuristic_cs}\end{eqnarray}
Structure is suppressed for scales with $k>k_{\rm m}$, which enter the horizon when $c_a^2=1$ \cite{marsh2010}. This wave number $k_{\rm m}$ is smaller (corresponds to larger length scales) as $m_{a}$ decreases. The effect saturates at the smaller scale $k_{\rm J}=a (16 \pi G \rho_m)^{1/4}m_a^{1/2}$. Therefore, like massive standard-model neutrinos or warm DM (e.g. Refs.~\cite{1982PhRvL..48.1636B,bode2001}), axions exhibit suppressed structure on small scales. The effect has a completely different origin, however, resulting from the macroscopic `wavy' properties of axions, unlike massive neutrinos, which display suppressed structure because of their large free-streaming velocity during structure formation. 

\begin{figure*}
\begin{center}
\includegraphics[width=0.49\textwidth]{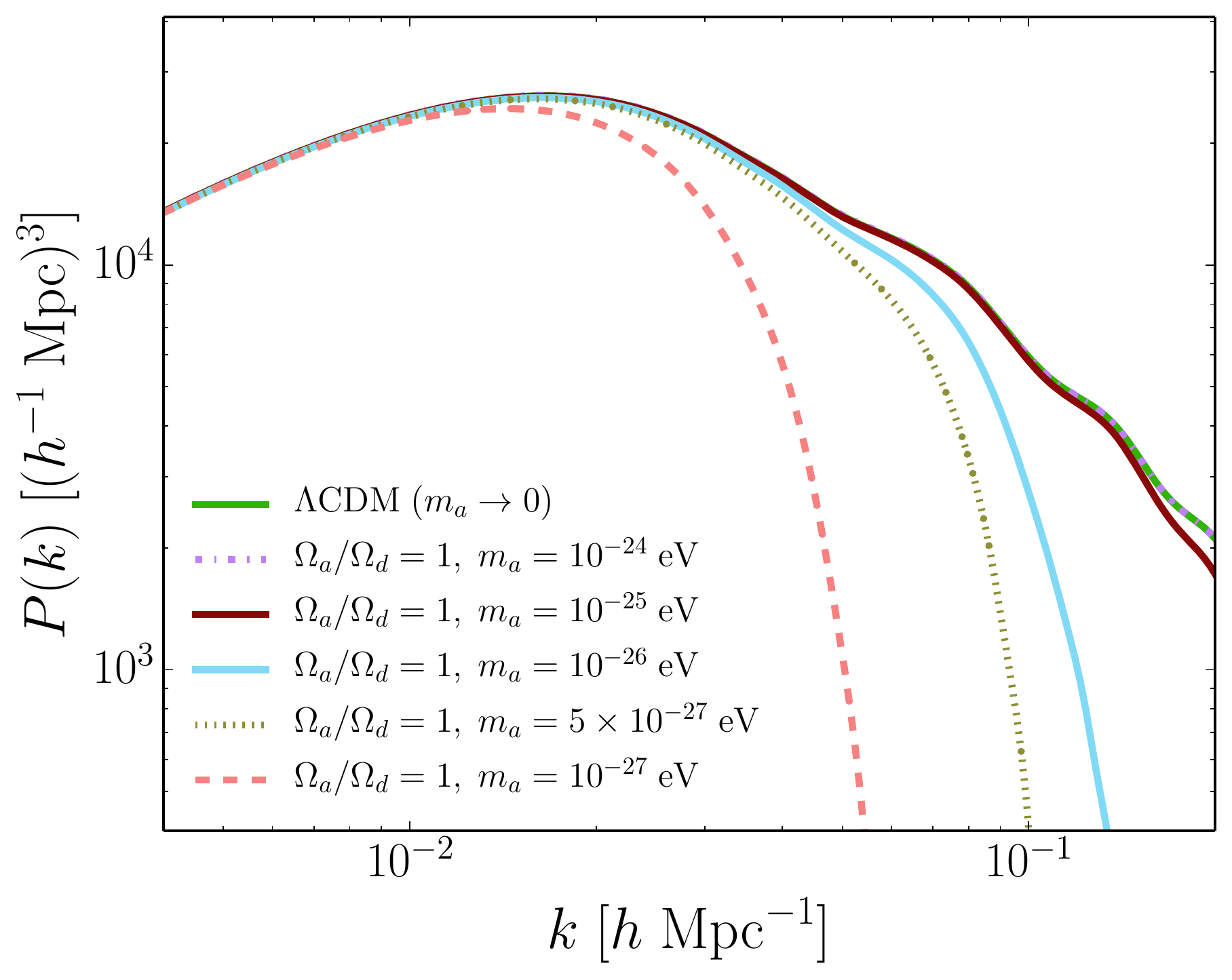}
\includegraphics[width=0.49\textwidth]{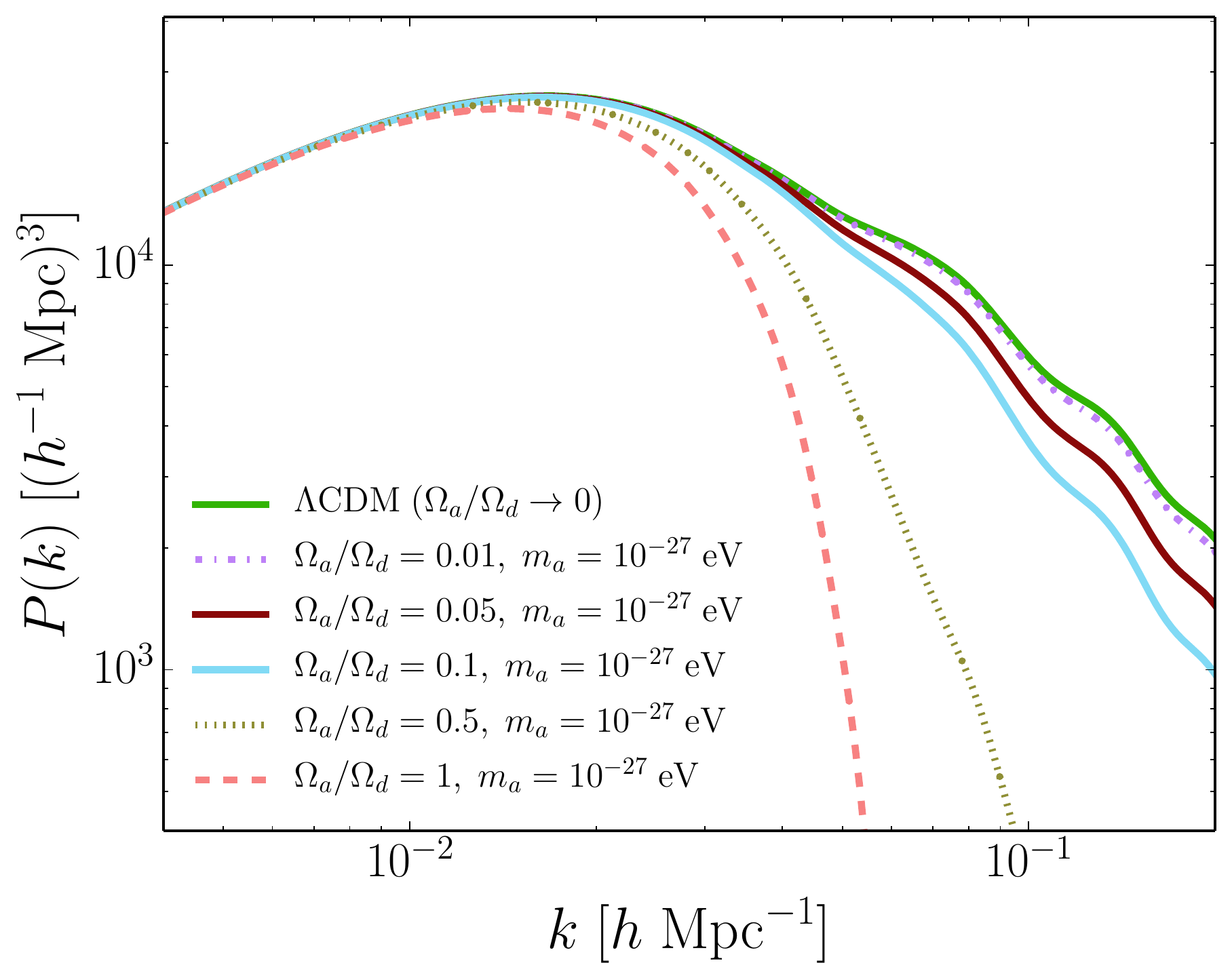} 
\caption{Adiabatic matter power-spectra generated with the modified \textsc{camb} described in Sec. \ref{axion_fluid_pert}, with varying axion mass and energy-density fraction $\Omega_a/\Omega_d$ at fixed total dark-matter density fraction $\Omega_{d}$. Power is suppressed for modes that enter the horizon when the axion sound speed $c_{s}\sim 1$. \label{matterspectra_dm_examples}}
\end{center}
\end{figure*}

The suppression of small-scale power in the matter power-spectrum is shown in Fig.~\ref{matterspectra_dm_examples}. For illustrative purposes we show the theoretical linear matter power-spectrum computed at $z=0$. Current measurements of the matter power-spectrum on linear scales, $k\lesssim 0.1\, h\text{Mpc}^{-1}$, and at various redshifts are consistent with $\Lambda$CDM, within experimental errors \cite{blake2011,blakeetal:2011}. By inspection of Fig.~\ref{matterspectra_dm_examples} one can therefore estimate the rough constraining ability of the matter power-spectrum to probe ULA masses $m_a\lesssim 10^{-25}\text{ eV}$ as dominant components of the DM.

Figure ~\ref{matterspectra_dm_examples} is obtained using a version of the Boltzmann code \textsc{camb} \cite{camb}, modified to include axions in an effective fluid description, as discussed below. We see that the matter power-spectrum is suppressed at small scales. We see that lower values of $m_{a}$ or higher values of $\Omega_{a}/\Omega_{d}$ cause progressively more severe suppression, indicating that LSS data can be used to constrain ULA properties. The effect is present on linear scales $k\lsim 0.1~{\rm Mpc}^{-1}$, and so the linear power-spectrum can be used to impose tight constraints to ULAs when $m_{a}\lesssim 10^{-25}~{\rm eV}$.

\begin{figure}
\includegraphics[width=0.49\textwidth]{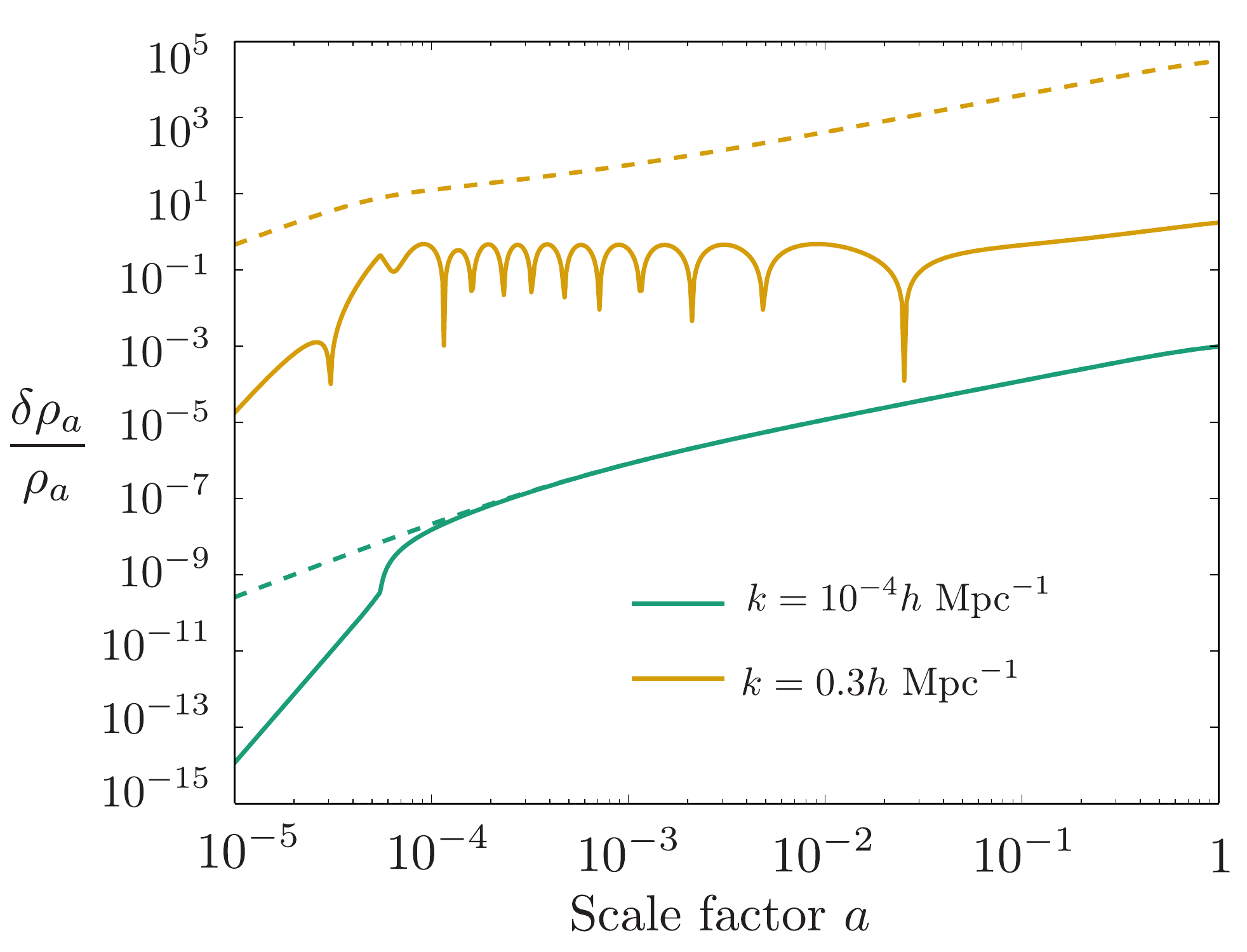}
\caption{Evolution of the fractional DM density-perturbation $\delta$ when $\Omega_{a}/\Omega_{d}=1$ (solid), for a ULA mass of $m_{a}=10^{-26}~{\rm eV}$ and a series of wave numbers $k$ (as shown in the figure), compared to standard CDM (dashed). The overall normalization of the mode amplitude is arbitrary here. The range of k-values encompasses different behaviors, with suppression of growth relative to CDM when $k\sim k_J(a)$, oscillation when $k>k_J(a)$ and growth as CDM when $k<k_J(a)$. This leads to an overall suppression of power for large-$k$ modes.}
\label{fig:overdensity_plot}
\end{figure}

We can gain some insight into the suppression of the power spectrum by examining the evolution of a variety of modes for a single ULA mass ($m_{a}=10^{-26}~{\rm eV}$), as shown in Fig. \ref{fig:overdensity_plot}. If $k<k_{\rm J}(a)$ at all times (as is the case if $k=10^{-4}h~{\rm Mpc}^{-1}$), the mode locks onto the CDM solution after an early period of DE-like behavior. 

If $k\sim k_{\rm J}(a)$ initially (as is the case if $k=0.1h~{\rm Mpc}^{-1}$), the mode shows suppressed growth initially, but has the same scaling with $a$ as the CDM case at late times, when $k>k_{\rm J}(a)$, yielding an overall suppression of power. Finally, if at early times, $k\gsim k_{\rm J}$ (as is the case for $k=0.3~{\rm Mpc}^{-1}$) the ULA perturbation oscillates rapidly until very late times ($a\sim 10^{-2}>a_{\rm osc}$), yielding a significant suppression of small-scale power. This illustrates why the matter power-spectrum is suppressed on small scales (as in Fig. \ref{matterspectra_dm_examples}) at the level of the mode evolution as a function of scale factor $a$. We discuss the detailed impact of altered mode evolution on cosmological observables in Sec.~\ref{sec:observables}.

The ULA hypothesis may have additional implications for cosmological structure formation. These include cored density profiles in dwarf-spheroidal galaxies \cite{hu2000,harko_d,sim_bec,bkain,2012MNRAS.422..135R,marsh2013b,schive2014}, suppressed number densities of Milky Way satellites \cite{marsh2013b} (providing a possible solution to well-known discrepancies between small-scale observations and the $\Lambda$CDM model, reviewed in Ref.~\cite{weinberg2013}), vortices/caustics in DM halos  \cite{2012MNRAS.422..135R,sikivie2010}, altered reionization due to delayed high-redshift galaxy formation \cite{Bozek:2014uqa}, and pulsar-timing searches for gravitational wave emission caused by coherently oscillating density profiles in DM halos \cite{khmelnitsky2013}. These techniques all depend on the nonlinear physics of ULAs in DM halos. For the rest of this work, we restrict our attention to the linear theory of ULA perturbations, which we now develop.

We begin in Sec. \ref{earlytime_eom} by describing the exact evolution of the scalar field in terms of fluid variables. In Sec. \ref{initcsec}, we discuss the initial conditions used in \textsc{camb} for the
combined system of ULAs, baryons, neutrinos, photons, CDM perturbations, self-consistently including the metric perturbation. Further details of the initial condition derivation are given in Appendix \ref{icinit}. We then derive, in Sec. \ref{latetime_eom}, the effective fluid EOMs in terms of the same fluid variables, valid in the coherently oscillating regime. Finally, in Sec. \ref{sec:camb_changes}, we summarize all the changes made to \textsc{camb} to compute cosmological observables for comparison with data in this work. During the preparation of this work, similar effective fluid methods have been independently developed and applied to novel coupled DM-DE systems \cite{Beyer:2014uqa}.

\subsection{Exact fluid equations for ULA perturbations}
\label{earlytime_eom}

The equations of motion (EOMs) for the Fourier modes of a perturbed scalar field $\phi=\phi_0(\tau)+\phi_1(\tau,\vec{k})$ [in synchronous gauge, with a Friedmann-Robertson-Walker (FRW) metric] are \cite{hu2000,perrotta1999}
\begin{align}
\ddot{\phi}_1+2\mathcal{H}\dot{\phi}_1+(m_a^2 a^2 +k^2)\phi_1&=-\frac{1}{2}\dot{\phi}_0\dot{\beta},
\label{eqn:field_eoms}
\end{align}
where $\beta$ is the trace of the scalar metric perturbation \cite{bertschinger1995}, $k$ is the comoving Fourier wave number of a perturbation, $a$ is the scale factor, and $\tau$ denotes conformal time. In the cosmological context, masses are always converted from units of ${\rm eV}$ to units of $h~{\rm Mpc}^{-1}$, where $h$ is the dimensionless Hubble constant today $h=H_{0}/(100~{\rm km}~{\rm s}^{-1}~{\rm Mpc}^{-1})$. There are four degrees of freedom coming from the perturbed scalar field: $\{\phi_0,\dot{\phi}_0,\phi_1,\dot{\phi}_1\}$. The components of the scalar-field energy-momentum tensor are found from these degrees of freedom in the usual way. 

In an arbitrary gauge, the components of the perturbed ULA energy momentum tensor are \cite{hu_covariant}:
\begin{align}
\delta\rho_a =&~ a^{-2}\left(\dot{\phi}_0\dot{\phi}_1 -\dot{\phi}_{0}^{2}A\right)+ m_{a}^2 \phi_0 \phi_1 \label{eqn:deltarho} \, ,\\
\delta P_a=&~a^{-2}\left(\dot{\phi}_0 \dot{\phi}_1 -\dot{\phi}_{0}^{2}A\right)- m_{a}^2\phi_0 \phi_1 \label{eqn:deltap} \, ,  \\
(\rho + P)(v_a-B) =&~a^{-2}k \dot{\phi}_0 \phi_1 \,   ,\label{comove_find}
\end{align}
where $A$ and $B$ are the scalar potential and vector longitudinal perturbations in the chosen gauge, respectively, to the metric tensor. A scalar field has no anisotropic stress at linear order in perturbation theory \cite{hu_covariant}. 

Using these definitions, one can exactly map the EOMs and four degrees of freedom onto those of a generalized DM (GDM) fluid, as shown in Ref. \cite{hu1998b}. The homogeneous (background) evolution is specified by the density $\rho_a$ and the equation of state $w_a$:
\begin{eqnarray}
\dot{\rho}_a &=&-3\mathcal{H}\rho_a(1+w_a)\, . 
\label{eqn:dot_rho} \\
w_{a}&=&\frac{P_{a}}{\rho_{a}}\label{wdef}.
\end{eqnarray} There are two degrees of freedom in the homogeneous scalar field equations, and so there is also an equation of motion for $P_{a}$ (and thus $w_{a}$).

After performing a gauge transformation, the GDM sound speed for the ULA is derived easily in the ULA comoving gauge, where the ULA perturbation $\phi_{1}$ vanishes. In this gauge, the ULA sound speed is easily seen to be \cite{hu_covariant}
\begin{equation}
c_{a}^{2}=\frac{\delta P}{\delta \rho}=1.\end{equation}
The GDM fluid EOMs in synchronous gauge then yield
\begin{align}
\dot{\delta}_{a}=&-ku_{a}-\left(1+w_{\rm a}\right)\dot{\beta}/2-3\mathcal{H}\left(1-w_{a}\right)\delta_{a}\nonumber \\-&9\mathcal{H}^{2}\left(1-c_{\rm ad}^{2}\right)u_{\rm a}/k,\label{eoma}\\
\dot{u}_{a}=&~2\mathcal{H}u_{a}+k \delta_{a}+3\mathcal{H}\left(w_{a}-c_{\rm ad}^{2}\right)u_{a}
\label{eomb},\end{align}where $\delta_{a}=\delta \rho_{a}/\rho_{a}$, and the adiabatic sound speed is
\begin{equation}
c_{\rm{ad}}^2\equiv \frac{\dot{P}_{a}}{\dot{\rho}_{a}}= w_{a}-\frac{\dot{w}_{a}}{3\mathcal{H}\left(1+w_{a}\right)}\label{adiabat_cs}.
\end{equation}
The dimensionless ULA heat flux is $u_{a}=(1+w_{a})v_{a}$. Equivalent fluid equations for a scalar field are obtained in Refs. \cite{beandore,lewisweller}. It is straightforward to show that this system is equivalent to the scalar field EOM, Eq.~(\ref{eqn:field_eoms}). These ULA EOMS are numerically solved along with the perturbed Einstein, fluid, and Boltzmann equations, in a modified version of \textsc{camb}, in order to compute CMB anisotropies and the matter power-spectrum.

We also need the contribution of ULA fluid variables to the source terms for the Einstein equations. In synchronous gauge, this is:
\begin{eqnarray}
\delta P_{a}&=&\rho_{a}\left[\delta_{a}+3\mathcal{H}(1-c_{\rm ad}^{2})v_{a}/k\right]\label{dpdef},\\
\delta \rho_{a}&=&\rho_{a}\delta_{a},\\
\left(\rho_{a}+P_{a}\right)v_{a}&=&\rho_{a}u_{a}.\label{dqdef}
\end{eqnarray}

\subsection{Initial conditions}
\label{initcsec}
To start off \textsc{camb} for any particular set of cosmological initial conditions, one needs a power series solution for all the fluid and metric variables, as the (non stiff) integrator used in \textsc{camb} can not be started at conformal time $\tau=0$, when the homogeneous densities of baryons, photons, DM, and neutrinos all diverge. \textsc{camb} begins the evolution of all modes when they are well outside the horizon ($x=k\tau\ll 1$) so we seek an expansion in powers of $x$. The relevant mode for our discussion is the adiabatic mode.\footnote{Note that we have also derived the power-series solutions for isocurvature modes, including the ULA isocurvature mode. We will discuss these and the associated observables in a future paper.} The power-series solution for this case is stated in Ref. \cite{bucher2000}, ignoring the contribution of ULAs to the cosmic energy density. 

We reproduce this solution using the eigenmode method of Refs. \cite{cambnotes,doran_wetterich,tristan_notes} in Appendix \ref{icinit}. We also confirm that this power-series solution is valid up to corrections of order $(k\tau)^{4}$ for metric and standard fluid perturbations, and $\tau/\tau_{\rm eq}$ for the ULA variables themselves, even when the contribution of ULAs to the energy density is included (here $\tau_{\rm eq}$ is the conformal time at matter-radiation equality). The initial conformal time for \textsc{camb} is already chosen such that these parameters are sufficiently small to obtain the required precision for comparison with all existing cosmological data of interest, and so we can safely neglect these corrections to the usual adiabatic initial conditions. We also require that the integration begins at an initial scale factor $a_{\rm init}<100 a_{\rm osc}$, where we set $\phi_{0}(\tau)={\rm constant}$ and $\dot{\phi_{0}}(\tau)=0$. In the adiabatic mode, ULA perturbations do not evolve or grow at leading order and early times, but this changes later when $m_{a}\gg 3H$, and the ULA begins to coherently oscillate, a regime we treat using an effective fluid approximation.
\begin{figure}
\includegraphics[width=0.49\textwidth]{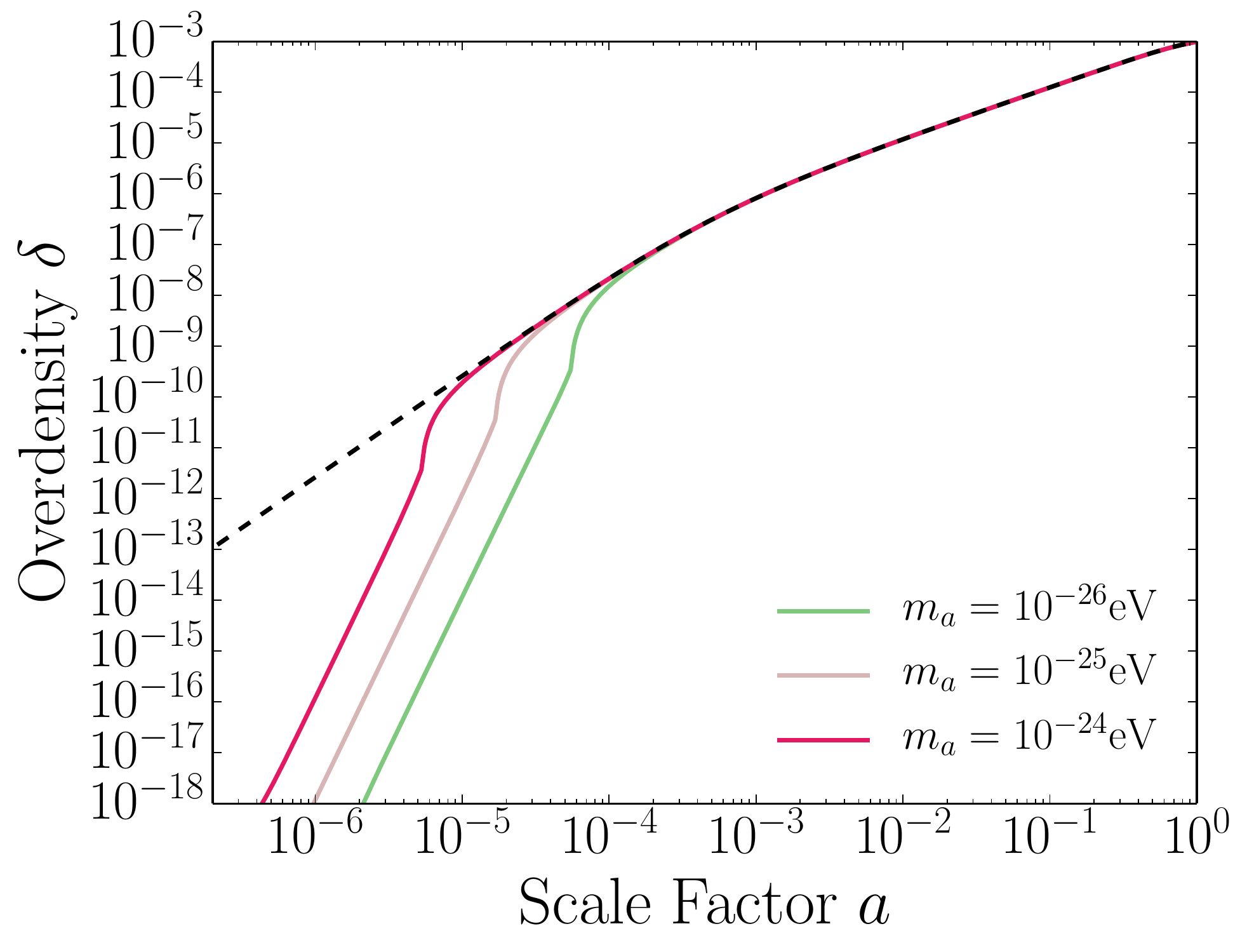}
\caption{Evolution of the fractional dark-matter density perturbation with wave number $k=10^{-4}h~{\rm Mpc}^{-1}$ for the $3$ different ULA masses indicated compared to the standard CDM case (dashed). For these ULA masses, $k<k_m$ always, and so soon after $a>a_{\rm osc}$, the mode behaves just as CDM.}
\label{fig:overdensity_initial}
\end{figure}
\subsection{Effective fluid equations for ULAs}
\label{latetime_eom}
Using the EOMs from Sec. \ref{earlytime_eom} with the initial conditions just discussed, and choosing the initial conformal time so that $\tau_{\rm init}\ll \tau_{\rm osc}$ and $\rho_{a}\ll \rho_{\gamma}, \rho_{a}\ll \rho_{\nu}, \rho_{a}\ll \rho_{m}$, we use \textsc{camb} to evolve the full system when $a<a_{\rm osc}$. We solve independently for the background quantities $\phi$, $c_{\rm ad}^{2}$ and $w_{\rm a}$, and use the history of $w_{\rm a}$ and $\dot{w}_{\rm a}$ to correctly compute the perturbation evolution.The initial value for $\phi_{0}$ is chosen using the shooting method to obtain the correct relic density via Eq.~(\ref{homorelic}) and the numerical solution for $\phi(a)$. 

The homogeneous ULA fields remain roughly frozen at their initial values until the mass overcomes the Hubble friction, at which point they coherently oscillate with decaying amplitude and frequency $m_{a}$. At times when $m_{a}\gg 3H$ these oscillations give rise to a large separation of time scales and direct integration of the scalar field EOMs becomes computationally prohibitive, even for modest ULA masses ($m_{a}\gsim 10^{-27}~{\rm eV}$). 

To address this difficulty, we use the WKB method to obtain an effective fluid approximation for perturbations, averaging over the fast-time scale in the problem and writing evolution equations for the fluid variables averaged over the oscillation time scale $m_{a}^{-1} $\cite{khlopov_scalar,branden,peebles1988,nambu1990,ratra1991a,ratra1991b,hu2000,hwang_noh_a,axiverse2009,hwang_noh_b,marsh2010,sim_bec,Noh:2013coa}. The behavior of the system is that of a fluid with the asymptotic behavior shown in Eq.~(\ref{heuristic_cs}), leading to suppressed structure growth on scales $k\gg k_{\rm m}$, with a dramatic cutoff when $k\gg k_{\rm J}$.  Precisely, in an arbitrary gauge, the EOM for a scalar-field perturbation is \cite{hu_covariant}
\begin{widetext}
\begin{equation}
\ddot{\phi}_{1}=-2\mathcal{H}\dot{\phi}_{1}-\left(k^{2}+a^{2}m_a^{2}\right)\phi_{1}+\left(\dot{A}-3\dot{H}_{L}-kB\right)\dot{\phi}_{0}-2Aa^{2}m_a^{2}\phi_{0}=0 \, ,\label{sf_gauge}
\end{equation}
\end{widetext}
where $H_{L}$ is the scalar perturbation to the spatial curvature. Following Refs. \cite{hwang_noh_a,hwang_noh_b}, we make the \textit{ansatz} that $\phi_0(\tau)=[\phi_{+}(\tau)\cos{(m_a\tau)}+\phi_{-}(\tau)\sin{(m_a\tau)}]/a^{3/2}$ and $ \phi_{1}=\delta \phi_{+}(k,\tau)\cos{(m_a\tau)}+\delta \phi_{-}(k,\tau)\sin{(m_a\tau)}$. We choose the ``comoving gauge" defined with respect to the oscillation-averaged fluid [that is, we set $v=B$ in Eq.~(\ref{comove_find}), which requires that $\delta \phi_{-} (k,\tau)\phi_{+}(m_a,\tau)=\delta \phi_{+}(k,\tau)\phi_{-}(m_a,\tau)$].

Substituting our \textit{ansatz} into Eqs.(\ref{eqn:rhoa})-(\ref{eqn:deltap}) and Eq.~(\ref{sf_gauge}), and assuming that metric perturbations vary only on conformal time scales $\tau\sim\mathcal{H}^{-1}\gg m_a^{-1}$, we obtain equations which can be grouped by powers of $\mathcal{H}/m_a$. We find that to leading order in $\mathcal{H}/m_a$, and when $a\gg a_{\rm osc}$,
\begin{equation}
c_{a}^{2}\equiv\frac{\delta P}{\delta \rho}=\frac{k^2/(4m_a^2a^2)}{1+k^{2}/(4m_a^{2}a^{2})},\label{cslate} \end{equation}
which smoothly interpolates between the asymptotic regimes given in Eq.~(\ref{heuristic_cs}). Going back to synchronous gauge [and taking average values over the fast time scale, that is, $w_{a}\simeq 0$ and $c_{\rm ad}\simeq0$, both easily obtained from the solution for $\phi_0(\tau)$, Eq.~(\ref{wdef}), and Eq.~(\ref{adiabat_cs})], the effective fluid equations for ULAs (when $a\gg a_{\rm osc}$) are
\begin{eqnarray}
\dot{\delta}_{a}&=&-ku_{a}-\frac{\dot{\beta}}{2}-3\mathcal{H}c_{a}^{2}\delta_{a}-9\mathcal{H}^{2}c_{a}^{2}u_{a}/k,\label{dens_late}\\
\dot{u}_{a}&=&-\mathcal{H}u_{a}+c_{a}^{2}k\delta_{a}+3c_{a}^{2}\mathcal{H}u_{a}.\label{v_late}
\end{eqnarray} 

To compute the evolution of ULA perturbations in \textsc{camb}, we use Eqs.~(\ref{eoma})-(\ref{eomb}) when $a<a_{\rm osc}$ together with the numerical background evolution of $\rho_a,w_a$. At late times when $a\geq a_{\rm osc}$ we use Eqs. (\ref{dens_late})-(\ref{v_late}) , with $\rho_a\propto a^{-3},w_a=0$. To be sure that this sudden transition does not produce numerical artifacts in the modified \textsc{camb} output, we verified that results are insensitive to changes in the exact matching time of order $\delta \tau=10 m^{-1}$. We also checked the code against a version of \textsc{camb} that directly solves for the perturbed scalar field, and for masses as high as $m_{a}\sim 10^{4}H_{0}$, found agreement between the exact and effective fluid treatments. The approximation improves at higher $m_{a}$ values, as the transition happens over shorter and shorter intervals compared to the whole of cosmic time. Since this mass is deep into the coherent oscillation regime today, we are confident that our approximations are valid over the full mass range considered, as discussed further in Sec. \ref{sec:camb_changes}.

\subsection{Summary of changes to CAMB and key physical effects}
\label{sec:camb_changes}
We self-consistently include the effect of ULAs on the homogeneous expansion history by numerically solving Eq.~(\ref{homo_eom}), including the ULA energy density in the computation of $H$ using the Friedmann equation. Using a shooting method, the initial value $\phi_{0}$ is chosen to obtain the desired input value of $\Omega_{a}/\Omega_{d}$ to a precision of $10^{-4}$. Additionally, we include the contributions of ULAs to $\mathcal{H}$ everywhere in \textsc{camb} that the Hubble expansion rate is needed, including the \textsc{RecFast} \cite{Seager:1999bc} recombination module itself and the calculation of the visibility function. Early-time ($m\leq 3H$) evolution of perturbations is followed using the equations of Sec. \ref{earlytime_eom}, with initial conditions set as discussed in Sec. \ref{initcsec} and Appendix \ref{icinit}. Late-time ($m\geq 3H$) evolution is followed using the equations of Sec. \ref{latetime_eom}. 

We now discuss the evolution of specific modes (output by our modified version of \textsc{camb}) in several cases of interest, in order to highlight some of the physical effects driving the behavior of the observable power spectra discussed in Sec.~\ref{sec:observables}. As already discussed in Sec. \ref{model}, Fig.~\ref{fig:overdensity_plot} shows the behavior of a range of modes for ULAs with $m_{a}=10^{-26}~{\rm eV}$. We see there that if ULAs constitute all the DM and the perturbation wavelength is smaller than or of order the ULA Jeans scale, linear structure growth is arrested until a later time. 

Evolution of a DM density perturbation with $k=10^{-4}h~{\rm Mpc}^{-1}$ is shown in Fig.~\ref{fig:overdensity_initial}. For this large-scale mode ($k\ll k_m$) and a large (CDM-like) value of $m_{a}$, we expect the ULA to behave as CDM. Once $a\gsim a_{\rm osc}$, the initial conditions are forgotten and the mode locks onto the universal CDM-like behavior. For higher $m_{a}$, $a_{\rm osc}$ is lower and CDM-like behavior begins earlier.

\begin{figure}
\includegraphics[width=0.49\textwidth]{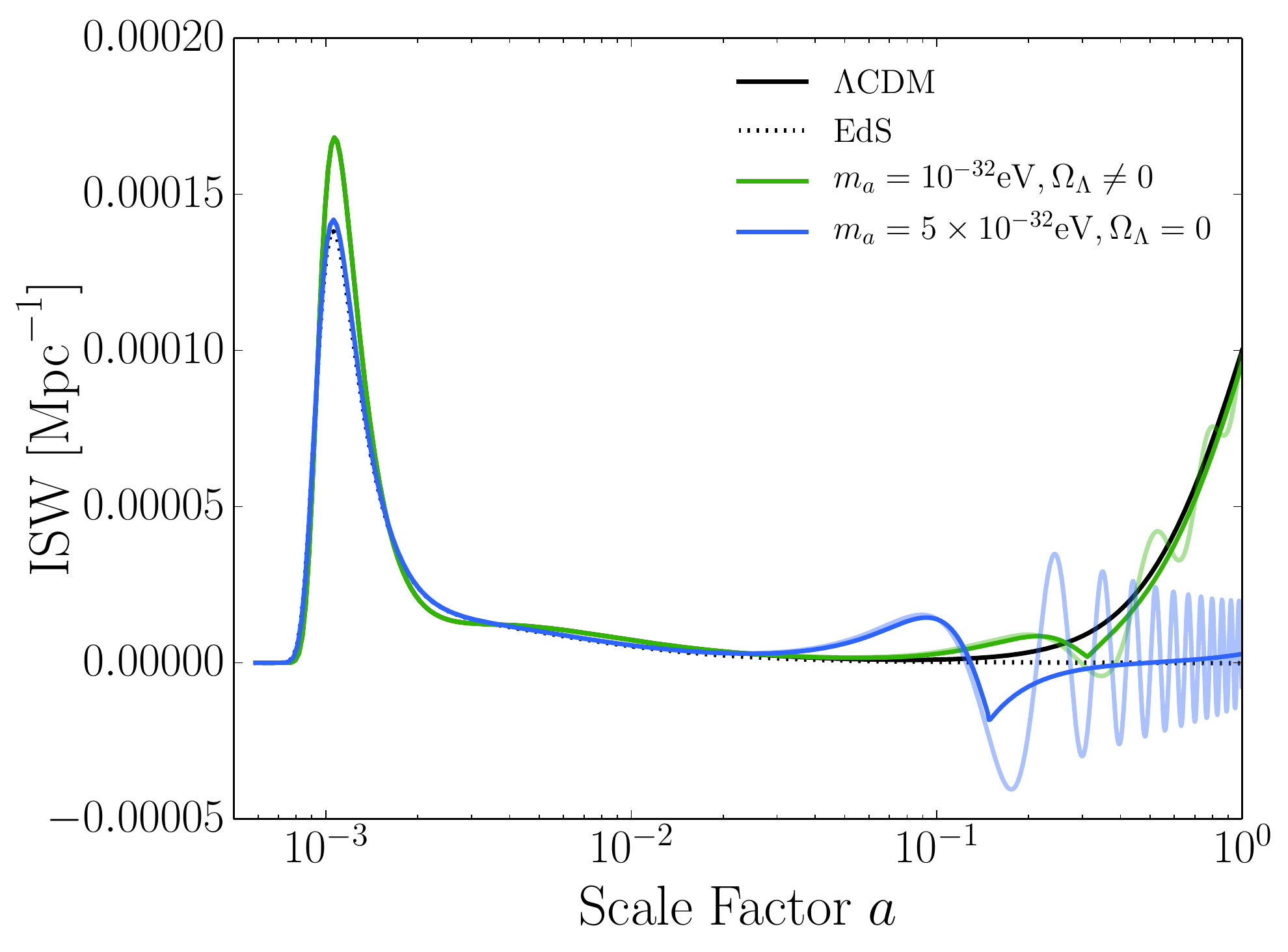}
\caption{Evolution of the integrated Sachs-Wolfe (ISW) source term \cite{camb} for a mode with $k=10^{-4}h~{\rm Mpc}^{-1}$. The overall amplitude is arbitrary. Dark colored curves are generated using the modified \textsc{camb} described in the text. Lighter curves are generated using direct numerical integration of scalar-field perturbation EOMs. Green curves show the effect of choosing $\Omega_{a}/\Omega_{d}=0.1$ (with all other parameters set to $\Lambda$CDM values) with $m_{a}=10^{-32}~{\rm eV}$. Blue curves are obtained assuming $\Omega_{\Lambda}=0$, $\Omega_{m}=1$
and $\Omega_{a}/\Omega_{d}=0.1$ with $m_{a}=5\times 10^{-32}~{\rm eV}$.}
\label{fig:ISW_plot}
\end{figure}

In Fig. \ref{fig:ISW_plot}, we show the behavior of the integrated Sachs-Wolfe (ISW) source term (see Ref. \cite{camb} for a definition) for a long-wavelength mode ($k=10^{-4}h~{\rm Mpc}^{-1}$) in $\Lambda$CDM and Einstein-deSitter (EdS) cosmologies as well as cosmologies which include ULAs with rather low masses ($10^{-32}~{\rm eV}-5\times 10^{-32}~{\rm eV}$), treating ULA perturbations using the effective fluid formalism and modified \textsc{camb} described above. 
The EdS cosmology is defined by the values $\Omega_{m}=1$, $\Omega_{\Lambda}=0$).

When low-mass ($m=10^{-32}~{\rm eV}$) ULAs replace some of the DM, there is an enhancement of the ISW effect due to the early DE-like behavior of ULAs. When $a>a_{\rm osc}$, these ULAs begin to behave as CDM, leading the ISW source term to reconverge to the $\Lambda$CDM behavior. The small deviation from $\Lambda$CDM behavior for scales that enter the horizon when $a<a_{\rm osc}$ will drive the CMB constraint for comparable ULA masses, as we discuss further in Secs. \ref{sec:observables} and \ref{results}.

As another example, we set $\Omega_{m}=1$, $\Omega_{\Lambda}=0$, and $\Omega_{a}=0.1$, with a higher ULA mass of $m_{a}=5\times 10^{-32}~{\rm eV}$. Because of their early DE-like behavior, these ULAs initially enhance the ISW source term. The higher $m_{a}$ (and lower $a_{\rm osc}$) value, however, causes CDM-like behavior to set in earlier than the preceding case.  ISW source term then closely tracks the EdS case, with a nearly vanishing late-time ISW effect.

For both ULA parameter sets in Fig. \ref{fig:ISW_plot}, we compare mode evolution in the effective fluid treatment with that obtained by directly numerically integrating the EOMs of scalar-field perturbations, using a code described in \cite{marsh2011b}. As expected, the onset of CDM-like behavior in the ULAs corresponds to the onset of coherent oscillation in the scalar-field perturbation, and occurs earlier for higher $m_{a}$ values. Averaged over time scales greater than $\sim m_{a}^{-1}$, the behavior in the effective fluid treatment agrees with the full evolution of the scalar field for both cases. This is one of several tests we used to verify that the effective fluid treatment agrees with the full scalar-field evolution.

\section{Cosmological Observables}
\label{sec:observables}
\begin{figure*}[htbp!]
\begin{center}
\includegraphics[width=0.49\textwidth]{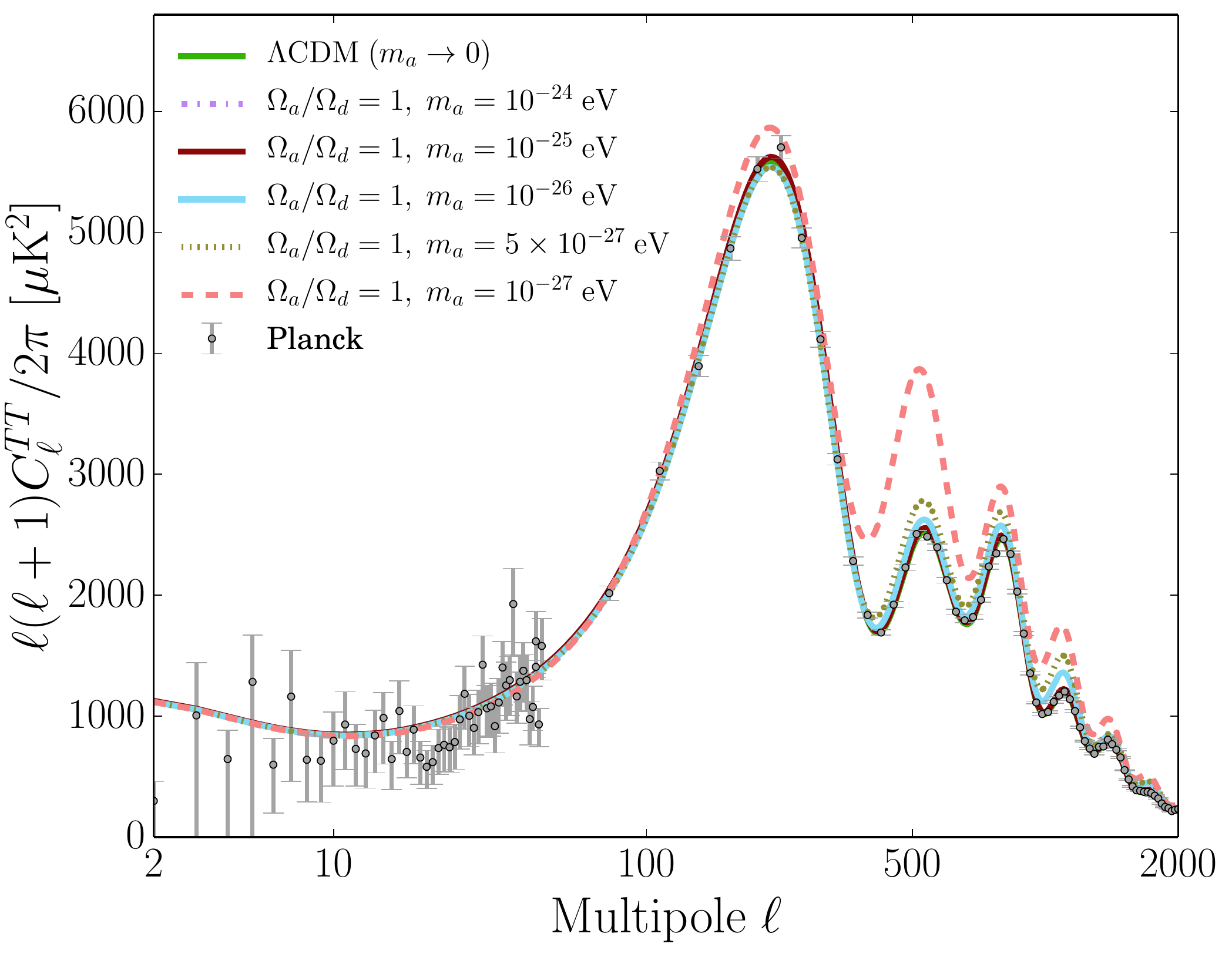}
\includegraphics[width=0.49\textwidth]{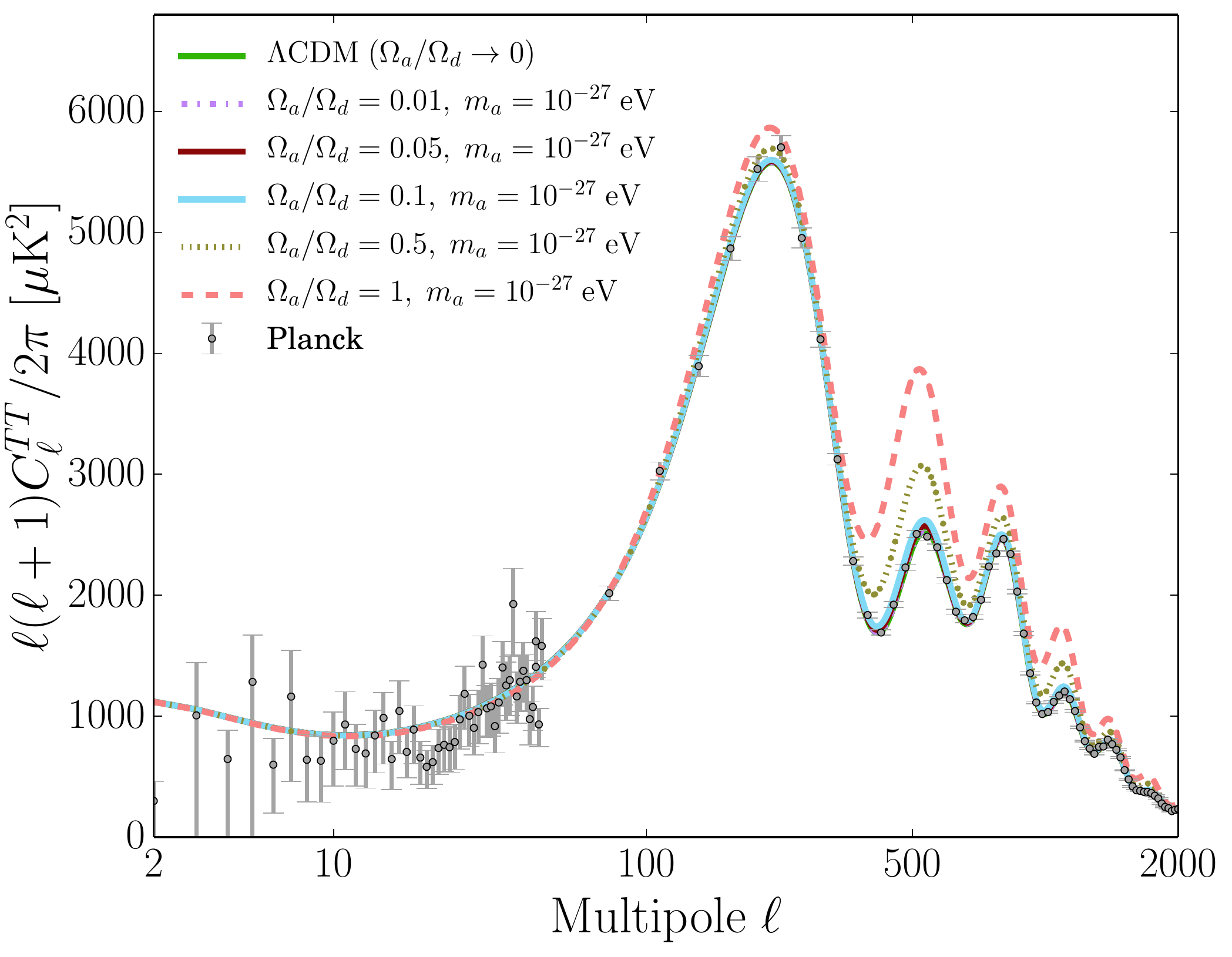} 
\caption{CMB temperature power-spectrum with varying ULA mass and energy-density fraction $\Omega_a/\Omega_d$. Here, as in Fig.~\ref{matterspectra_dm_examples}, we introduce ULAs as a fraction of the dark matter, holding $\Omega_d=\Omega_a+\Omega_c$ fixed. Since ULAs have $w_a=-1$ for some time during the radiation era this changes the ratio of matter to radiation and alters the relative heights of the CMB acoustic peaks. For dark-matter like ULAs with the highest ULA masses, and lowest fractions, $C_\ell^{TT}$ becomes indistinguishable from $\Lambda$CDM, with the $\Lambda$CDM curve lying directly underneath the ULA curve. \label{cmbspectra_dm_examples}}
\end{center}
\end{figure*}
Before using CMB and galaxy-clustering data to search for ULAs, we explore the observables and estimate the expected level of constraints. All power spectra are computed using the modified version of \textsc{camb} described in Sec.~\ref{axion_fluid_pert}.

The overall behavior of the adiabatic matter power-spectrum in the presence of a ULA playing the role of DM can be understood using two simple physical effects \cite{khlopov_scalar,branden,peebles1988,nambu1990,ratra1991a,ratra1991b,hu2000,amendola2005,hwang_noh_a,arvanitaki2010,hwang_noh_b,marsh2010,marsh2011b,sim_bec}. The first is that the ULA equation-of-state transitions from the DE-like $w_{a}\approx-1$ for $a\ll a_{\rm osc}$ to the DM-like $w(a)\approx 0$ for $a\gg a_{\rm osc}$. This leads to new nontrivial behavior of the ratio of CDM+ULA energy density to the radiation energy-density, shifting the redshifts of equality, recombination, and decoupling. The second observable effect of ULAs is the scale-dependent sound speed of the ULA fluid, which leads to suppressed clustering power on small scales. The amplitude of both effects increases with the fraction of matter composed of ULAs, and is more dramatic for lower ULA mass, as seen already in Fig. \ref{matterspectra_dm_examples}. 

We now discuss the effects of a ULA on the CMB, when it is either DM- or DE-like. We also discuss the effects of a DE-like ULA on the matter power-spectrum, as well as its effect on the observable galaxy power-spectrum.
\subsection{The CMB}
In Fig.~\ref{cmbspectra_dm_examples} we show the two-point temperature auto-correlation power spectrum, $C_\ell^{TT}$, for the same models as Fig.~\ref{matterspectra_dm_examples}, where ULAs are introduced as a fraction of the DM, holding $\Omega_d=\Omega_a+\Omega_c$ fixed. Introducing a fraction of DM that has $w=-1$ for some period of cosmic history changes the matter-to-radiation ratio compared with the same ratio in a pure CDM Universe. This changes the structure of the acoustic peaks of the CMB. The change is most severe for the lightest ULAs where $w=-1$ for longer, and increases with the fraction of DM in ULAs.

With $m_a>H(z_{\rm eq})\sim 10^{-27}~{\rm eV}$ the ULAs behave as matter throughout the matter-dominated era and so this leaves the large scale, low $\ell$, of the CMB power unchanged, as the late-time growth and expansion rate imprinted by the ISW effect is not altered. Since the expansion rate is not altered, the angular size of the sound horizon is also not changed much, and so the location and size of the first acoustic peak remains unaltered also for $m_a\gtrsim 5 \times 10^{-27}\text{ eV}$. Indeed the constraining power of WMAP1 in Ref.~\cite{amendola2005} cuts out at around this mass scale. Without accurate measurements of the higher acoustic peaks, only the lightest ULAs that oscillate in the matter era and change the ISW plateau or the distance to the last scattering surface could be constrained by WMAP.  Looking at the second third and fourth acoustic peaks, however, which are well measured in \textit{Planck}, ACT and SPT data, we see that the CMB can distinguish slightly larger masses of $m_a\sim 10^{-26}\text{ eV}$ at a fraction of around $\Omega_{a}/\Omega_{d}=0.05$. We therefore expect $\sim 10\%$-level sensitivity to the ULA DM mass fraction for all masses $m_a\lesssim 10^{-26}\text{ eV}$.

We have so far considered the effects of introducing ULAs to the DM that are heavy enough to leave the large angle CMB unchanged. What about the lighter ULAs that do alter the low-$\ell$ CMB temperature power spectrum\cite{marsh2011b}? ULAs with $m_a<10^{-27}\text{ eV}$ have $a_{\rm osc}>a_{\rm eq}$, therefore in order to keep the physical condition that matter-radiation equality be unchanged so that there are bound objects formed on small scales, in all the following examples we choose to keep $\Omega_c h^{2}=0.120$ fixed. In order to see the effects on the CMB of introducing the lightest ULAs, we discuss various cases holding other parameters fixed. 
\begin{figure*}[htbp!]
\begin{center}
\includegraphics[width=0.49\textwidth]{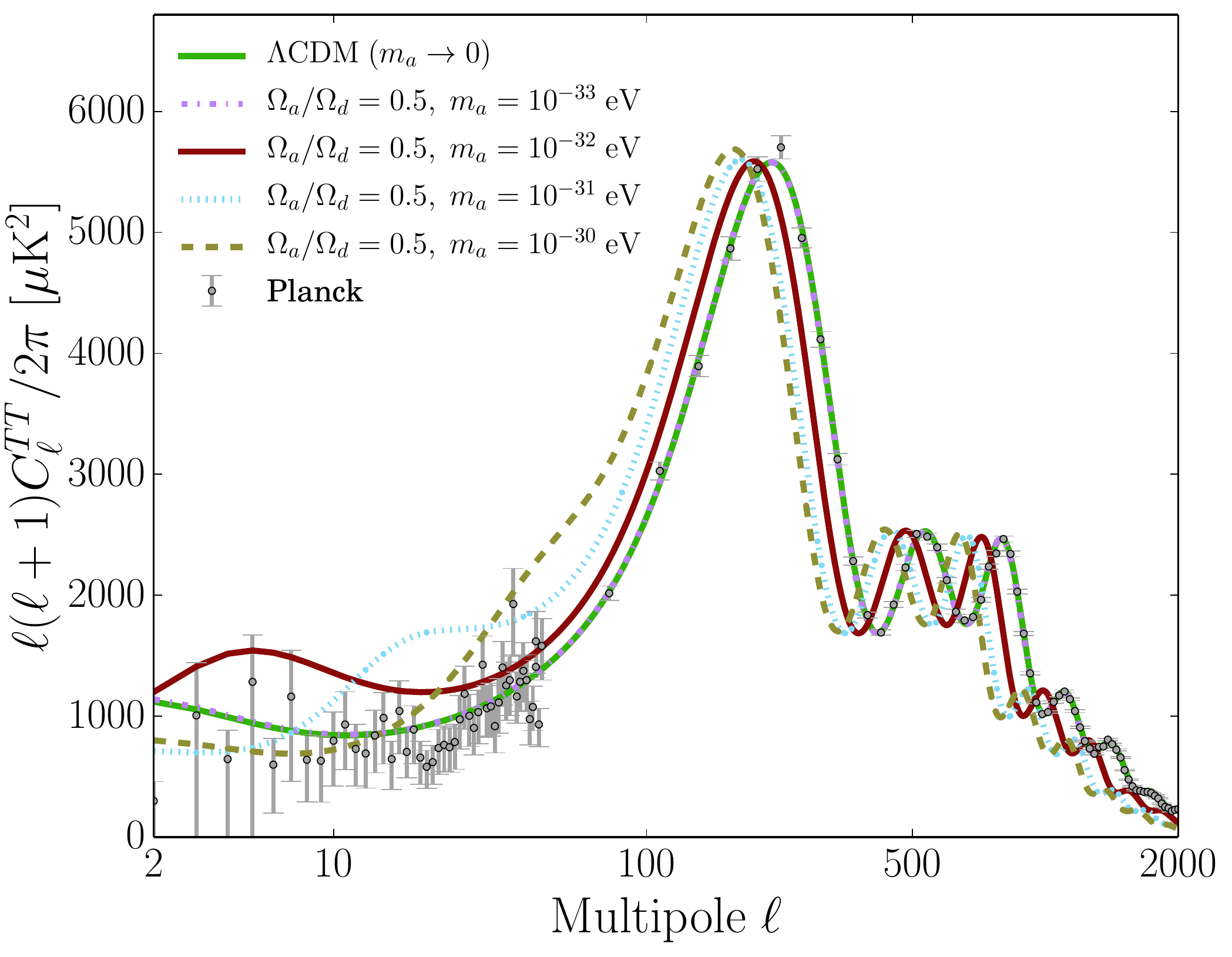}
\includegraphics[width=0.49\textwidth]{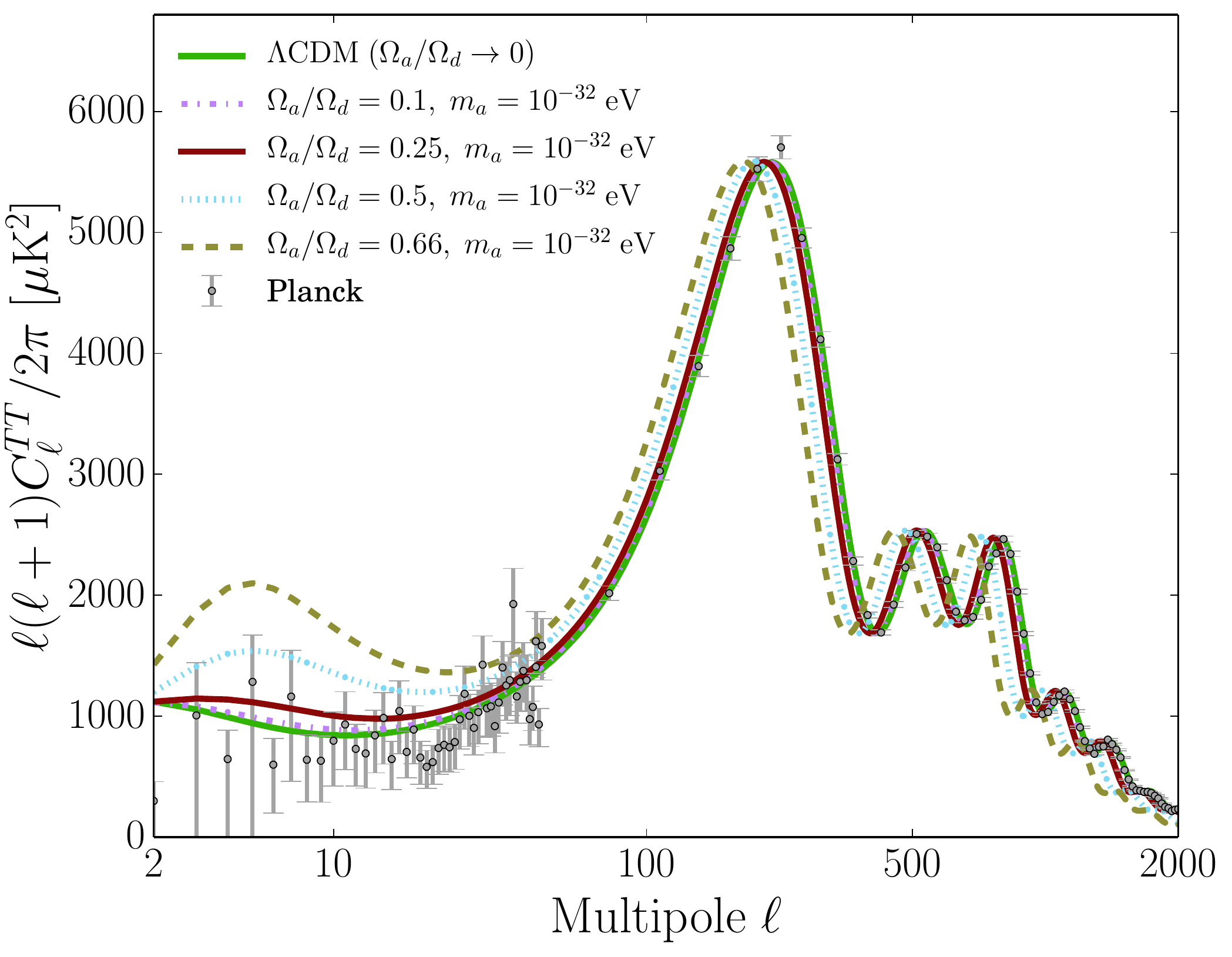} 
\caption{CMB temperature power-spectrum with varying ULA mass and energy-density fraction $\Omega_a/\Omega_d$. Here, we introduce the lightest ULAs as a fraction of the dark energy, holding $\Omega_c h^2$ and $H_0$ fixed so that maintaining flatness while introducing ULAs reduces $\Omega_\Lambda$. The lightest ULAs transition to matter-like behaviour late in the lifetime of the Universe and can contribute to the dark energy. The visible effects come from the change in the age of the Universe, which changes the angular size of the sound horizon, and in changing the integrated effect of dark energy, which changes the amplitude of the ISW plateau. For dark-energy like ULAs with the lowest ULA masses, and lowest fractions, $C_\ell^{TT}$ becomes indistinguishable from $\Lambda$CDM, with the $\Lambda$CDM curve lying directly underneath the ULA curve. \label{fig:cmbspectra_de_examples_h0}}
\end{center}
\end{figure*}

In Fig.~\ref{fig:cmbspectra_de_examples_h0} we introduce $\Omega_a h^2\neq 0$ holding $H_0$ (and thus also the fractional density of DM) fixed. As we are also holding $\Omega_c h^2$ fixed, introducing ULAs in this way reduces the amount of DE. The ULAs introduced act as DE while $a<a_{\rm osc}$. This is during the matter or $\Lambda$ era, and so we refer loosely to the lightest ULAs as ``DE-like". Since the scale of structure suppression for these ULAs is comparable to the scale of structure suppression for $\mathcal{O}({\rm eV})$ mass neutrinos they could also be said to be ``neutrino-like", or ``HDM-like" \cite{amendola2005,marsh2011b}, although we will find the analogy to DE more useful here.

In the left panel of Fig.~\ref{fig:cmbspectra_de_examples_h0} we fix $\Omega_{a}/\Omega_{d}$ and vary the ULA mass. For the fiducial cosmology shown, taking $\Omega_{a}/\Omega_{d}=0.5$ reduces $\Omega_\Lambda$ from $\Omega_\Lambda=0.68$ to $\Omega_\Lambda=0.42$, so ULAs make up a little over a third of the DE density. The integrated contribution of DE is changed in the ULA cosmology, which has a number of effects. The age of the Universe is smaller in the ULA cosmologies, being reduced from $13.8\times 10^9$ years in the fiducial cosmology to $11.5\times 10^9$ years with $m_a=10^{-30}\text{ eV}$. This reduces the distance to the surface of last scattering, and so increases the angular size of the sound horizon, $\theta_A$, shifting the locations of the CMB acoustic peaks to lower $l$. Since the integrated effect of DE is altered, the ISW plateau is also changed relative to $\Lambda$CDM. The lightest ULA we consider has $m_a=10^{-33}\text{ eV}$, and is so close to $\Lambda$ in the evolution of the energy density that it has no discernible effects on the CMB, regardless of how much of the energy density it makes up, as long as flatness is maintained. In the right panel of Fig.~\ref{fig:cmbspectra_de_examples_h0} we fix $m_a=10^{-32}\text{ eV}$ and vary the ULA relic-density, so varying $\Omega_\Lambda$ at fixed $H_0$.

For low mass ULAs, the ULA relic-density is degenerate with the value of $\theta_A$ at fixed $H_0$. We now explore the effect of ULAs on the CMB holding $\theta_A$ fixed by varying $H_0$. Compared to Fig.~\ref{fig:cmbspectra_de_examples_h0} this will shift the locations of the acoustic peaks back towards their $\Lambda$CDM locations and shift the ULA effects largely into the ISW. We hold the $l$ value of the first acoustic peak in $C_{l}^{\rm TT}$ (and thus also of the higher acoustic peaks) fixed, which requires reducing $H_0$ at fixed $\Omega_c h^2$ and $\Omega_a h^2$. For example, with $m_a=10^{-32}\text{ eV}$ and $\Omega_{a}/\Omega_{d}=0.25$, $H_0$ is reduced from $67.15\text{ km s}^{-1}\text{Mpc}^{-1}$ to $50.15\text{ km s}^{-1}\text{Mpc}^{-1}$ to maintain constant $\theta_A$. As $H_0$ is lowered at fixed $\Omega_c h^2$ and $\Omega_a h^2$ in order to maintain flatness eventually one finds $\Omega_\Lambda<0$. We exclude such situations by prior. They can lead to a collapsing Universe at $a\leq 1$, and will always collapse in the future. They are ruled out by any reasonable prior on $H_0$. Not all values of $\theta_A$, $\Omega_{a}/\Omega_{d}$ and $m_a$ are therefore consistent with our prior. In Fig.~\ref{fig:cmbspectra_de_examples_theta} we show a selection of models where varying $H_0$ can be used to fix the $l$ values of the acoustic peaks. From this the DE-like nature of the lightest ULAs is clear: they alter the shape of the ISW plateau of the CMB and effects on small scales can be absorbed by lowering $H_0$.

From the preceding discussion of DE-like ULAs it should be clear that the CMB can constrain ULAs of this type with $m_a\gtrsim 10^{-32}\text{ eV}$. Changes to $C_{\ell}^{TT}$ are large for $\Omega_{a}/\Omega_{d}>0.1$ and require extreme values of $H_0$, which suggests constraints at least at the level $\Omega_{a}/\Omega_{d}\sim 10^{-2}$ taking into account all degeneracies, consistent with the results of Ref.~\cite{amendola2005}. Even for the lightest mass we consider, $m_a=10^{-33}\text{ eV}$, which behaves almost indistinguishably from a cosmological constant, $\Omega_{a}/\Omega_{d}$ is constrained to be less than unity. Consider taking all the DM to be CDM, and all the DE to be this ULA. In the $\Lambda$CDM cosmology one has $\Omega_c\lesssim 3 \Omega_{DE}$ at a high level of confidence, which gives $\Omega_{a}/\Omega_{d}\lesssim 0.75$. This provides an approximate upper bound to $\Omega_{a}/\Omega_{d}$ even for the lightest DE-like ULAs.

\subsection{The matter power-spectrum revisited}\label{mps:revisit}

We now turn to the effect of DE-like ULAs on the matter power-spectrum, as well as the more subtle effect of ULAs on the galaxy power-spectrum, which requires an approximate treatment of scale-dependent bias. The matter power-spectrum, $P(k)$, is defined from the matter overdensity, $\delta_m$, and is related to the observed galaxy power-spectrum, $P_{\rm gal}(k)$, by the linear bias, $b$ as 
\begin{equation}
P_{\rm gal}(k)= b^2 P(k) \, .
\end{equation}
Galaxies are assumed to follow the total matter-density in a prescribed manner, which fixes the form of $b(k)$ which is fit from simulations and included in the likelihood when using galaxy power-spectrum data \cite{blake2011,blakeetal:2011,blaketal}.

On large scales, CDM and galaxies both cluster and have the same linear growth. To a first approximation the bias is constant and relates the amplitudes of the power spectra. ULAs, however, have scale-dependent growth and do not end up in collapsed structures on all scales. Clearly, the galaxy field is uncorrelated with the ULA density field on scales where ULAs do not form structure. On these scales, galaxies can only trace whatever component of the matter is still clustered. If $P(k)$ is the total matter power-spectrum including ULA perturbations, then specifying what portion of the matter fluctuations the galaxies trace on a given scale amounts to specifying a scale-dependent bias, $b(k)$. We will treat the problem of scale-dependent bias by asking the question ``when do we include ULAs as part of the ``matter'"in the matter power-spectrum?''

The importance of this issue for DE-like ULAs can be illustrated with a simple example. This will demonstrate an approximate way to treat the problem, which we will adopt here. A full solution to the problem, following Ref.~\cite{LoVerde:2014pxa}, is deferred for future study.
\begin{figure}[tp!]
\includegraphics[width=0.49\textwidth]{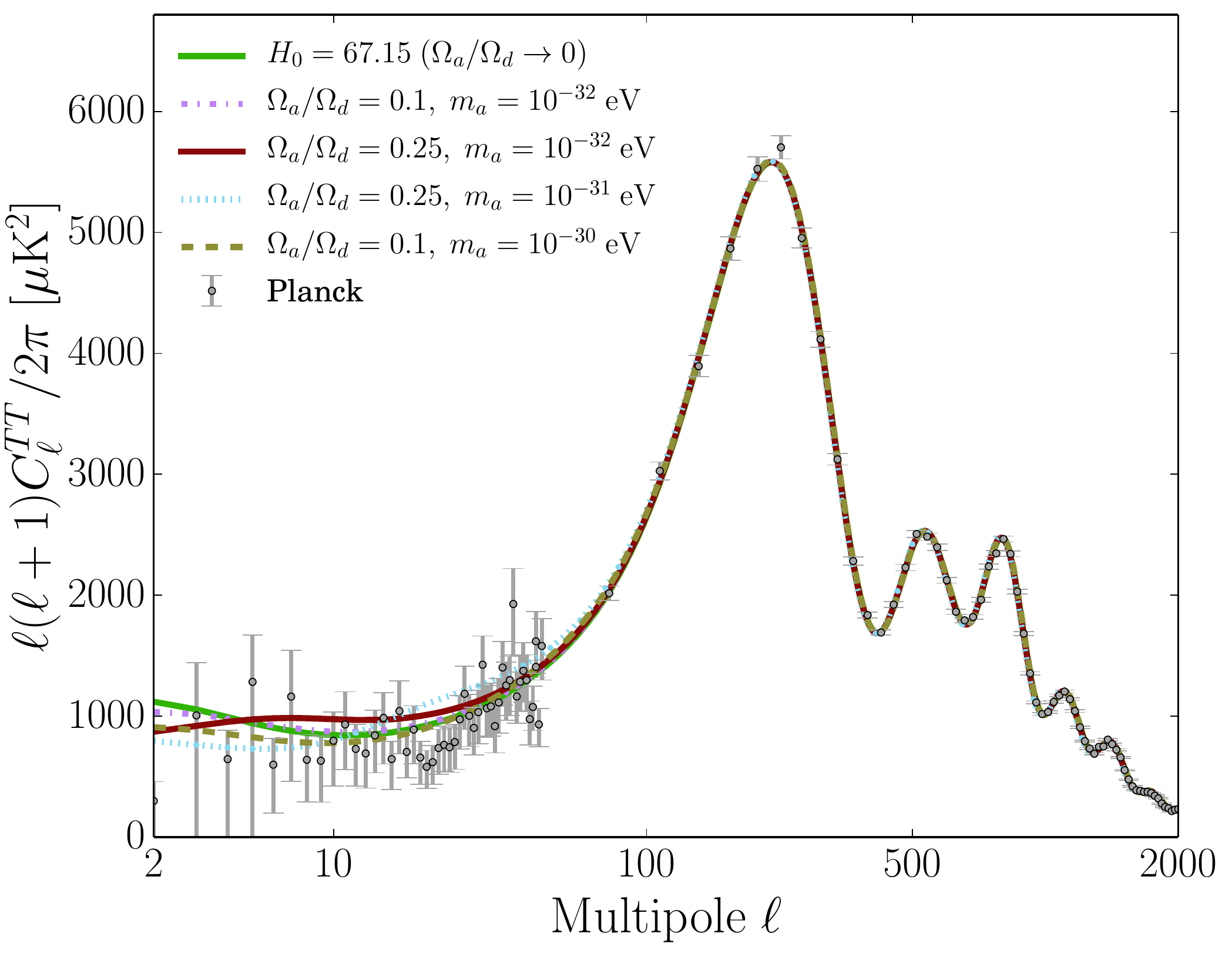}
\caption{CMB temperature power-spectrum with varying ULA mass and energy-density fraction $\Omega_a/\Omega_d$. Here, we introduce the lightest ULAs as a fraction of the dark energy, holding $\Omega_c h^2$ and $\theta_A$ fixed, which requires varying $H_0$. With the angular size of the sound horizon fixed, the ULAs only affect the CMB by altering the shape of the ISW plateau. Low values of $H_0\sim 60\text{ km s}^{-1}\text{Mpc}^{-1}$ were necessary in these examples to keep $\theta_A$ fixed, and certain cosmologies cannot be brought to fixed $\theta_A$ while maintaining an expanding Universe with $\Omega_\Lambda \geq 0$. \label{fig:cmbspectra_de_examples_theta}}
\end{figure}

For high $m_{a}$ values, ULAs behave as DM on large scales. For these values, we wish to include ULAs in the matter density so that for sufficiently high $m_{a}$ they can completely replace the CDM and fit the observed $P(k)$. This suggests the definition
\begin{align}
\delta \rho_m &= \delta \rho_c + \delta \rho_b + \delta\rho_a \, , \nonumber\\
\bar{\rho}_m & = \rho_c+\rho_b + \rho_a \, , \nonumber\\
\delta_m & = \delta \rho_m/\bar{\rho}_m \,. \nonumber\\
\label{eqn:rhom_def_wrong_bias}
\end{align}

On the other hand, when $m_a$ is small, ULAs do not cluster on any of the scales observed in a galaxy survey. Consider the extreme case of $m_a<H_0$. Such a ULA does not cluster on any subhorizon scales, so that $\delta\rho_a\approx 0$. We can replace $\rho_\Lambda$ with $\rho_a$, while holding $\rho_c$ fixed at its $\Lambda$CDM value. The left panel of Fig.~\ref{fig:cmbspectra_de_examples_h0} demonstrates that replacing $\Lambda$ by a ULA with $m_a=10^{-33}\text{ eV}<H_0$ produces no observable effect on the CMB. This parameter choice has exactly the same CMB temperature and lensing spectra, since the gravitational potentials, which are physical observables, always self-consistently feel the entire energy-density content regardless of what is included in the definition of ``matter". The evolution of the potential is not affected by replacing $\Lambda$ by a ULA with $m_a<H_0$. 

Using the definition in Eqs.~(\ref{eqn:rhom_def_wrong_bias}), we find that the shape of $P(k)$ is not changed relative to $\Lambda$CDM, since $\delta\rho_a\approx\delta\rho_\Lambda=0$. The amplitude, on the other hand, changes by a factor of $[\Omega_m/(\Omega_m+\Omega_\Lambda)]^2\sim\mathcal{O}(0.1)$ because of the increase in $\bar{\rho}_m$ for this definition. The change in $P(k)$ would unfairly penalize the $m_a<H_0$ cosmology in the likelihood relative to $\Lambda$CDM despite their physical equivalence on all observable times and scales.\footnote{One could try to restore agreement with the data by increasing $\Delta^2_{\mathcal{R}}$ to absorb this suppression. The primordial power-spectrum $\Delta^2_{\mathcal{R}}$, however, is also constrained by CMB data. If $m_a<H_0$ the suppression can be absorbed into the large-scale (constant) bias, but this is not be the case for all $m_{a}$.} In order to treat the lightest ULAs consistently with $P(k)$ data we adopt an ULA-mass dependent definition $\rho_m$ when computing $P(k)$.

A simple prescription is motivated by the band-limited nature of the data. Galaxy power-spectrum data from any given survey is only available down to some minimum wave number $k_{\rm obs}$ set by the size of the survey. Clearly if ULAs do not cluster on any $k>k_{\rm obs}$ then to some approximation the galaxy density field on those scales should not be correlated to the ULA density field and so we should exclude ULAs from definition of the matter density on these scales. This can be achieved by estimating the scale at which ULAs cease to cluster as being the horizon size when oscillations began, $k_{\rm osc} = a_{\rm osc} H(a_{\rm osc})$, and excluding ULAs from the matter density if $k_{\rm osc}<k_{\rm obs}$. This suggests that we can define the matter density in the following way
\begin{align}
\delta \rho_m  &= \Theta (a_{\rm osc}-a_{\rm bias}) (\delta \rho_c+\delta \rho_b) \,  \nonumber \\
&+ \Theta (a_{\rm bias}-a_{\rm osc}) (\delta \rho_c+\delta \rho_b+\delta \rho_a) \, ,
\label{eqn:rhom_def_correct_bias}
\end{align}
where $\Theta (x)$ is the Heaviside function, $k_{\rm obs}=a_{\rm bias}H(a_{\rm bias})$, and similarly for the average density, $\bar{\rho}_m$. For our ULA cosmologies we compute $a_{\rm osc}(m_a)$ from the Klein-Gordon equation and so specifying $a_{\rm bias}$ gives the desired, simple, mass-dependent prescription for $\rho_m$. 
\begin{figure*}[t!]
\begin{center}
\includegraphics[width=0.49\textwidth]{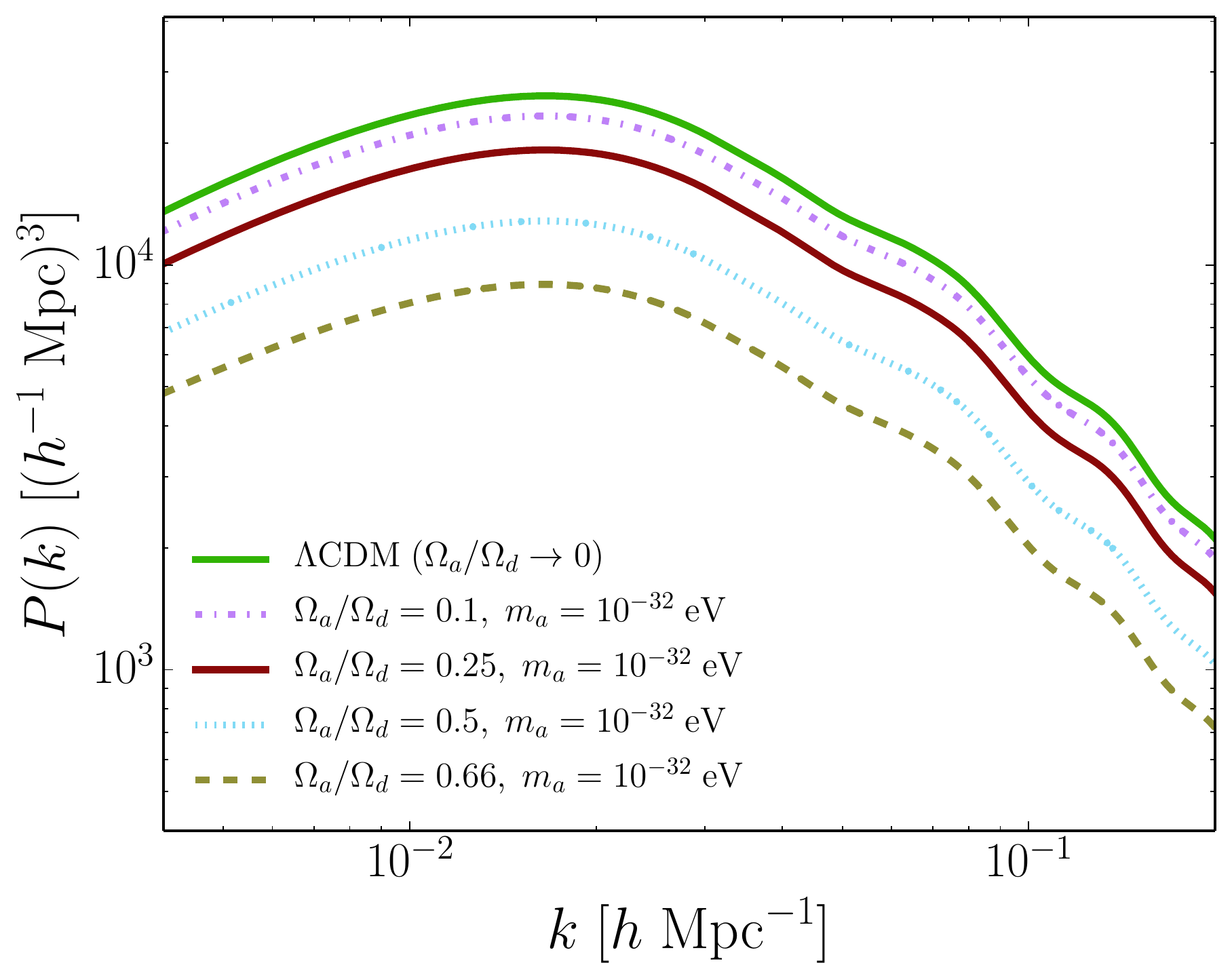}
\includegraphics[width=0.49\textwidth]{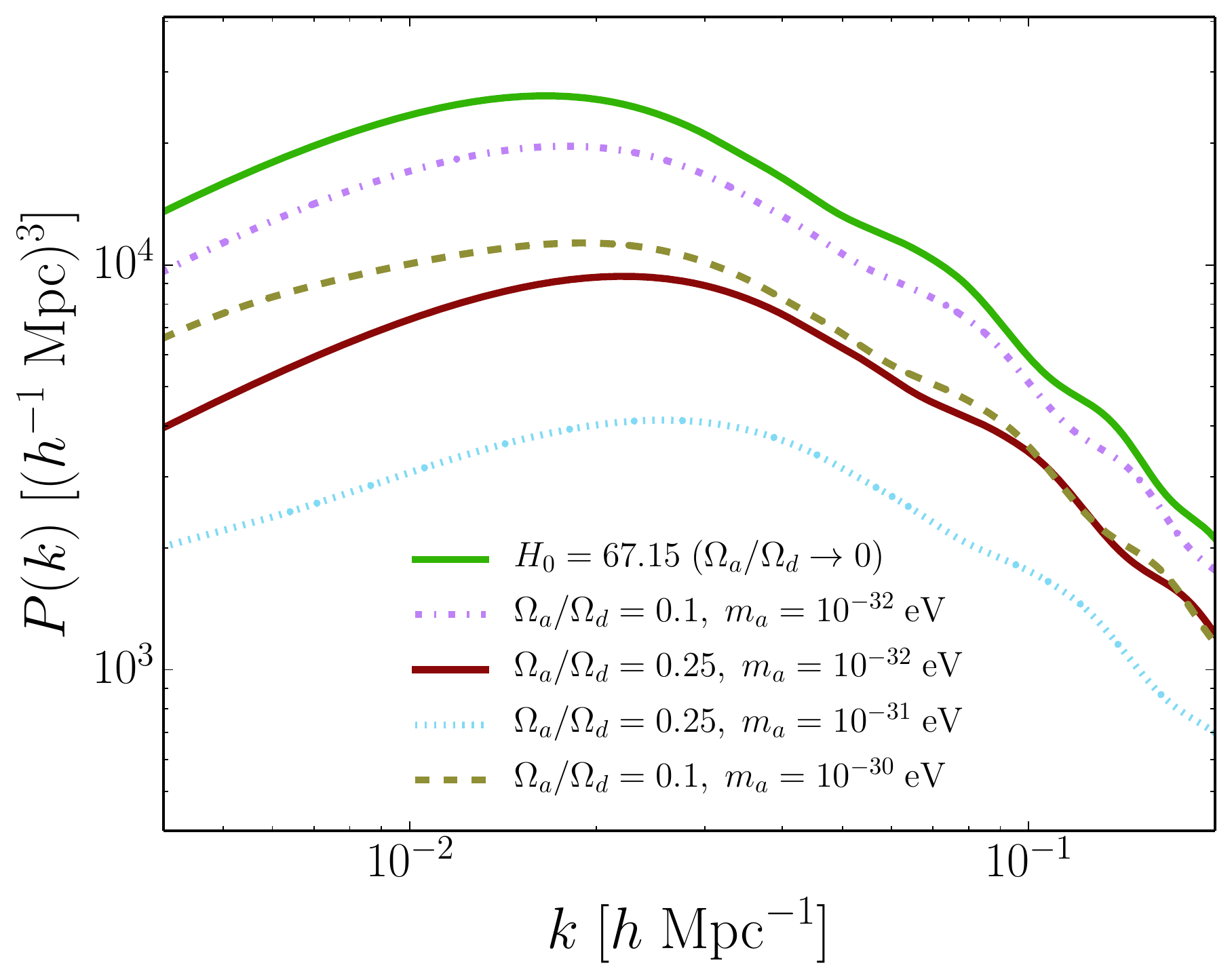} 
\caption{Matter power-spectrum with varying ULA mass and energy-density fraction $\Omega_a/\Omega_d$. Here we introduce the lightest ULAs as a fraction of the DE, holding $\Omega_c h^2$ fixed. No masses considered in this example cluster on scales where there are data and we exclude them from the matter density used to define $P(k)$ [taking $a_{\rm bias}=a_{\rm eq}$ in Eq.~(\ref{eqn:rhom_def_correct_bias})]. {\em Left Panel:} Fixed $H_0$, reducing $\Omega_\Lambda$ to maintain flatness. On scales shown the shape of $P(k)$ is unchanged, and the only effect comes from the reduction in the age of the Universe giving less growth time. {\em Right Panel:} Fixed $\theta_A$, which requires reducing $H_0$ as ULAs are introduced. Measurements of $P(k)$ will clearly rule out these extremely low values of $H_0$. \label{fig:matterspectra_de_examples}}
\end{center}
\end{figure*}

The value of $k_{\rm obs}$ for WiggleZ, which we use, is close to $k_{\rm eq}$, and no galaxy survey to date has observed scales $k\lesssim k_{\rm eq}$. For simplicity we therefore take $a_{\rm bias}=a_{\rm eq}$ as our benchmark. When we obtain constraints in Sec.~\ref{results} we will test the effect of this prescription by comparing constraints with $a_{\rm bias}=a_{\rm eq}$ and $a_{\rm bias}=1$, where $a_{\rm bias}=1$ only excludes the `most $\Lambda$-like', $m_a<3H_0$, ULAs. 

We will only ever use $P(k)$ data in conjunction with CMB data. Therefore if the CMB (through the late-time ISW effect) already provides strong constraints on all masses in the range where $a_{\rm osc}>a_{\rm eq}$ ($m_a\lesssim 10^{-27}\text{ eV}$), then constraints from CMB+$P(k)$ should be the same for any $a_{\rm bias}>a_{\rm eq}$. This will be the case if remaining effects in $P(k)$ for any choice of $a_{\rm bias}>a_{\rm eq}$ are small [relative to experimental error bars on $P(k)$] for $a_{\rm osc}>a_{\rm eq}$ within the limits on $\Omega_a$ set by the CMB. We verify later that our choice of $a_{\rm bias}$ has little effect on our constraints.

Our prescription, Eq.~(\ref{eqn:rhom_def_correct_bias}), is an approximate way to treat the bias for ULA cosmologies. It is, however, an improvement upon just blindly including both standard CDM and structure-suppressing species in the matter density. It is a definition of ``matter" to only include those components that were redshifting with the dominant matter at equality. Such a definition is necessary due to our wide mass prior, and is consistent with existing prescriptions for neutrinos and clustering DE \cite{camb}. 

A full treatment of scale-dependent bias would fix the form of $b(k)$ based on the transfer function, relating the perturbations in each component to the perturbations in the total-matter field. Such a treatment is appropriate, but by no means standard, in WDM and neutrino cosmologies (see, e.g., Refs.~\cite{smith2011,LoVerde:2014pxa,Castorina:2013wga}). In neutrino cosmologies the effects are small since $\Omega_\nu h^2$ is small within the limits on neutrino mass set by the CMB, and so it is reasonable in this case to define bias with respect to just the CDM \cite{Castorina:2013wga}.

For clustering DE cosmologies, there is a default prescription in \textsc{camb} to ignore the clustered component of DE (in the definition of matter) when computing the galaxy-clustering power-spectrum. This is reasonable, as DE clustering is still included in the potentials which determine the physical effects of clustering DE on the CMB, as well as the trajectories of DM particles and halos that show up in the matter power-spectrum. The validity of this default prescription requires that DE not cluster on the same scales as galaxies do. Such simple assumptions should be tested systematically in future work. 

In Fig.~\ref{fig:matterspectra_de_examples}, we show some examples of $P(k)$ for DE-like ULAs when $a_{\rm bias}=a_{\rm eq}$. For the DE-like ULAs this bias prescription excludes them from the definition of the matter density in $P(k)$, so all effects are indirect via the expansion rate and the potentials to which the CDM and baryons respond.

In the left panel of Fig.~\ref{fig:matterspectra_de_examples}, $H_0$ is held fixed. It shows the same models as the right panel of Fig.~\ref{fig:cmbspectra_de_examples_h0}. With $H_0$ fixed the epoch of equality is unchanged, leaving the $P(k)$ peak unmoved. The ULAs in this example do not cluster on any of the scales observed or shown, and so potentials for the CDM and baryons, and thus the shape of $P(k)$ is unaffected. Nevertheless, a constraint to ULAs in this mass range can be obtained from the matter power-spectrum. This occurs because the age of the Universe is reduced, as ULAs do not behave as $\Lambda$ for all of cosmic history. The time available for the growth of perturbations is thus lower, decreasing $P(k)$ relative to the $\Lambda$CDM-case at all scales.

In the right panel of Fig.~\ref{fig:matterspectra_de_examples}, the value of $\theta_{A}$ (and thus the angular scale of all the acoustic peaks) is held fixed. It shows the same models as Fig.~\ref{fig:cmbspectra_de_examples_theta}. The models here would not be heavily disfavored by the CMB alone. The extremely low values of $H_0\sim 50~\text{km s}^{-1}\text{Mpc}^{-1}$ necessary to fix $\theta_A$, however, are strongly disfavored by measurements of the matter power-spectrum $P(k)$. This demonstrates the well known complementarity of the CMB and matter power-spectrum. The matter power-spectrum contains information about the baryon acoustic oscillation (BAO) scale in galaxies, and so can be used to probe $H_0$ in conjunction with the CMB. These cosmologies require low $H_0$ but match the CMB (within the errors) otherwise. The CMB temperature power alone (we do not include lensing in our analysis) does not strongly constrain $H_0$. These low $H_0$ cosmologies are inconsistent with measurements of the matter power-spectrum. 

\begin{figure*}[t!]
\begin{center}
\includegraphics[width=0.49\textwidth]{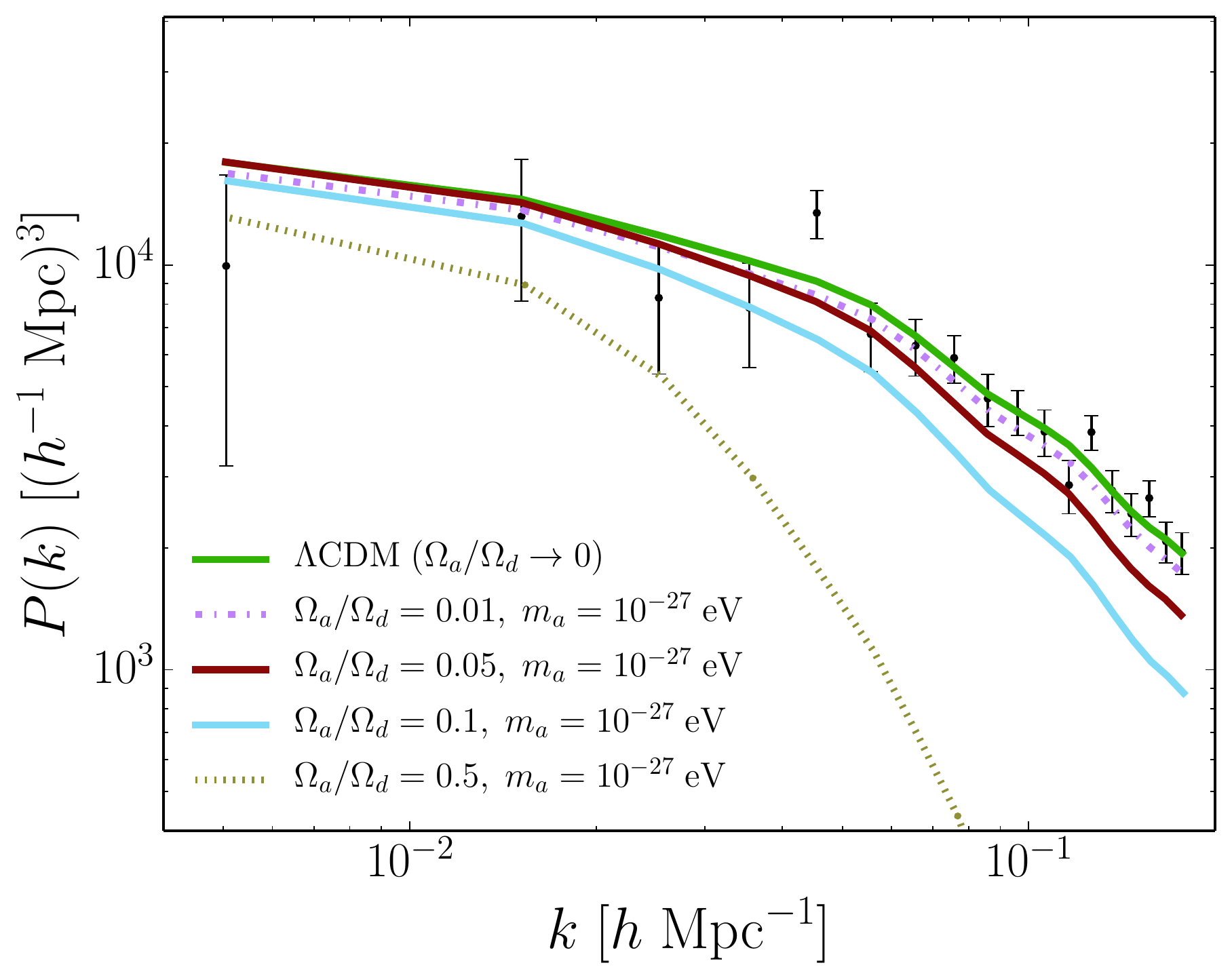}
\includegraphics[width=0.49\textwidth]{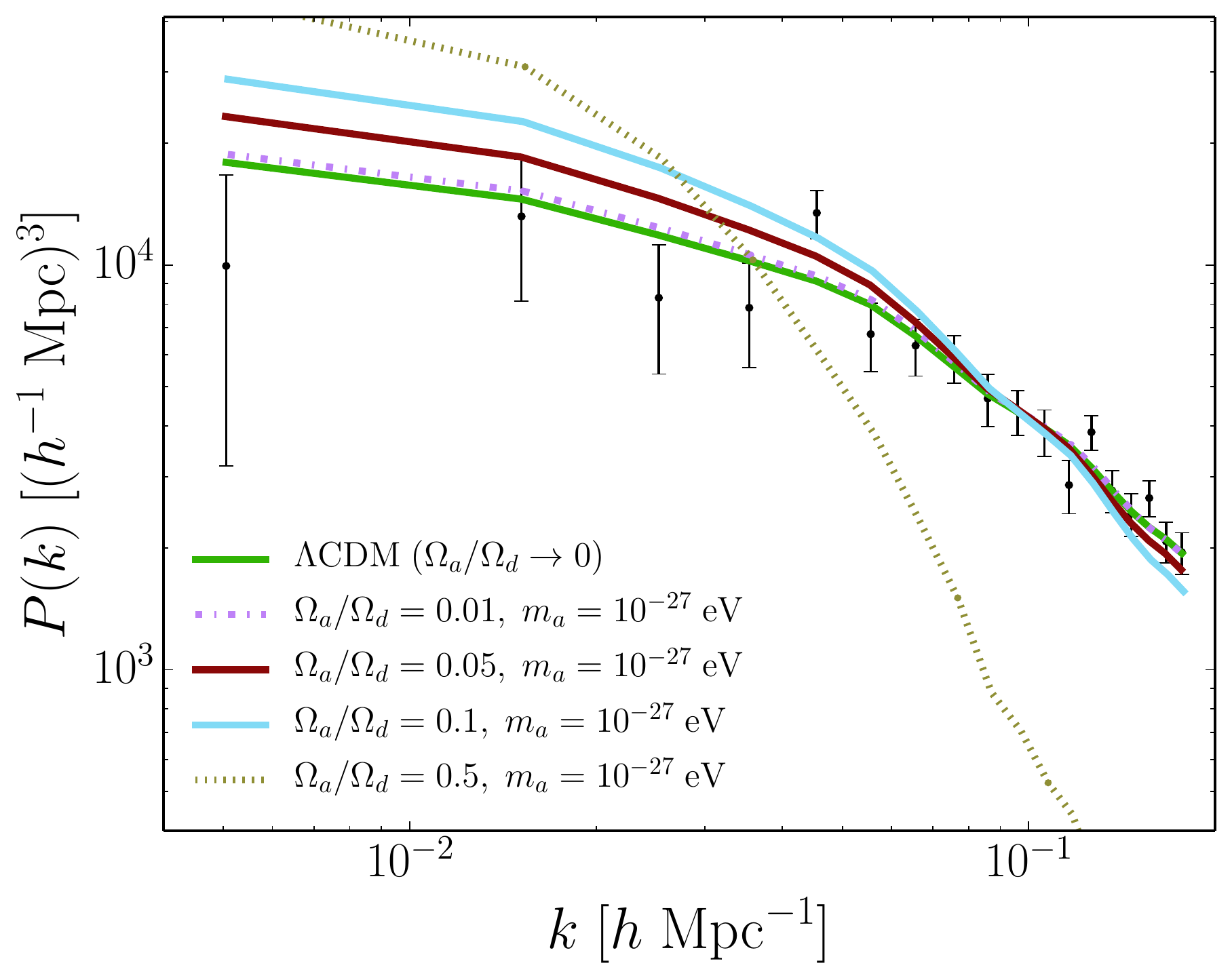} 
\caption{Theoretical galaxy clustering power spectra, varying the ULA mass fraction $\Omega_{a}/\Omega_{d}$ at fixed ULA mass $m_{a}=10^{-27}~{\rm eV}$. All non-ULA parameters are held at fiducial $\Lambda$CDM values. An example selection of WiggleZ data are shown from the ``9 hour region" \cite{blaketal,blake2011}. The theoretical curves have been computed at the same redshift ($z=0.6$) as the WiggleZ data, and multiplied by the window functions for the region in question. {\it Left panel:} Power spectra with galaxy bias fixed to its best-fit $\Lambda$CDM value. {\it Right panel:} Power spectra, marginalizing over bias in the course of parameter-space exploration. At high values of $\Omega_{a}/\Omega_{d}$, the preference is for higher values of the bias (to absorb the overall power suppression), explaining the upward trend in power at large scales in this panel.}
 \label{fig:wigglez_vs_axireality}
\end{center}
\end{figure*}

To develop some intuition for the sensitivity of LSS data to ULA parameters, it is useful to compare by eye the output of our modified \textsc{CAMB} with survey observables. To do this, biased theory power-spectra must be convolved with observational window functions (in particular, that of the WiggleZ survey used to obtain constrains in Sec. \ref{results}), as described in Refs. \cite{blaketal,blake2011,Parkinson:2012vd}. The results are shown in Fig. \ref{fig:wigglez_vs_axireality} using the appropriate binning. We see that for $m_{a}\lsim 10^{-25}~{\rm eV}$, we expect $\sim 1\%$-level constraints to the ULA mass fraction $\Omega_{a}/\Omega_{d}$.

\section{Results}
\label{results}
 \subsection{Data sets}
In order to map out the allowed regions in ULA parameter-space, we make use of several data sets. We use \textit{Planck} temperature data \cite{planckfull, plancklikelihood}, as well as WMAP large-scale CMB polarization data \cite{bennett/etal:2012}. In addition, we add small-scale data from the Atacama Cosmology Telescope (ACT) \cite{dasetal:2013} and the South Pole Telescope (SPT) \cite{spt}, as included in the \textit{highL} likelihood within the {\em Planck} public likelihood code.

In addition to the CMB data, we include matter power-spectrum data, from the WiggleZ survey \cite{blaketal,blake2011,Parkinson:2012vd}. We use the full shape of the matter power-spectrum. The shape also includes the information about the BAO. In order to avoid double-counting we do not separately use the WiggleZ measurement of the BAO peak scale. The BAO are complementary to the CMB data in measuring $H_0$, providing additional constraining power on the lightest ULAs (Fig.~\ref{fig:matterspectra_de_examples}). We restrict our analysis to wave numbers of $k \lsim 0.2 h^{-1}\mathrm{Mpc}$ and do not include nonlinear scales from the WiggleZ data. We make this choice because the \textsc{HaloFit} \cite{Smith:2002dz} prescription for computing nonlinear power used in \textsc{camb} has not been calibrated using simulations of ULA DM, and incorrect modeling of the matter power-spectrum on nonlinear scales could lead to spurious constraints.
\subsection{Sampling}
The degeneracy structure in the eight-dimensional parameter space including a wide prior on $m_{a}$ is complex and highly non-Gaussian. In order to fully explore this parameter space we had to go beyond the standard Metropolis-Hastings MCMC cosmological parameter estimation.

We make use of the \textsc{MultiNest} \cite{multinest} nested-sampling package implemented in the December 2013 version of \textsc{CosmoMC} \cite{cosmomc}, combined with our modified version of \textsc{camb} to compute the power spectra. 
This is in contrast to the existing constraints on ULAs in Ref.~\cite{amendola2005}, where a grid-based likelihood and analytic approximations for the power spectra were used. We allow \textsc{MultiNest} to search for multiple nodes within the likelihood. For the vanilla $\Lambda$CDM model, the two methods of standard MCMC and nested sampling agree extremely well in the derived cosmological parameters. 

We speed up the \textsc{CosmoMC} exploration of the space by fixing the foreground parameters for the CMB data to their best-fit values. We tested this assumption by unpinning foreground parameters and examining all possible pairings of ULA and foreground parameters. In no case were there degeneracies that change any of our conclusions. More specifically, we computed the correlation coefficient of the axion parameters with a coarsely sampled run over the full parameter space. The correlation coefficient between the axion parameters and the Poisson amplitude of the Planck $100$-GHz data is $c_\mathrm{corr} < 0.2$; for all other parameters there is less than 10\% correlation between the primary and foreground parameters. 

We checked for the dependence of the CMB results on the fixed foreground model assumption by finding the best fit primordial and foreground parameters given axion parameters which are fixed at the best-fit positions in the medium mass bin. We then fixed the foregrounds to these newly determined best-fit values of the foregrounds (rather than the best fit from the Planck results) and found the best-fit ULA parameters in that case. We see shifts of less than $0.7\sigma$ for the ULA parameters.

We ultimately vary $\Delta^{2}_{\mathcal{R}}$, $n_{s}$ , $\Omega_{b}h^{2}$, $\Omega_{c}h^{2}$, $\tau_{\rm re}$, $m_{a}$, $\Omega_{a}h^{2}$, and the \textsc{CAMB/CosmoMC} parameter $\theta_{\rm MC}$. The value of $\theta_{\rm MC}$ closely tracks that of $\theta_{A}$ under the assumption that ULAs behave entirely as DM \cite{Hu:1994uz}. It is not physical for low values of $m_{a}$, when ULAs are dark-energy like, but is a useful tool to efficiently step in $H_0$, a derived parameter. Hence in Table I we only quote constraints on $H_0$ and not on $\theta_{A}$. We assume zero spatial-curvature ($\Omega_{k}=0$) and determine the cosmological constant $\Omega_{\Lambda}$ accordingly. The Hubble constant $H_{0}~\mathrm{[km/s/Mpc]}$ is a derived parameter, as is the initial axion field displacement, $\phi_i$.

\begin{figure}[t!]
\begin{center}
\includegraphics[width=0.5\textwidth]{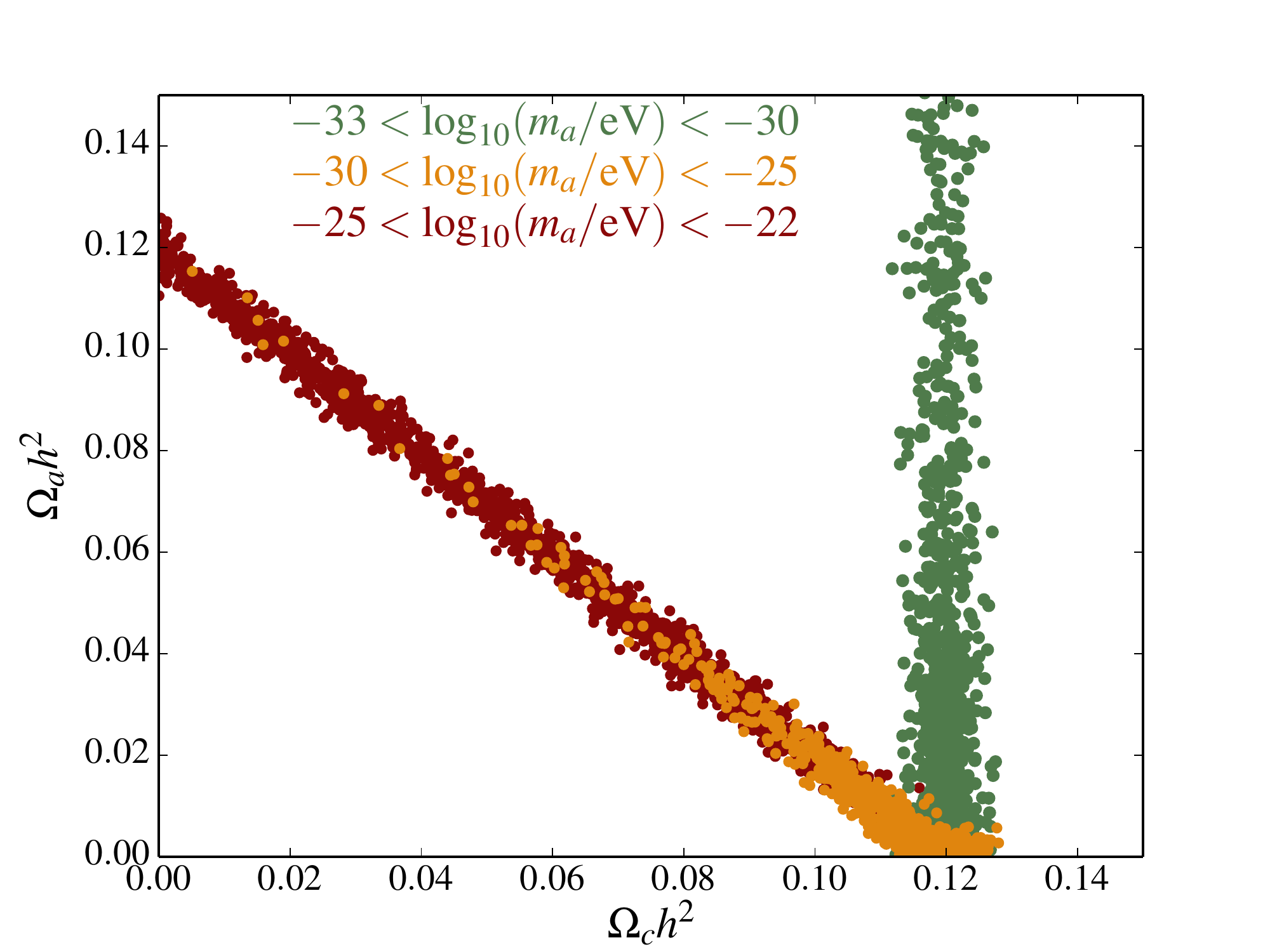}
\caption{Mass-dependent degeneracy of axions and CDM. Points are shown for a \textsc{MultiNest} chain and colored by $m_{a}$. If axions are light ($m_a < 10^{-30}~\mathrm{eV}$), they behave as dark energy. Therefore while the CDM density is unchanged as $\Omega_ah^2$ increases, the dark-energy density $\Omega_\Lambda$ is reduced (see Fig.~\ref{fig:1d_like}). If axions are heavy ($m_a > 10^{-25}~\mathrm{eV}$), they behave as dark matter, and so there is a perfect degeneracy between $\Omega_ch^2$ and $\Omega_ah^2$. For $m_{a}$ in the intermediate range range, the axion energy-density is constrained to be small. \label{degeneracy}}
\end{center}
\end{figure}

ULAs are degenerate either with CDM (for large $m_{a}$) or DE (for low $m_{a}$); this results in a mass-dependent degeneracy between the $\Omega_{a}h^{2}$ and $\Omega_{c}h^{2}$, illustrated in Fig.~\ref{degeneracy} for our \textsc{MultiNest}-sampled chains. We show the point density of the chains sampled in three regions of the ULA mass, and color the points by the value of the mass, in three bins. Very low masses are not degenerate with CDM, and $\Omega_a h^2$ can be large independent of $\Omega_c h^2$. Heavy axions are indistinguishable from CDM and there is a perfect degeneracy between $\Omega_a h^2$ and $\Omega_c h^2$. For intermediate-mass axions $\Omega_a h^2$ is constrained (although it still lies along the degeneracy line for high-mass axions) and $\Omega_c h^2$ remains close to its $\Lambda$CDM value. 

This mass-dependent degeneracy makes computing a covariance matrix difficult in a normal MCMC scenario. In particular, the bimodality of the $m_a-\Omega_ah^2$ plane consists of two regions where the axion density relative to the total density is poorly constrained. These walls in the distribution present significant challenges, as MCMC chains starting in either region can become `blocked' in the highly probable regions, which are separated by a well-constrained ``valley" for intermediate-mass axions. For a standard MCMC, therefore, this valley is hard to traverse. 

Nested sampling is far better suited to exploring likelihood surfaces like this, and so we choose 
to use \textsc{MultiNest} instead of standard Metropolis-Hastings MCMC techniques. We are still limited computationally, however, by the number of live-points used by \textsc{MultiNest}. Properly sampling the constrained valley in a global exploration of our mass range proved prohibitive, and using standard techniques, we could not obtain accurate constraints in the two-dimensional space $(m_a,\Omega_a/\Omega_d)$ in the constrained valley even using nested sampling.  

Our solution to this problem is to break the parameter space into three regions: 
\begin{align}
-33 <& \log_{10}(m_a/\mathrm{eV} )< -30  \quad \text{(low mass)} \, , \nonumber \\
-30 <& \log_{10}(m_a/\mathrm{eV}) < -25 \quad \text{(med. mass)} \, , \nonumber \\
-25 < &\log_{10}(m_a/\mathrm{eV} )<-22 \quad \text{(high mass)} \, . \label{eqn:local_chain_regions}
\end{align}
We term these ``local chains," and they are demarcated by the dashed vertical lines in Fig.~\ref{fig:money_plot}. We perform a \textsc{MultiNest} run with 500 live points and a tolerance of 0.3 in each region, satisfying the criterion $\Delta \ln \mathscr{L} = 0.1$, where $\mathscr{L}$ is the likelihood. This typically results in $\sim 100 000$ likelihood evaluations for each region. This ensures that each region is well sampled in the local chains. In addition, we check that splitting the chain in two parts and computing constraints with different parts of the chain produces results consistent at the $\sim 0.1$-$0.2\sigma$ level.

In order to combine the information from multiple regions together to form a chain across the full space, we do a coarse global \textsc{MultiNest} run over the entire mass range; we call this the ``global" chain. We use this global chain to re-weight the output from the individual regions as follows. We first convert the global chain into a single chain where each point has equal density (to ensure a valid relationship between likelihood and point density). To make a single chain we first divide the weight of each step by the maximum global weight (and so in that way turns the weights into fractional weights, and keeps the information from the MCMC sampling). We then throw a random number and accept this new point (and writes it with weight one) to the single chain if it that random number is less than the normalized weight.

The single global chain is then binned in the $(m_a,\Omega_a/\Omega_d)$ plane and we use the point density in two-dimensional bins as a posterior with which to re-weight the individual (separately computed and hence statistically independent) local chains. We perform an interpolation of the points in the 2D mass-fraction plane for the individual, local chains to obtain a re-weighting coefficient from the global 2D histogrammed point density. Following this two-dimensional importance sampling \cite{hobson2010bayesian}, the local chains are combined to form a ``master chain," which is processed as usual, and the global chain is not used again, as the local chains are no longer independent from the global. The master, combined chain is now well sampled in the full parameter-space, and the proper relative likelihood applies across the full range of ULA masses. This 2D importance sampling from the coarse global chain allows us to keep global information about the relationship between mass and fraction, but achieves better sampling in the three regions.

\subsection{Priors}
\label{sec:priors}

The most conservative prior to place on the unknown parameter $m_{a}$ is a Jeffreys prior, which is uniform in logarithmic space. We bound this as
\begin{equation}
 -33< \log_{10}{(m_{a}/\text{ eV})} < -22 \quad \text{(global chain)} \, ,
 \end{equation}
and correspondingly for each local chain of Eq.~(\ref{eqn:local_chain_regions}). We recall that this is also the preferred theoretical prior for axions in the string landscape \cite{axiverse2009}.
 
We impose flat priors on the axion and matter energy-densities. Alternatively, we could have imposed a uniform prior on the initial axion misalignment angle $\phi_{i}$ \cite{hertzberg2008} resulting in a density prior $P(\Omega_{a}h^2) \propto 1/(\sqrt{\Omega_{a}h^2})$. We do not use this prior, and choose to be consistent in our treatment of baryon, CDM and axion densities. To ensure that we probe all the way down to axion mass-fractions of $\Omega_a/(\Omega_a + \Omega_c) = 10^{-4},$ we allow $ \Omega_ah^2, \Omega_ch^2$ to vary in the range $10^{-5}\to 0.3.$ As a test for prior dependence, we tried an alternate procedure, in which the chains were importance sampled with uniform priors in $\Omega_{a}/\Omega_{d}$ or $\ln{(\Omega_{a}/\Omega_{d})}$. There is a weak prior dependence in that chains importance sampled uniformly in $\ln \Omega_a/\Omega_d$ give less weight to the top of the ``U" in the low- and high-mass regions. The bounds on the axion fraction in the highly constrained  intermediate mass range are unchanged by our choice of prior. 
\subsection{Cosmological parameter constraints}
\begin{table*}[t] 
\caption{\small{Constraints on the cosmological parameters in the axion model in the tightly constrained (data-driven) mass range $-32\leq \log_{10}{(m_a/{\rm eV})} \leq -25.5$.} The one-sided limits are upper 95\% bounds, while the error bars quoted represent the upper and lower 95\% errors. The lower limit should be the central value minus the error bar. \label{table:likes}}\begin{center}
\begin{tabular}{|c|c|c|}
\hline
Parameter & Planck + highL+lowL+WP (CMB) & CMB+ WiggleZ \\
\hline
\hline
$\Omega_ah^2$ & $< 0.0058$ & $< 0.0062$  \\
$\Omega_ch^2$ &$ 0.119_{-0.008}^{+0.005}$ & $0.121^{+0.004}_{-0.005}$\\
$\Omega_{a}/\Omega_{d}$ & $<0.048$ & $<0.049$ \\
$\phi_{i}/M_{\rm pl}$ & $ 0.073^{+0.1482}_{-0.058} $ & $0.089^{+0.239}_{-0.073}$ \\
$\log(10^{10}A_{s})$ & $ 3.092 \pm {0.046}$ & $3.091 \pm {0.046}$ \\
$n_{s}$ &  $0.959 \pm 0.012$& $ 0.956 \pm 0.011$ \\
$\tau_{\rm re} $ & $0.091\pm 0.025 $ & $0.089 \pm 0.025$ \\
$100\Omega_{b} h^{2} $ & $2.212^{+0.043}_{-0.045} $ & $2.201\pm{0.046}$ \\
$H_{0}~[\mathrm{km/s/Mpc}]$ & $ 67.3^{+2.4}_{-3.5}$ & $ 66.2 _{-4.9}^{+2.4} $ \\
\hline
\end{tabular}
\label{table:lcdm_params}
\end{center}
\end{table*}

\begin{figure}
\begin{center}
$\begin{array}{@{\hspace{-0.2in}}l}
\includegraphics[scale=0.5]{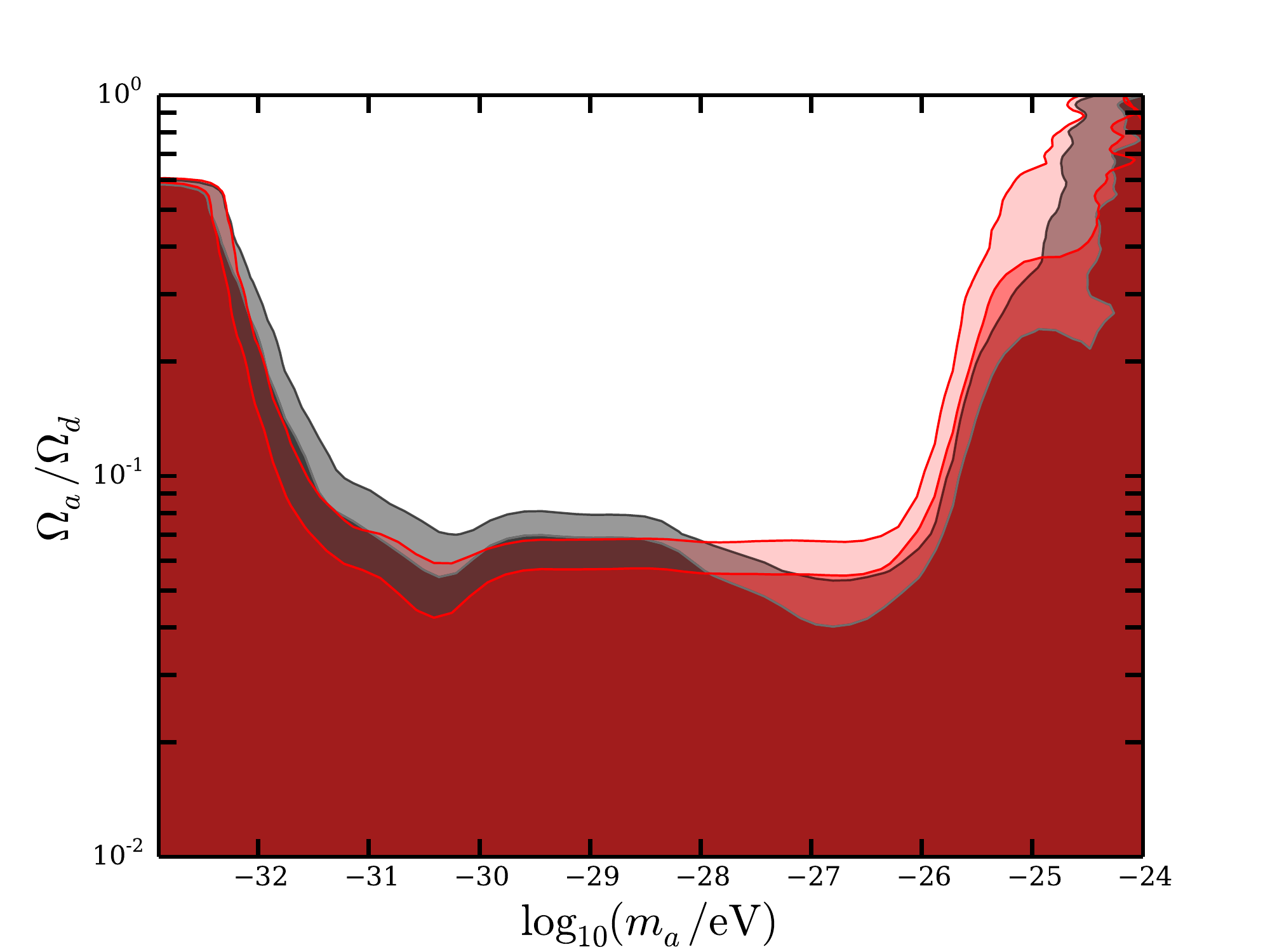} \\
 \end{array}$
\caption{Marginalized $2$ and 3$\sigma$ contours in the $m_a-\Omega_a/\Omega_d$ plane for both the CMB-only \textit{(red)} and CMB+WiggleZ \textit{(black)} combinations of data sets. We obtain constraints of $\Omega_{a}/\Omega_{d}\leq 0.03-0.05$ at $95\%$ confidence level over some seven orders-of-magnitude in $m_{a}$. The high-mass fluctuations/dips in the plane are due to sampling of the space rather than true features in the data set. For ultralight axions (ULAs) with masses $m_a\lesssim 10^{-20}\unit{eV}$,
small-scale structure formation is 
suppressed \cite{Frieman:1995pm,Coble:1996te,hu2000,amendola2005,marsh2010,park2012} on astronomically observable length scales.
\label{fig:cont_2d}}
\end{center}
\end{figure}

Our main results are constraints in the plane $(m_a,\Omega_a/\Omega_d)$, shown in Fig.~\ref{fig:cont_2d} (and Fig.~\ref{fig:money_plot} on a linear scale), marginalized over all other cosmological parameters. We display $2$ and $3\sigma$ exclusion regions for the CMB and CMB+WiggleZ combinations of data sets.\footnote{We have checked that the 2 and 3$\sigma$ constraints are robust to a variety of tests: they are unaffected by priors, binning, and sampling methodology. The 1$\sigma$ constraint, on the other hand, showed some sensitivity to these tests due to the flatness of the likelihood near the $\Omega_a\rightarrow 0$ boundary and being sample-size limited in this region. Thus we do not show the 1$\sigma$ constraint. On physical grounds it is clear that it should extend from $\Omega_a=0$ upwards for all masses.} Examining Fig.~\ref{fig:cont_2d}, we see that $\Omega_a/\Omega_d\lesssim 0.07$ across the highly-constrained region $-32\lesssim\log_{10}(m_a/\text{eV})\lesssim -25.5$]. Properly marginalizing over all $m_{a}$ values in this region, we obtain the precise constraint $\Omega_{a}/\Omega_{d} \leq0.048$ at $95\%$ confidence.

\begin{figure*}
\begin{center}
$\begin{array}{@{\hspace{-0.2in}}l@{\hspace{-0.2in}}l}
\includegraphics[scale=0.45]{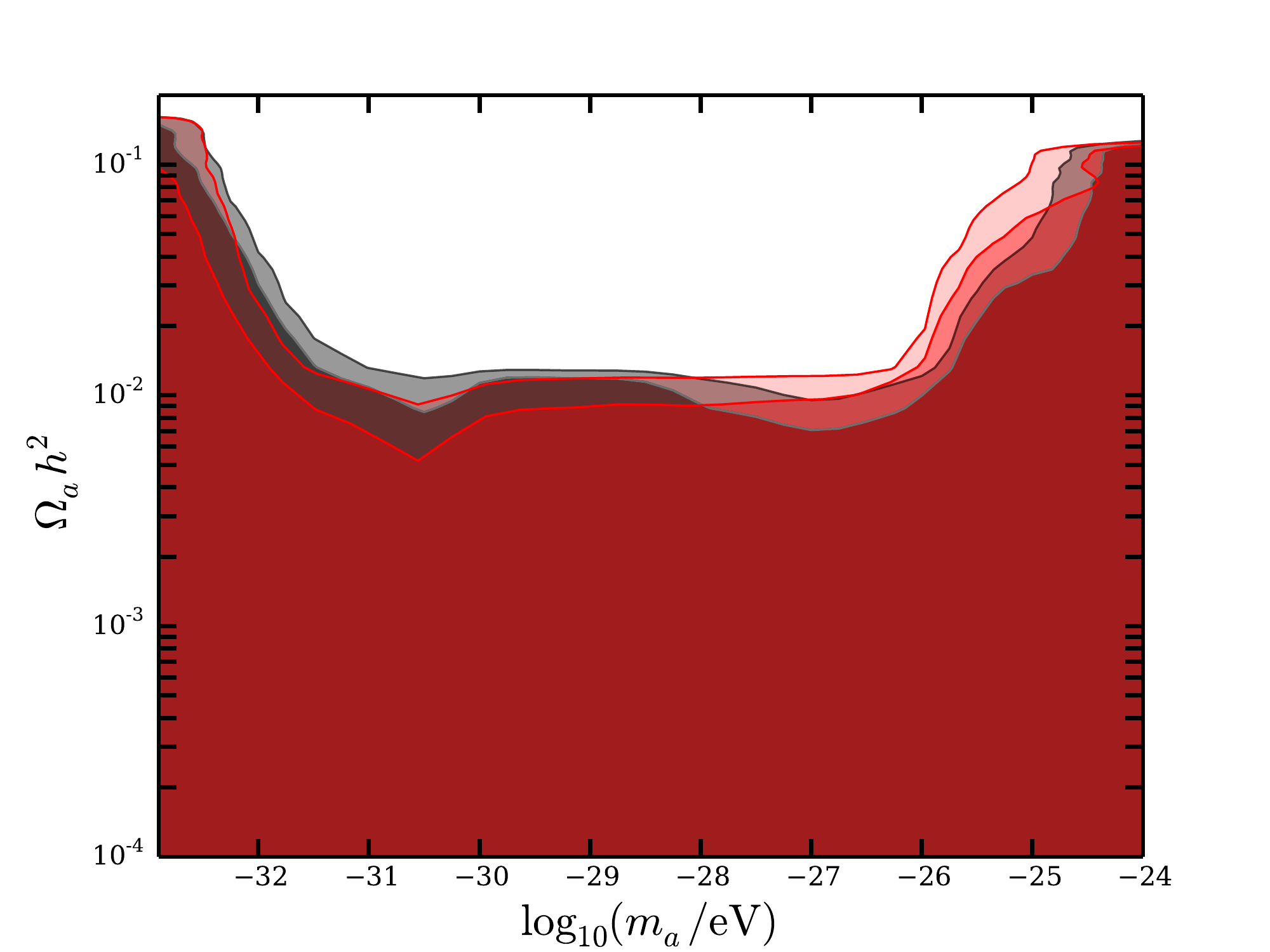} &
\includegraphics[scale=0.45]{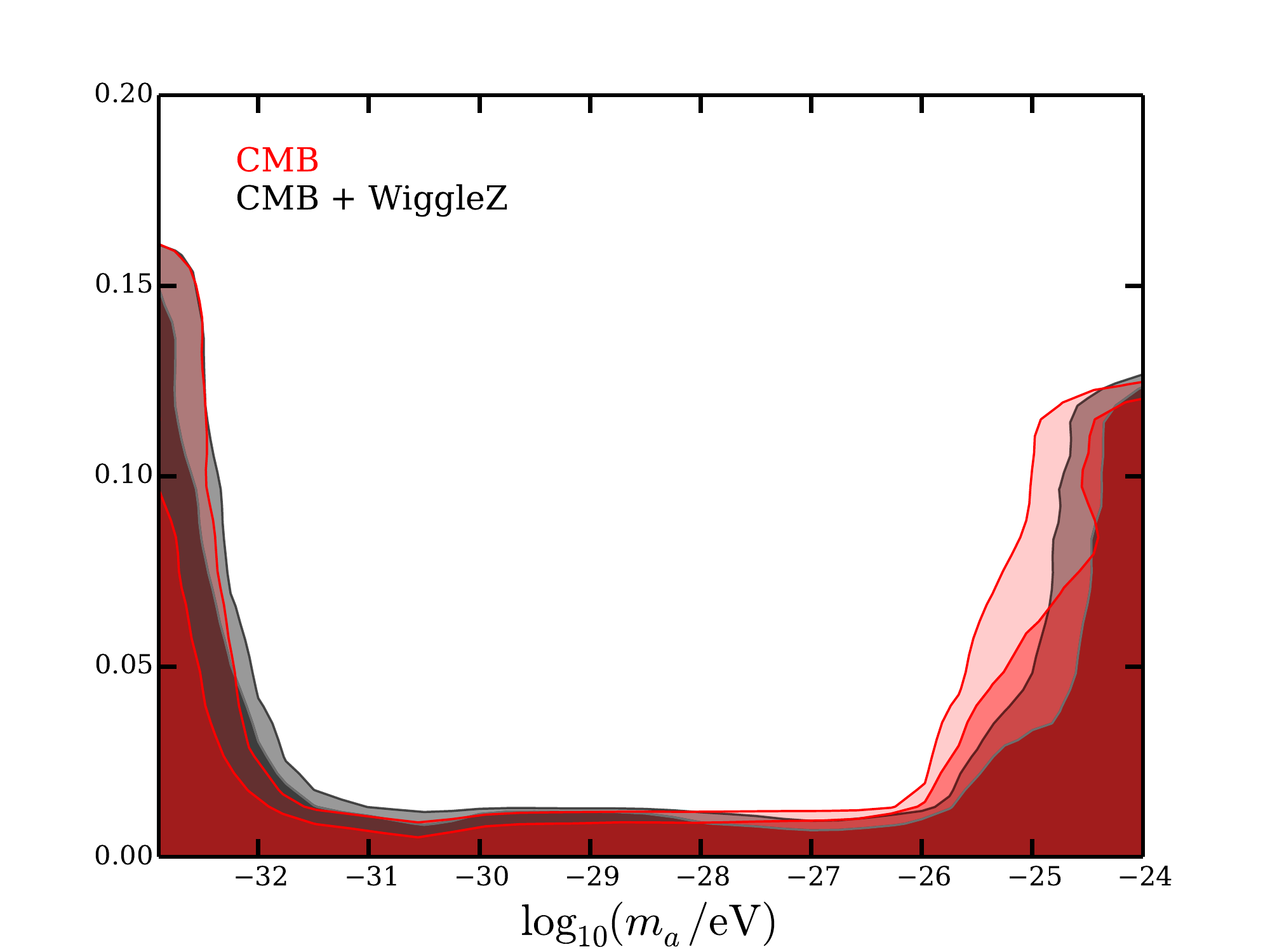}\\
 \end{array}$
\caption{Marginalized $2$ and 3$\sigma$ contours in the $m_a - \Omega_a h^{2}$ plane for both the CMB-only and CMB+WiggleZ combinations of data sets. The left panel shows the contours with the axion density shown on logarithmic scale, while the right hand side shows the same contours on a linear scale. We obtain constraints of $\Omega_{a}h^{2}\leq 0.006$ at $95\%$ confidence level over some seven orders of magnitude in axion mass $m_{a}$\label{fig:cont_2d_ox}. Color code is as in Fig. \ref{fig:cont_2d}.}
\end{center}
\end{figure*}

Another way of viewing the results is in the $m_{a}-\Omega_{a}h^{2}$ plane. The resulting constraints are shown in Fig. \ref{fig:cont_2d_ox}. We see that across the highly constrained region [$-32\lesssim\log_{10}(m_a/\text{eV})\lesssim -25.5$], ULAs can contribute a mass fraction bounded as $\Omega_{a}h^{2}\lsim0.010$ at $95\%$ confidence. Properly marginalizing over all $m_{a}$ values in this region, we obtain the precise constraint $\Omega_{a}h^{2} \leq0.0058$ at $95\%$ confidence. Our results have placed percent-level constraints on a possible ULA contribution to the DM energy-density over some six orders of magnitude in ULA mass, with looser constraints extending even further in mass. 

The constrained regime [$-32\lesssim\log_{10}(m_a/\text{eV})\lesssim -25.5$] spans across our individual mass regions of Eq.~(\ref{eqn:local_chain_regions}), showing that its existence and size was not biased by our sampling procedure. The tightly constrained region is a data-driven feature and the marginalized constraint on $\Omega_ah^2$ in this region is independent of the ULA mass-prior. The cosmological parameter constraints in the constrained region are quoted in Table~\ref{table:lcdm_params}.

The inclusion of WiggleZ data affects the constraints in a few key ways. In the mass range $-28 < \log_{10}(m_a/{\rm eV}) < -25$, the WiggleZ data are sensitive to damping in the matter power-spectrum on small scales, hence the constraints on the axion fraction tighten. The limits on the fraction at lower masses are actually weakened very slightly by the inclusion of galaxy-clustering data. This could partly be due to the differences in clustering preferred by LSS relative to CMB data \cite{joezuntz,planck_sz,battye}. CMB data favor higher $\Omega_m$ and lower $\sigma_8$ than clustering data and can thus tolerate a larger $\Omega_{a}/\Omega_{d}$ (where $\sigma_{8}$ is the variance of the matter power-spectrum on $8 h^{-1}~{\rm Mpc}$ scales). This could help reconcile the difference between the CMB and LSS power spectra. We will explore this issue further in future work. 

Our LSS constraints are likely to be overly permissive, due to the fact that there is some constraining power on scales for which $a_{\rm osc}>a_{\rm bias}$ (defined in Sec. \ref{mps:revisit}). We ran an exploratory MCMC run in the highly-constrained region with $a_{\rm bias} = 1$, and found that the LSS constraints tightened by $\sim 30\%$, and that the edges of the allowed wings moved out by roughly an order of magnitude in mass. In any case, the CMB constraints are more robust and stringent, so we defer a detailed treatment of scale-dependent bias to future work. 
The apparent feature at $\log_{10}(m_a/\mathrm{eV}) = -30.5,$ present in both data sets is weakly dependent on the binning procedure. In this region the shape of the ISW signal from DE-like axions has a nontrivial shape, and can play a role fitting low-$\ell$ anomalies in CMB data.

\begin{figure*}[t!]
\begin{center}
\includegraphics[width=0.9\textwidth]{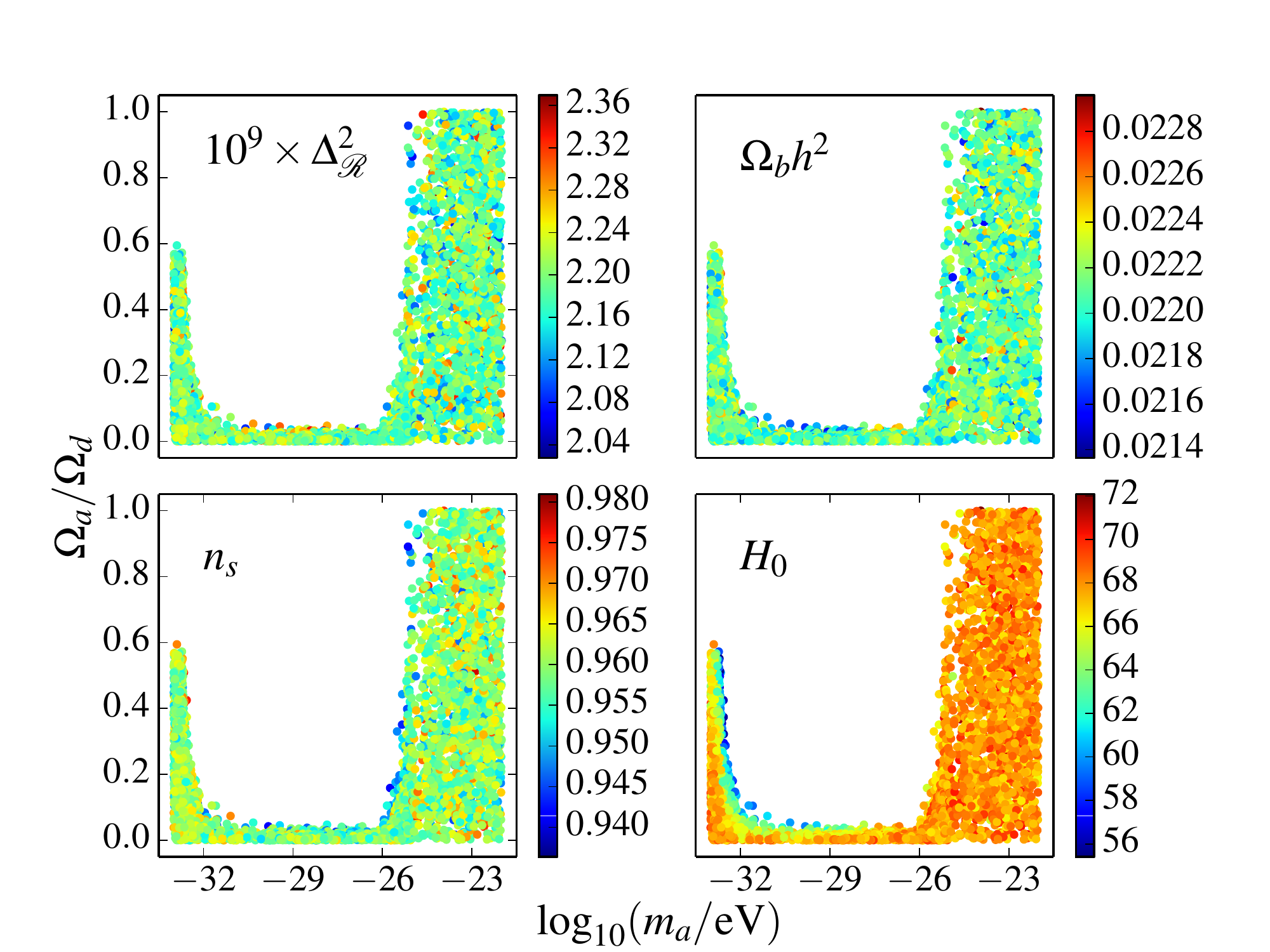}
\caption{Degeneracies between the axion parameters and other cosmological
parameters. Color indicates value of indicated parameter, as shown by color bar to the right of each panel. Axion parameters are independent of the baryon density, as well as the normalization and tilt of the primordial power spectrum, as can be seen in the {\textit top right}, {\textit top left}, and {\textit bottom left} panels of the
plot. In the {\textit bottom right} panel, we see that axion parameters can be degenerate with the Hubble constant $H_{0}$ (in ${\rm km}~{\rm s}^{-1}/{\rm Mpc}$) in the dark-energy like part of parameter space, at low values of $m_{a}$, where allowed values can drop to $H_{0}\lesssim 60~{\rm km}~{\rm s}^{-1}/{\rm Mpc}$.\label{param:degeneracy}}
\end{center}
\end{figure*}

In Fig.~\ref{param:degeneracy} we show sample points from our \textsc{MultiNest} chains in the $(m_a,\Omega_a/\Omega_d)$ plane colored by various other cosmological parameters. There is no significant degeneracy between axion parameters and $\Omega_bh^2$ or $n_s$. A mild degeneracy with $H_0$ is observed, with points on the edge of our constraints at low mass favoring lower $H_0$. 

\begin{figure*}
\begin{center}
$\begin{array}{@{\hspace{-0.2in}}l@{\hspace{-0.2in}}l}
\includegraphics[scale=0.5]{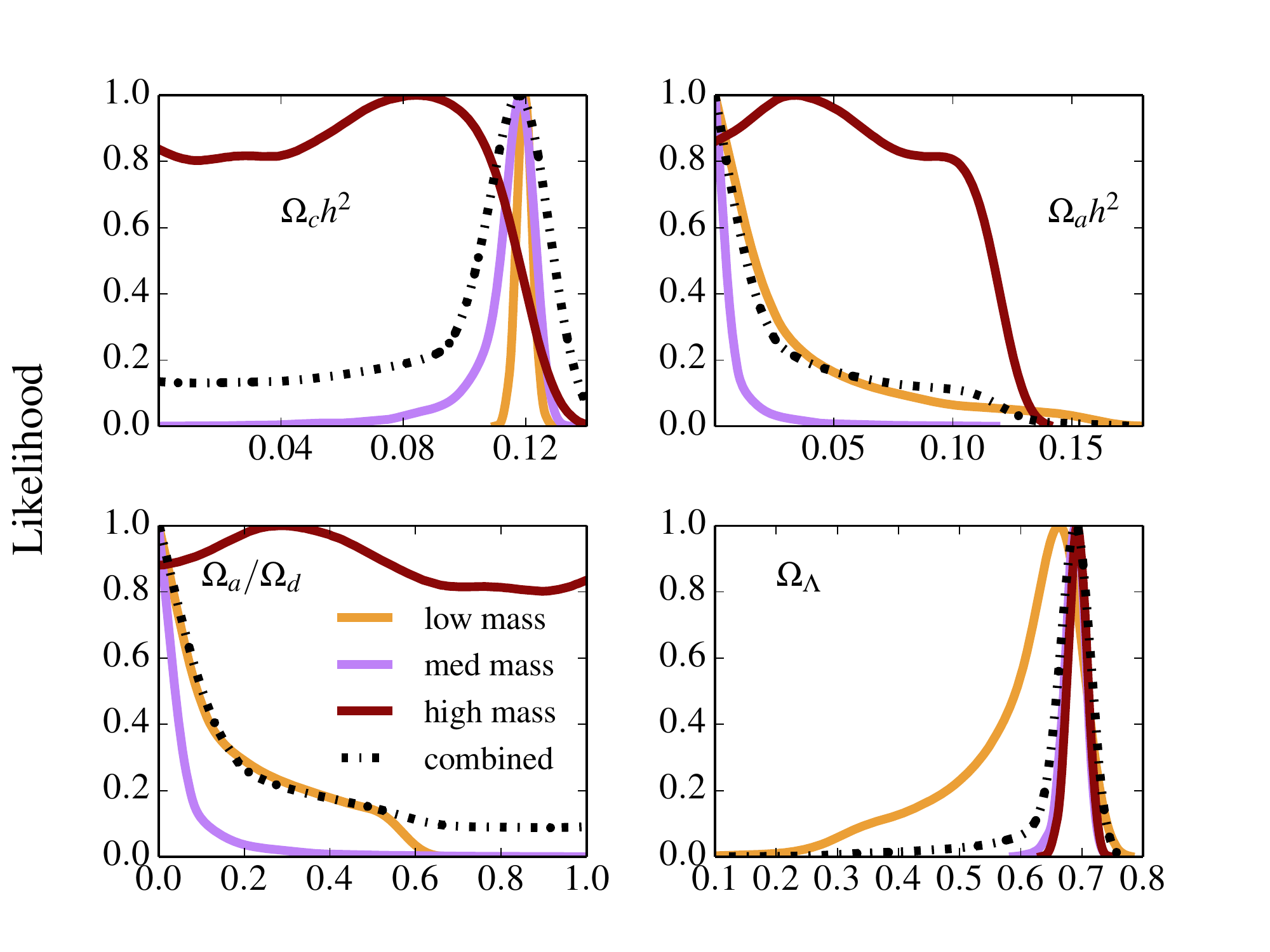} &
\includegraphics[scale=0.5]{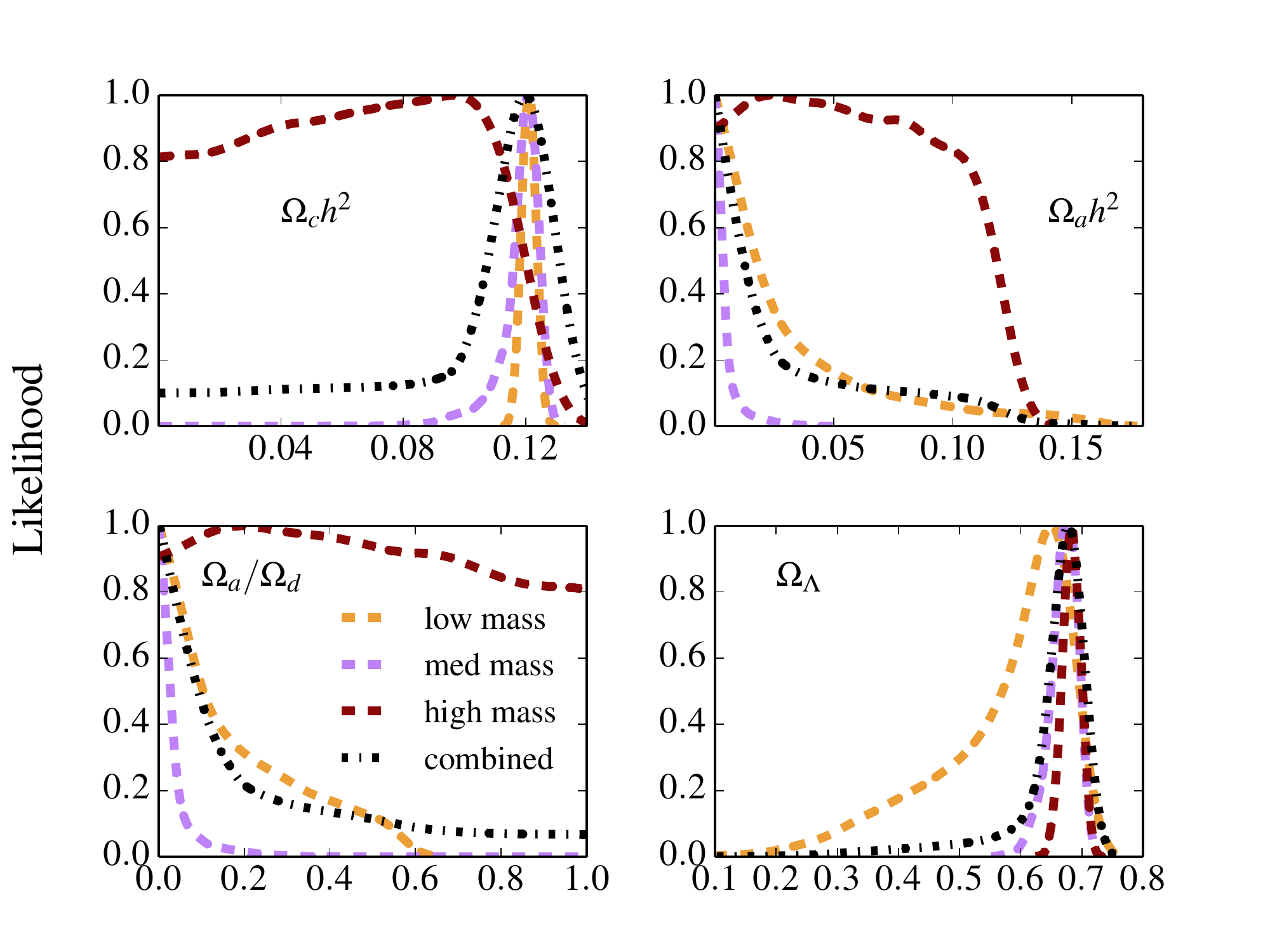}\\
 \end{array}$
\caption{Marginalized one-dimensional constraints on the axion parameters for various data sets. \textit{Left panel:} the solid lines show the constraints when considering only CMB data, while the dashed lines (\textit{right panel}) show the constraints when adding in WiggleZ data. In both panels the parameter constraints are shown for the axions sampled in separated mass bins. The black dot-dashed lines indicate the constraints obtained when combining the chains from the individual runs, weighted by the global run.\label{fig:1d_like}}
\end{center}
\end{figure*}

Figure \ref{fig:1d_like} shows one-dimensional marginalized constraints on various parameters. The constraints in each local mass range (low, medium, high) are shown to demonstrate the physical effects of ULAs of different masses. In the high-mass regime, ULAs are degenerate with CDM. Both $\Omega_ah^2$ and $\Omega_c h^2$ can therefore go to zero, with upper bounds close to the $\Lambda$CDM constraint on $\Omega_c h^2$. In the high-mass regime $\Omega_\Lambda$ is unchanged from its $\Lambda$CDM value near 0.68. In the low-mass regime, ULAs are degenerate with DE, and so $\Omega_\Lambda$ can become small compared to its $\Lambda$CDM value, while $\Omega_c h^2$ remains sharply peaked near $\Omega_ch^2=0.12$. In the medium-mass regime, ULAs are neither degenerate with CDM nor DE and $\Omega_a h^2$ in constrained to be small. The constraints from the CMB (left panel) and CMB+WiggleZ (right panel) are qualitatively similar, with WiggleZ adding additional constraining power in the medium-mass regime. 

\subsection{Local limits}
\begin{figure*}
\begin{center}
$\begin{array}{@{\hspace{-0.2in}}l@{\hspace{-0.2in}}l}
\includegraphics[scale=0.45]{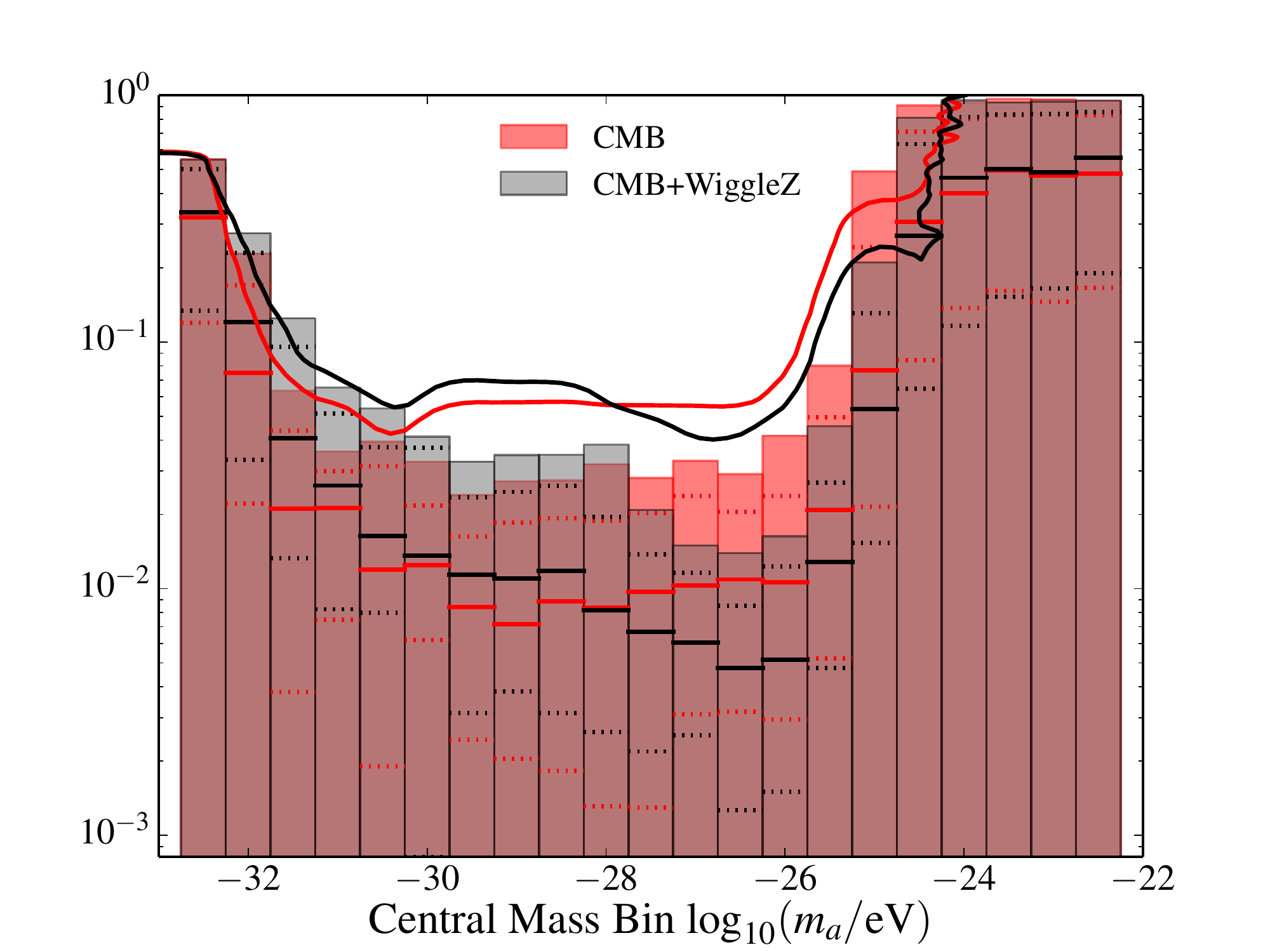} &
\includegraphics[scale=0.45]{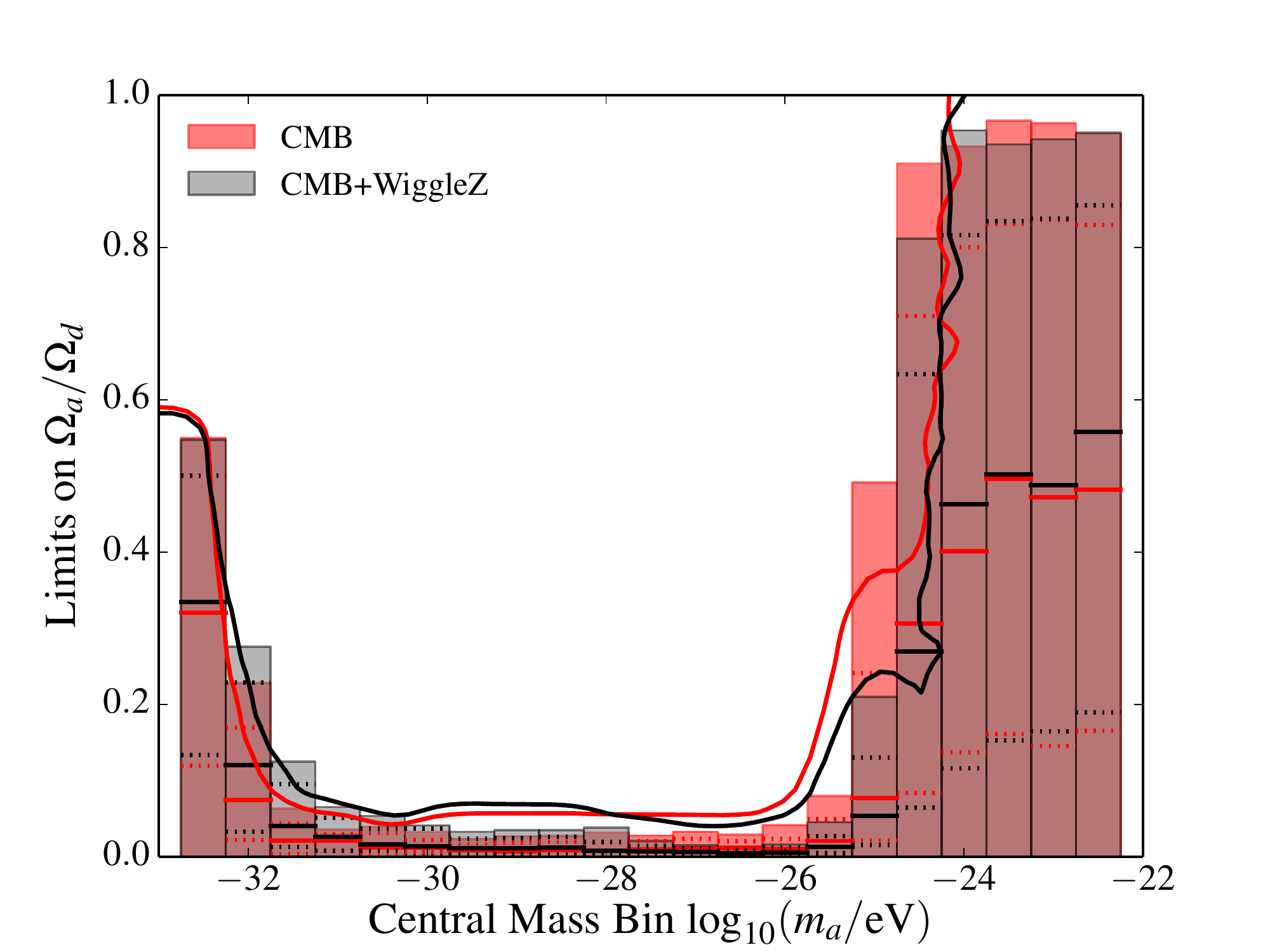}\\
 \end{array}$
\caption{Comparison of marginalized 95\% contours and locally defined one-dimensional limits on the axion fraction. The chains are binned in mass bins of $\Delta \log_{10}(m_a/{\rm eV}) = 0.5.$ The bars give the $95\%$ upper percentile of the axion fraction. The solid line in each bar shows the location of the 50\% percentile of the chain, and the two dotted lines show the 84\% and 16\% percentiles respectively.   \label{fig:bayes_freq}}
\end{center}
\end{figure*}

The marginalized two-dimensional $m_a-(\Omega_{a}/\Omega_{d})$ plane allows one to visualize the degeneracy between the fraction and mass concretely. While a global limit on the axion fraction (as a function of $m_{a}$) is interesting, one might also ask a related question - in a narrowly defined mass bin, what are the limits on the fraction, and how do these compare to the constraints in the two-dimensional $m_a -(\Omega_{a}/\Omega_{d})$ plane?

We compare the one-dimensional limit computed over a range of masses [and within a mass bin of $\Delta  \log_{10}(m_a/{\rm eV})= 0.5$] to the marginalized, two-dimensional, global contours in Figure~\ref{fig:bayes_freq}. The mass-binned method is quasi-frequentist, while the full two-dimensional contours are fully Bayesian. We see that the 95\% constraints closely agree between these two methods. This is further evidence that we have adopted a consistent methodology to sample and constrain the challenging ULA parameter space.

While the global chain constraints are computed for chains that have been added and re-weighted using the prescription described above (and are indicated by the solid lines), the individual constraints in a mass bin (indicated by the bar chart) do not take the relative prior volume into account. The one-dimensional limits are thus tighter than the full $n$-dimensional case in the tightly constrained mass range, as the extra $n-1$ degrees of freedom have been integrated out, while the marginalized two-dimensional contours have only integrated out $n-2$ degrees of freedom. It is, however, not surprising that the limits are still largely consistent between the two treatments of the chains.

\subsection{Constraining the axion decay constant}

\begin{figure}[htbp!]
\begin{center}
$\begin{array}{@{\hspace{0.0in}}l}
\includegraphics[width=0.50\textwidth]{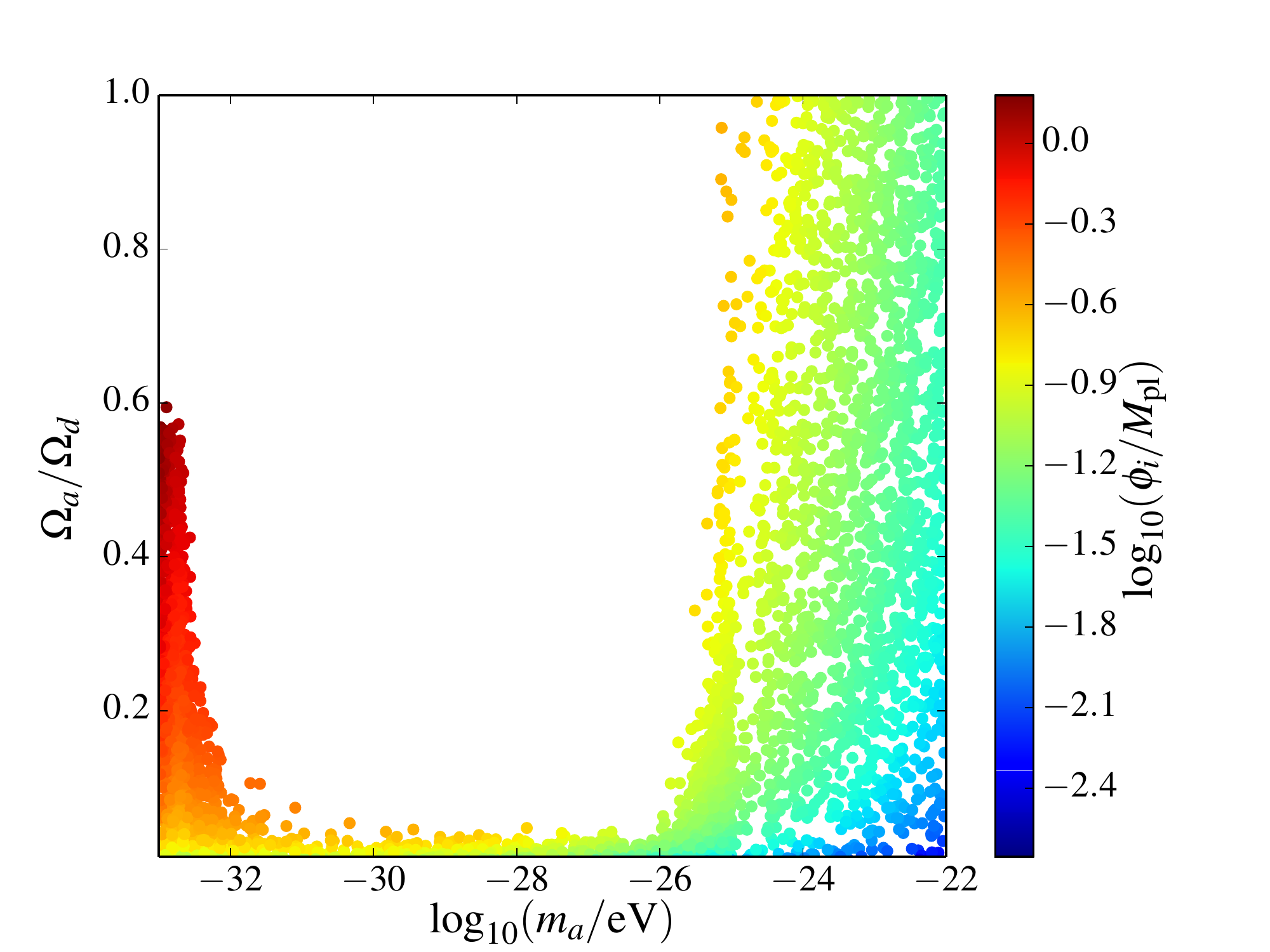}\\[0.0cm]
 \end{array}$
\caption{The $m_a-\Omega_a/\Omega_d$ parameter space showing sample points for the CMB-only data, colored by the initial field displacement $\phi_i/M_{pl}$. All points satisfy $\phi/M_{pl}<\pi$ and so are consistent with sub-Planckian decay constants, $f_a<M_{pl}$, and the Weak Gravity Conjecture. Most points satisfy $\phi_i/M_{pl}<1$ and so are consistent with $f_a<M_{pl}$. Regions with $\phi_i/M_{pl}<0.01$ are consistent with a GUT-scale decay constant with no need for additional production mechanisms. \label{fig:phi} }
 \end{center}
\end{figure}

Finally, we investigate the significance of our constraints for the axion decay constant, $f_a$, tuning of initial conditions, and models of axion production. In Fig.~\ref{fig:phi}, we plot points from a \textsc{MultiNest} chain in the $m_a-\Omega_a/\Omega_d$ plane colored by the value of the initial field displacement $\phi_i/M_{pl}$. As already discussed, $\phi_i$ is a derived parameter in our chains, found by using a shooting method to obtain the correct axion relic-density from the vacuum realignment mechanism. 

For any fixed value of $f_a$, we can divide the plane up according to the value of $\phi_i$. Regions with $\phi_i/f_a<1$ are consistent with the $m_a^2\phi^2$ approximation to the potential with no need for anharmonic effects or other additional production mechanisms. On the other hand, regions with $\phi_i/f_a<10^{-3}$ might be said to be tuned, like the anthropic window for the QCD axion.

In most of the plane the initial field displacement is small in Planck units, and can therefore be accommodated within the $m_a^2\phi^2$ approximation for the axion potential with sub-Planckian decay constant, $f_a<M_{pl}$. In particular, this applies to the constrained intermediate mass region. Regions where $\phi_i/M_{pl}<0.01$ are consistent with $f_a\lesssim 10^{16}\text{ GeV}$ with no need for additional production mechanisms. For $\phi_i/M_{pl}>0.01$ a larger value of $f_a>10^{16}\text{ GeV}$, anharmonic effects, multiple degenerate axions, or other production mechanisms are necessary to obtain the larger values of the relic density \cite{Kamionkowski:2014zda}.

The only region favoring $\phi_i/M_{\rm pl}>1$ is at low ULA mass [$m_{a}\lsim 10^{-32}~{\rm eV}$ and large density fraction ($\Omega_a/\Omega_d\gsim 10^{-1}$)]. In this regime, ULAs drive todayÕs accelerated cosmic expansion. Even so, all sample points respect the bound $\phi_i/M_{pl}<\pi$ and so everywhere we are consistent with $f_a<M_{pl}$ for the simple choice of a cosine potential and small anharmonic corrections. Our results are therefore consistent with the WGC described in Sec. \ref{model}.

\section{Discussion and Conclusions}
\label{conclusions}

It has become clear that certain particles and fields in cosmology supply us with a powerful portal into fundamental physics. Recent developments in neutrino physics are a prime example, with future high resolution measurements primed to measure the neutrino mass hierarchy with extraordinary precision \cite{DeBernardis:2009di}. The presence of ultralight axions in cosmology can also lead to constraints on new mass scales in particle physics, as well as on the dynamics of the early Universe.

In this paper we have presented the first ever cosmological search for ultralight axions using a fully self-consistent Boltzmann code, modern Bayesian statistical methods (including nested sampling), as well as state-of-the-art CMB and LSS data. We have derived constraints in the eight-dimensional parameter space of $\{\Delta_{\mathcal{R}}^{2},n_s,H_{0},\tau_{\rm re},\Omega_b h^2, \Omega_c h^2, \Omega_a h^2, m_a\}$, exploring all possible degeneracies, as well as those including foregrounds. 

We have presented these constraints marginalized down to one or two-dimensional spaces. Our main results are shown in Figs.~\ref{fig:cont_2d},~\ref{fig:1d_like}, and \ref{fig:bayes_freq}, as well as in Table~\ref{table:likes}. We show that axions in the mass range $10^{-32}\text{ eV}\leq m_a\leq 10^{-25.5}\text{ eV}$ must contribute $\Omega_a/\Omega_d<0.048$ at 95\% confidence (CMB only) and  $\Omega_a/\Omega_d<0.049$ at 95\% confidence (CMB + WiggleZ). Large fractions are allowed outside this regime: for $m_a\lesssim 10^{-32}\text{ eV}$ axions become indistinguishable from dark energy, while for $m_a\gtrsim 10^{-25.5}\text{ eV}$ axions become indistinguishable from CDM. For the case of CMB+WiggleZ data, this turnover from the constrained to the dark-matter like region occurs at a higher mass, as we can see in Fig.~\ref{fig:cont_2d}.

This interesting and challenging axion parameter space required the use and development of new techniques. In order to solve for the affect of axions on the cosmological observables in a fully consistent manner, we developed code to solve not only for the background but for the perturbations in the axions. To that end, we modelled axions as a perfect fluid with an equation of state and a sound speed, modifying \textsc{CAMB} to consistently account for axions.

Sampling the axion space is challenging. The unusually shaped parameter space caused standard Metropolis-Hastings chains to get  stuck in the middle region of intermediate mass, preventing them from climbing the ``walls" of the U-shaped distribution in the axion mass-axion fraction plane. We were able to improve sampling by using \textsc{Multinest}. The final chains, however, were under-sampled in precisely the intermediate regime, as \textsc{Multinest} is designed to find the largest-volume allowed regions. We tackled the problem by performing \textsc{Multinest} runs restricted to three mass ranges, and then combined the chains using information from a global, more a coarsely sampled run to weight the individual, ``local" chains. This allowed us to closely probe all regions of interest while including information about the relative probabilities of the three separate mass ranges explored.

There are many open avenues to extend our analysis. Preliminary investigations of CMB lensing data suggest it will be possible to increase the constraint on $m_a$ by an order of magnitude or more using the $\ell\sim 1000$ measurement of the lensing potential power-spectrum by the ACT \cite{dasetal:2013} and SPT collaborations \cite{spt}. Galaxy lensing data will complement the CMB deflection data \cite{marsh2011b}. Lensing data will impose $\sim 1\%$-level constraints on the axion energy-density using well-understood linear physics. These constraints will strengthen cruder and more systematic-limited constraints from galaxy formation and reionization \cite{Bozek:2014uqa}. Including isocurvature perturbations will allow us to place constraints on the energy scale of inflation independently of the $B-$mode polarization. Axion-type isocurvature is sensitive to extremely low-scale inflation inaccessible to searches for tensor modes. The combination of more accurate $E-$mode polarization measurements from {\em Planck} in the interim and AdvACT \cite{Calabrese:2014gwa} will place the strongest bounds on isocurvature and lensing. We do not include additional constraints on the BAO angular scale from SDSS \cite{reid/etal:2012}, but leave a detailed comparison of constraints from different probes of LSS to future work. One might also consider including more varied inflation scenarios with axions, for example changing the shape of the primordial power-spectrum. No additional modifications to our version of \textsc{camb} will be required to explore these possibilities.

We have not included various other extended sets of well-motivated cosmological parameters, which may have interesting degeneracies with axions. These include curvature, $\Omega_k$, dark-energy equation-of-state variables, $(w_0,w_a)$, and extended neutrino-sector parameters, $(N_{\rm eff},\Sigma m_\nu)$. The version of \textsc{camb} developed for this work will require additional modifications to accommodate these parameters. 

Cosmological observations are now narrowing in on the minimum neutrino mass scale consistent with oscillation experiments, and so the degeneracies of axion and neutrino parameters is particularly interesting \cite{marsh2011b}. Some tensions between CMB and LSS-derived parameters may be resolved by neutrino mass (see Ref.~\cite{Beutler:2014yhv} and references therein), but perhaps ultralight axions offer better resolutions than neutrinos to these tensions. 

We have used precision cosmological data to search for ultralight axions. Although we have found no evidence for axions yet, our results place strong and robust constraints to axion parameter space. Axions are well-motivated dark-matter candidates in string theory and particle physics. We have probed ranges of axion parameter space inaccessible to other searches. We have developed powerful computational tools to allow our analysis to be extended and applied to future data. Our techniques demonstrate the power of cosmological data not only to indicate the existence of dark matter and dark energy, but also to constrain the detailed physics of the dark sector.

\begin{acknowledgments}
DJEM acknowledges the hospitality of the Department of Astronomy at Princeton University, the Institute for Advanced Study, the BIPAC, Oxford, and KICP Chicago. DG and RH thank the Perimeter Institute for hospitality, where some of this work was completed. We are grateful for useful discussions with L.~Amendola, C.~Burgess, F-Y.~Cyr-Racine, R.~de Putter, R.~Flauger, A.~Liddle, O.~Mena, P.~Pani, D.~Parkinson, J.~Patterson, A.~Pontzen, L.~Price, B.~Reid, P.~Scott, K.~Sigurdson, and T.~L.~Smith. We especially thank Y. Ali-Ha\"{i}moud, E.~Calabrese and J.~Dunkley for carefully reading the manuscript and providing useful feedback.\\ \\
DJEM's research at Perimeter Institute is supported by the Government of Canada through Industry Canada and by the Province of Ontario through the Ministry of Research and Innovation. DG was supported at the Institute for Advanced Study by the National Science Foundation (AST-0807044) and NASA (NNX11AF29G). DG is funded at the University of Chicago by a National Science Foundation Astronomy and Astrophysics Postdoctoral Fellowship under Award NO. AST-1302856. This work was supported in part by the Kavli Institute for Cosmological Physics at the University of Chicago through grant NSF PHY-1125897 and
an endowment from the Kavli Foundation and its founder Fred Kavli. PGF was supported by STFC, BIPAC and the Oxford Martin School. RH is supported by a Spitzer fellowship at Princeton.
 \end{acknowledgments}

\renewcommand{\theequation}{A\arabic{equation}}
\setcounter{equation}{0} 
\appendix
\begin{widetext}
\renewcommand{\theequation}{A\arabic{equation}}
\section{Suppression of clustering by Axions- Jeans, Hubble and de Broglie}
\label{hubble_debroglie}

A heuristic understanding of the suppression of clustering in an expanding Universe containing axions is possible, using a simple argument that relates the Jeans scale to the de Broglie scale using only the Hubble expansion. Consider a particle of mass $m$ moving with the Hubble flow, $H$, separated by a distance $r$ from an observer. In the observer's frame of reference the particle is moving with a velocity
\begin{equation}
v = H r \, .
\end{equation}
According to the observer, this velocity gives the particle a de Broglie wavelength
\begin{equation}
\lambda_{\rm dB} = \frac{1}{mv} = \frac{1}{mHr} \, .
\label{eqn:deBroglie_hubble}
\end{equation}

As the particle moves further away from the observer and the distance $r$ increases, the speed at the which the particle is moving relative to the observer also increases. The de Broglie wavelength therefore decreases, and the particle can be localized on smaller scales. The particle can only be localized within the celestial sphere of radius $r$ when the following inequality is obeyed
\begin{equation}
r\geq\lambda_{\rm dB}\, ,
\label{eqn:deBroglie_ineq}
\end{equation} 
Substituting for $\lambda_{\rm dB}$ in Eq.~(\ref{eqn:deBroglie_hubble}) we find that a particle moving with the Hubble flow can be localized on all scales $r$ satisfying
\begin{equation}
r\gtrsim (mH)^{-1/2} \, .
\end{equation}

Identifying the wave number $k=\pi/r$ we find that the above inequality is saturated at
\begin{equation}
k_\star = \pi \sqrt{mH} \, .
\end{equation}
For all $k\gtrsim k_\star$ quantum mechanics prevents an observer from localizing a particle moving with the Hubble flow, and so the clustering of particles is forbidden at large wave number. When $m$ is small and $k_\star$ is cosmologically observable, this leads to an observable suppression of power relative to the case where $m$ is large. 

It is now simply a numerological fact to observe that $k_\star\sim k_{\rm J}$ for axion DM, where $k_J$ is the Jeans scale. Is this a coincidence? A simple argument suggests not. The Jeans scale is derived by taking the nonrelativistic limit and short-long time-scale separation of the Klein-Gordon equation, transforming into fluid form and identifying the sound speed from the pressure term. The de Broglie wavelength emerges from the wave-like properties of the Schr\"{o}dinger equation. The Schr\"{o}dinger equation is, however, also a description of the same limits of the Klein-Gordon equation \cite{Widrow:1993qq}. Transforming between the Schr\"{o}dinger and fluid pictures introduces the quantum pressure that is responsible for the sound speed and thus the Jeans scale. This suggests that the two interpretations are related, if not equivalent.

The difference between our heuristic derivation of $k_\star$ above and the Jeans analysis is that $H$ only appears in the Jeans analysis along with the correct power of $k$ after applying the Poisson equation. The Jeans analysis thus depends on perturbation theory while our heuristic argument depends only on the background expansion. In a modified theory of gravity it is therefore possible that the two scales $k_\star$ and $k_{\rm J}$ will not coincide.
\section{Derivation of power series initial conditions}
\label{icinit}
Series solutions for the fluid+Einstein equation system laid out in Refs. \cite{bertschinger1995,bucher2000} may be obtained by
applying power series expansions in $\tau$ and $x=k\tau$ to the system. This expansion is valid for super-horizon modes, a valid assumption since the \textsc{camb} code begins mode integration well outside the horizon. These equations are derived in the tight-coupling regime, valid at early times. The  solution method originally used to obtain the power series solutions in Refs. \cite{bertschinger1995,bucher2000} is not specified, but these solutions are readily (if tediously) obtained using a linear eigenmode analysis, as first sketched in Refs. \cite{cambnotes,doran_wetterich} and discussed in Ref. \cite{tristan_notes}. Here we review this analysis, including the evolution of the scalar field in a mixed matter-radiation background and its influence on the gravitational field through Einstein's equations. Here we compute and state values for all the other fluid and metric variables as a function of the dimensionless conformal time $\tau_{\rm b}$.

\subsection{Framework for obtaining series solution to Einstein+fluid system}
The synchronous gauge axion EOMs in terms of fluid variables are stated in Sec. \ref{axion_fluid_pert}. All the other fluid equations and Einstein equations are given by well-known expressions in Refs. \cite{bertschinger1995,bucher2000}, with additional axion source terms given by Eqs.~(\ref{dpdef})-(\ref{dqdef}). If the full system of differential equations can be written in the form
\begin{eqnarray}
\frac{d\vec{U}_{\vec{k}}}{d\ln x}=\left(\underline{A}_{0}+\underline{A}_{1}x+...\underline{A}_{n}x^{n}\right)\vec{U}_{k}\label{sa1}
\end{eqnarray}
where $x=k \tau$, $k$ is the wave number and $\tau$ is the conformal time and $\vec{U}_{k}$ is the Fourier transform of the vector of all fluid+metric variables of interest, then the space of solutions is spanned (to lowest order) by the eigenvectors $\vec{U}_{k}^{\alpha}$ (with eigenvalue $\alpha$) of $\underline{A}_{0}$:\begin{eqnarray}
\vec{U}_{k}(\tau)=\sum_{\alpha} c_{\alpha} x^{\alpha}\vec{U}_{k}^{\alpha}.\label{sa2}
\end{eqnarray} Here $c_{\alpha}$ are coefficients setting the contribution of each eigenmode to the solution, and are chosen so that fluid variables match initial conditions. As we shall see, lowest-order solutions often yield zero values for certain variables, and we desire an expansion that yields the first nonzero components for all fluid quantities. Around each eigenmode, we can extend each eigenmode to a solution $\mathcal{U}_{\vec{k}}^{\alpha}(\tau)$ including higher order corrections: \begin{eqnarray}
\mathcal{U}_{\vec{k}}^{\alpha}(\tau)=U_{\vec{k}}^{\alpha}x^{\alpha}+U_{\vec{k},\left(1\right)}^{\alpha}x^{\alpha+1}+...U^{\alpha}_{\vec{k},\left(i\right)}x^{\alpha+i}+....\label{sa3}.
\end{eqnarray}
We derive the corrections to the lowest-order solution by applying Eq.~(\ref{sa1}) to the \textit{ansatz}, Eq~(\ref{sa3}), obtaining \cite{doran_wetterich}:
\begin{eqnarray}
\left[\left(\alpha+1\right)\mathcal{I}-\underline{A}_{0}\right]\vec{U}^{\alpha}_{\vec{k},\left(1\right)}&=&\underline{A}_{1}\vec{U}^{\alpha}_{\vec{k}}\label{u1eqa},\\
\left[\left(\alpha+2\right)\mathcal{I}-\underline{A}_{0}\right]\vec{U}^{\alpha}_{\vec{k},\left(2\right)}&=&\underline{A}_{1}\vec{U}^{\alpha}_{\kappa,\left(1\right)}+\underline{A}_{2}\vec{U}^{\alpha}_{\vec{k}},\label{u2eqa}\\
\left[\left(\alpha+3\right)\mathcal{I}-\underline{A}_{0}\right]\vec{U}^{\alpha}_{\vec{k},\left(3\right)}&=&\underline{A}_{1}\vec{U}^{\alpha}_{\vec{k},\left(2\right)}+\underline{A}_{2}\vec{U}^{\alpha}_{\vec{k},\left(1\right)}\nonumber \\&+&\underline{A}_{3}\vec{U}^{\alpha}_{\vec{k},\left(1\right)},\label{u3eqa}\\
\left[\left(\alpha+4\right)\mathcal{I}-\underline{A}_{0}\right]\vec{U}_{\vec{k},\left(4\right)}^{\alpha}&=&\underline{A}_{1}\vec{U}^{\alpha}_{\vec{k},\left(3\right)}+\underline{A}_{2}\vec{U}^{\alpha}_{\vec{k},\left(2\right)}+\underline{A}_{3}\vec{U}^{\alpha}_{\vec{k},\left(1\right)}\nonumber \\&+&\underline{A}_{4}\vec{U}^{\alpha}_{\vec{k}}.\label{u4eqa}
\end{eqnarray} Here $\mathcal{I}$ is the identity matrix in the space of all fluid variables. The solutions to this linear system can yield higher-order corrections to the time-evolution of the fluid variables for each eigenmode. 

\subsection{Fluid and Einstein equations in convenient variables for eigenmode analysis}
We work in coordinates where the scale factor at equality $a_{\rm eq}=1/4$ by definition and $\tau_{\rm b}\equiv \mathcal{C} \tau$ with $\mathcal{C}^{2}=4\pi G \rho_{\rm eq}a_{\rm eq}^{4}/4$ (where $\rho_{\rm eq}$ is the radiation energy-density at matter-radiation equality). For our purposes, `matter'-radiaiton equality is defined by the relationship:
\begin{equation}
\rho_{a}+\rho_{b}+\rho_{c}=\rho_{\gamma}+\rho_{\nu},\end{equation} where $\rho_\gamma$ and $\rho_{\nu}$ are the energy densities of photons and neutrinos, while $\rho_{b}$ and $\rho_{c}$ are the energy densities of baryons and CDM, respectively.

The solution to the Friedmann equation at early times ($\rho_{a}\ll \rho_{\rm m},\rho_{a}\ll \rho_{\rm rad}=\rho_{\gamma}+\rho_{\nu}$, $a\ll a_{\rm osc}$) is
\begin{eqnarray}
a&=&\tau_{\rm b}+K\tau_{\rm b}^{2},\label{homosol_early_a}\\
K&=&\left\{\begin{array}{ll}
\left(1-f\right)&\mbox{if $a_{\rm osc}\leq a_{\rm eq}$}\\
\frac{\left(1-f_{\rm NR}\right)}{\left(1-f_{\rm NR}\right)+f_{\rm NR}a_{\rm eq}^{3}/a_{\rm osc}^{3}}&\mbox{if $a_{\rm osc}>a_{\rm eq}$}.
\end{array}\right.\\
f_{\rm NR}&=&\Omega_{a}/(\Omega_{a}+\Omega_{m}).\label{homosol_early_b}\end{eqnarray} These are the same conventions for conformal time and expansion history employed in Ref. \cite{bucher2000}, facilitating ease of comparison with the expansions derived in that work. The one distinction between the early-time expansion history here and in Ref. \cite{bucher2000} is that we have self-consistently allowed for axions to make up such a high fraction $f_{\rm NR}$ of the nonrelativistic matter density today, that if $\tau\ll \tau_{\rm osc}$, the nonrelativistic matter density is reduced from what it would have been if there were no axions (since axions act like a cosmological constant at such early times). The axion-free case corresponds to the choice $K=1$. 

We use a dimensionless wave number $\kappa=k/C$ so that $x=k\tau=\kappa \tau_{\rm b}$, dimensionless velocities $\tilde{t}_{i}\equiv\theta_{i}/\left(\mathcal{C}\kappa x^{2}\right)$, and rescaled density contrasts $\tilde{\delta}_{i}\equiv\delta_{i}/x$. The axion velocity $u_{a}$ is already dimensionless, so we define $\tilde{u}_{a}=u_{a}/x^{2}$. We also define a metric velocity $\Theta\equiv \beta^{\prime}$, where the derivative $^{\prime}\equiv\kappa^{-1} d/d\tau_{\rm b}$. It is useful to rescale higher-order moments in the neutrino hierarchy using $\tilde{\sigma}_{\nu}\equiv\sigma_{\nu}/x$ and $\tilde{F}_{\nu}^{\left(3\right)}\equiv F_{\nu}^{\left(3\right)}/x^{2}$. We now reexpress the synchronous gauge fluid+Einstein equation system from Refs. \cite{bertschinger1995,bucher2000}, using the choice of variables just described and adding the axion EOMs and source terms of Eqs.~(\ref{eoma})$-$(\ref{eomb}) and Eqs.~(\ref{dpdef})$-$(\ref{dqdef}), obtaining a system solvable using Eqs.~(\ref{sa1})$-$(\ref{sa3}) and Eqs.~(\ref{u1eqa})$-$(\ref{u4eqa}): 
\begin{eqnarray}
\tilde{\delta}_{\gamma}^{\prime}&=&-\tilde{\delta}_{\gamma}-\frac{4}{3}\tilde{t}_{\gamma\rm b}x^{2}-\frac{2\Theta}{3},\\
\tilde{\delta}_{\nu}^{\prime}&=&-\tilde{\delta}_{\nu}-\frac{4}{3}\tilde{t}_{\nu}x^{2}-\frac{2\Theta}{3},\\
\tilde{\delta}_{c}^{\prime}&=&-\tilde{\delta}_{c}-\tilde{t}_{c}x^{2}-\frac{\Theta}{2},\\
\tilde{\delta}_{\rm b}^{\prime}&=&-\tilde{\delta}_{\rm b}-\tilde{t}_{\gamma\rm b}x^{2}-\frac{\Theta}{2},\\
\tilde{t}_{\gamma\rm b}^{\prime}&=&-2\tilde{t}_{\gamma \rm b}+\frac{\tilde{\delta}_{\gamma}}{4\left[\frac{3R_{b}\left(x/\kappa \right)\left(Kx/\kappa +1\right)}{R_{\gamma}}+1\right]}-\frac{
\frac{\left(2Kx/\kappa +1\right)}{\left(x/\kappa +1\right)}\tilde{t}_{\gamma}\frac{3R_{\rm b}}{R_{\gamma}}\left(x/\kappa \right)\left(xK/\kappa +1\right)}{\left[\frac{3R_{\rm b}}{R_\gamma}\left(x/\kappa \right)\left(xK/\kappa +1\right)+1\right]},\\
\tilde{t}_{\nu}^{\prime}&=&-2\tilde{t}_{\nu}+\frac{\tilde{\delta}_{\nu}}{4}-\tilde{\sigma}_{\nu},\\
\tilde{t}_{c}^{\prime}&=&-2\tilde{t}_{c}-\frac{\left(2xK/\kappa +1\right)}{\left(xK/\kappa +1\right)}\tilde{t}_{c},\\
\tilde{\sigma}_{\nu}^{\prime}&=&-\tilde{\sigma}_{\nu}+\frac{4 \tilde{t}_{\nu}x^{2}}{15}-\frac{3\tilde{F}_{\nu}^{\left(3\right)}x^{2}}{10}+\frac{2\Theta}{15}+\frac{8\left(R_{\gamma}\tilde{t}_{\gamma}+R_{\nu}\tilde{t}_{\nu}\right)}{5\left(1+xK/\kappa \right)^{2}}+\frac{24x\left(R_{c}\tilde{t}_{c}+R_{\rm b}\tilde{t}_{\rm b}\right)}{5\kappa \left(1+xK/\kappa \right)}\nonumber \\&+&\frac{16\pi Gx^{4}}{5C^{2}\kappa^{4}}\left(1+Kx/\kappa\right)^{2}\rho_{a}\tilde{u}_{a},\\
\tilde{F}_{\nu}^{\left(3\right)}&=&-2\tilde{F}_{\nu}^{\left(3\right)}+\frac{6\tilde{\sigma}_{\nu}}{7},\\
\Theta^{\prime}&=&-\frac{\left(2xK/\kappa +1\right)}{\left(xK/\kappa +1\right)}\Theta-\frac{6\left(R_{\gamma}\tilde{\delta}_{\gamma}+R_{\nu}\tilde{\delta}_{\nu}\right)}{\left(1+xK/\kappa \right)^{2}}-\frac{12x\left(R_{c}\tilde{\delta}_{c}+R_{\rm b}\tilde{\delta}_{\rm b}\right)}{\kappa \left(1+xK/\kappa \right)}-\frac{32\pi G a^{2}\rho_{a} x^{2} \tilde{\delta}_{a}}{C^{2}\kappa^{2}}\nonumber \\ &-&\frac{72\pi G a^{2}\rho_{a}x^{2}}{C^{2}\kappa^{2}}\tilde{u}_{a}\left(1-c_{\rm ad}^{2}\right)\left(\frac{1+2Kx/\kappa}{1+Kx/\kappa}\right),\label{fes1}\\
\eta^{\prime}&=&\frac{2x}{\left(1+xK/\kappa \right)^{2}}\left(R_\gamma\tilde{t}_\gamma+R_\nu\tilde{t}_\nu\right)+\frac{6x^{2}}{\kappa \left(1+xK/\kappa \right)}\left(R_{\rm b}\tilde{t}_{\rm b}+R_{\rm c}\tilde{t}_{c}\right)+\frac{4\pi G x^{5}}{C^{2}\kappa^{4}}\left(1+Kx/\kappa\right)^{2}\rho_{a}\tilde{u}_{a},\label{fes2}\\
\tilde{\delta}_{a}^{\prime}&=&-\tilde{\delta}_{a}-\left(1+w_{a}\right)\frac{\Theta}{2}-\frac{3\left(1+2Kx/\kappa\right)}{\left(1+Kx/\kappa\right)}\left(1-w_{a}\right)\tilde{\delta}_{a}-9\left(1-c_{a}^{2}\right)\tilde{u}_{a}\frac{\left(1+2Kx/\kappa\right)^{2}}{\left(1+Kx/\kappa\right)^{2}},\\
\tilde{u}^{\prime}_{a}&=&\frac{2\left(1+2Kx/\kappa\right)}{\left(1+Kx/\kappa\right)}\tilde{u}_{a}+\tilde{\delta}_{a}-2\tilde{u}_{a}+\frac{w_{a}^{\prime}\tilde{u}_{a}x}{1+w_{a}},\\
\mathcal{A}&=&\frac{\rho_{\rm a}}{a_{0}^{4}\Omega_{r}\rho_{\rm crit}}.
\end{eqnarray}In these expressions $^{\prime}=d/d\ln{x}$ and $a_{0}$ is the scale factor today under this convention:
\begin{equation}
a_{0}=\left\{
\begin{array}{ll}
\frac{\Omega_{m}}{4\Omega_{r}}\left \{\left(1-f_{\rm NR}\right)+f_{\rm NR}\left(\frac{a_{\rm eq}}{a_{\rm osc}}\right)^{3}  \right\}&\mbox{if $a_{\rm osc}>a_{\rm eq}$,}\\ 
\frac{\Omega_{m}}{4\Omega_{r}}&\mbox{if $a_{\rm osc}\leq a_{\rm eq}$}.\end{array}\right.
\end{equation}

The neutrino energy-density fraction is defined to be
\begin{equation}
R_{\nu}=\Omega_{\nu}/(\Omega_{\nu}+\Omega_{b}),\end{equation} and we assume for this work that all standard model neutrinos are massless. Conversely, the photon energy-density fraction (defined relative to the total energy density in relativistic species) is $R_{\gamma}=1-R_{\nu}$. 

Equation (\ref{fes1}) is obtained from a linear combination of the Einstein equations \cite{bertschinger1995,bucher2000}
\begin{eqnarray}
k^{2}\eta-\frac{1}{2}\frac{\dot{a}}{a}\dot{\beta}=-4\pi G a^{2}\delta \rho,\label{einstein_constraint}\\
\ddot{\beta}+2\frac{\dot{a}}{a}\dot{\beta}-2k^{2}\eta=-24\pi G a^{2} \delta P,\label{einstein_dynamic}
\end{eqnarray} where $\delta \rho$ and $\delta P$ are the total energy density and pressure perturbations, respectively. Eq.~(\ref{fes2}) is obtained
from the Einstein equation \cite{bertschinger1995,bucher2000}
\begin{equation}
k^{2}\dot{\eta}=4\pi G \sum_{i}\left(\overline{\rho}+\overline{P}\right)_{i}\theta_{i}+4\pi G u_{a},\end{equation}
where the sum on $i$ is over all the conventional fluid species. The axion energy-density $\rho_{\rm a}$ has some time dependence, which we compute below, where we also obtain the time evolution of the axion EOS $w_{a}$ and adiabatic sound speed $c_{\rm ad}$.

\subsection{Homogeneous scalar field evolution in a mixed (matter+radiation) Universe}
To obtain power series solutions for the initial conditions, we must compute the squared adiabatic sound speed $c_{\rm ad}^{2}$ and scale factor $w_{a}$ as a function of conformal time, using Eq.~(\ref{adiabat_cs}),
evaluating Eqs. (\ref{eqn:rhoa})$-$(\ref{eqn:pa}), and using the field evolution as specified by Eq.~(\ref{homo_eom}).  Since we are in the regime
 $\rho_{a}\ll \rho_{m},\rho_{a}\ll \rho_{r}, a\ll a_{\rm osc}$, we may use Eqs.~(\ref{homosol_early_a})$-$(\ref{homosol_early_b}) to evaluate the conformal Hubble parameter $\mathcal{H}$. Making a power series expansion in the dimensionless conformal time $\tau_{\rm b}$, we obtain the desired results from the solution for the homogeneous field $\phi_{0}\left(\tau_{\rm b}\right)$ :
 \begin{align}
 w_{a}=&-1+\frac{2m^{2}\tau_{\rm b}^{4}}{25\mathcal{C}^{2}}+\frac{4Km^{2}\tau_{\rm b}^{5}}{75\mathcal{C}^{2}}+...,\label{weq_early}\\
 c_{\rm ad}^{2}=-&\frac{7}{3}+\frac{10K\tau_{\rm b}}{9}-\frac{520K^{2}\tau_{\rm b}^{2}}{189}+\frac{3445K^{3}\tau^{3}_{\rm b}}{567}+\left(\frac{-151465K^{4}}{11907}+\frac{2m^{2}}{27\mathcal{C}^{2}}\right)\tau^{4}_{\rm b}+\left(\frac{870025K^{5}}{35721}+\frac{26Km^{2}}{405C^{2}}\right)\tau_{\rm b}^{5}+....,\\
 \rho_{a}=&\rho_{a}^{\left(0\right)}\left[1-\frac{3m^{2}\tau_{\rm b}^{4}}{50\mathcal{C}^{2}} -\frac{2Km^{2}\tau_{\rm b}^{5}}{25\mathcal{C}^{2}}+...\right],
 \end{align}
 where $\rho_{a}$ is the asymptotic value of $\rho_{a}$ when $a\ll a_{\rm osc}$. Converting to physical (dimensional) conformal time via the substitution \begin{equation}\tau_{\rm b}= \Omega_{m}H_{0}\tau/\left(4\sqrt{\Omega_{r}}\right),\label{contiki}\end{equation} we see that Eq.~(\ref{weq_early}) agrees with the early time evolution of the quintessence equation of state derived in Ref. \cite{cambnotes}.
 \subsection{Modes of the system}
 To obtain the normal modes of the system, we make an expansion in both $\tau$, and $x$, valid for super-horizon deep into radiation domination. Using the assignment $\vec{U}_{k}=\left\{\tilde{\delta}_{\gamma},\tilde{\delta}_{\nu},\tilde{\delta}_{c},\tilde{\delta}_{b},\tilde{t}_{\gamma b},\tilde{t}_{\nu},\tilde{t}_{c},\tilde{\sigma}_{\nu},\tilde{F}_{\nu}^{3},\Theta,\eta,\tilde{\delta}_{a},\tilde{u}_{a}\right\}$, we determine the matrices $\underline{A}_{0},\underline{A}_{1},\underline{A}_{2},\underline{A}_{3},\underline{A}_{4}$. To check that our machinery is consistent with past work, we begin by restricting attention to the case where there are no axion perturbations, and the expansion history is not adjusted for the reduced matter density at early times due to axions rolling slowly. In this case, using Eqs.~(\ref{u1eqa})-(\ref{u4eqa}), we recover exactly the growing adiabatic, baryon isocurvature, CDM (cold DM) isocurvature, neutrino density isocurvature, neutrino velocity isocurvature modes, as well as a set of decaying modes, as stated in Refs. \cite{bucher2000}. 
 
The familiar adiabatic mode has eigenvalue $\alpha=1$ and corresponds to the initial condition
\begin{eqnarray}
\delta_{\gamma}=\delta_{\nu}=\frac{4}{3}\delta_{\rm c}=\frac{4}{3}\delta_{\rm b},\\
\delta_{i}=(1+w_{i})\delta_{c}.
\end{eqnarray} where $\delta_{\gamma}$, $\delta_{\nu}$, $\delta_{\rm c}$, and $\delta_{\rm b}$ are the fractional energy over-densities in photons, neutrinos, CDM, and baryons respectively. Since at early times, $w_{\rm a}=-1$, the adiabatic condition for axions implies $\delta_{\rm a}=0$ initially. In synchronous gauge, the corresponding power series solution (valid at early times) is \cite{bertschinger1995,bucher2000}
\begin{eqnarray}
\delta_{\gamma}&=&\delta_{\nu}=-\frac{\left(\kappa\tau_{\rm b}\right)^{2}}{3},\\
\delta_{\rm c}&=&\delta_{\rm b}=-\frac{\left(\kappa\tau_{\rm b}\right)^{2}}{4},\\
\frac{\theta_{\gamma}}{\mathcal{C}\kappa}&=&\frac{\theta_{\rm b}}{\mathcal{C}\kappa}=-\frac{\left(\kappa\tau_{\rm b}\right)^{3}}{36},\\
\frac{\theta_{\nu}}{\mathcal{C}\kappa}&=&-\frac{\left(23+4R_{\nu}\right)\left(\kappa\tau_{\rm b}\right)^{3}}{36\left(15+4R_{\nu}\right)},\\
\theta_{\rm c}&=&0,\\
\sigma_{\nu}&=&\frac{2\left(\kappa\tau_{\rm b}\right)^{2}}{3\left(15+4R_{\nu}\right)},\\
F_{\nu}^{\left(3\right)}&=&\frac{4\left(\kappa\tau_{\rm b}\right)^{3}}{21\left(15+4R_{\nu}\right)},\\
\delta_{a}&=&0,\\
u_{a}&=&0.\\
\beta&=&\frac{\left(\kappa \tau_{\rm b}\right)^{2}}{2},\\
\eta&=&1-\frac{\left(5+4R_{\nu}\right)\left(\kappa \tau_{\rm b}\right)^{2}}{12\left(15+4R_{\nu}\right)},
\end{eqnarray}where the metric perturbations $\beta$ and $\eta$ are defined as described in Ref. \cite{bertschinger1995,bucher2000}, as are the fluid perturbations.  The dimensionless conformal time is defined in Eq.~(\ref{contiki}). 

 We also confirm that it is valid up to corrections of order $(k\tau)^{4}$ for metric and standard fluid perturbations, and $\tau/\tau_{\rm eq}$ for the axion variables themselves, even when the contribution of axions to the energy density is included. Corrections to $\delta_{a}$ appear at order $(k\tau)^{4}$ for metric and standard fluid perturbations, and $\tau/\tau_{\rm eq}$ for the axion variables themselves, even when the contribution of axions to the energy density is included. The overall normalization of the perturbations at this stage of the analysis is arbitrary, but is eventually set by the power spectrum $P_{\mathcal{R}}(k)$ for the gauge-invariant curvature inside \textsc{camb}.

 \end{widetext}


%
\end{document}